\documentclass{jpp}
\usepackage{graphicx,xcolor}
\usepackage{epstopdf, epsfig}
\usepackage{hyperref}

\shorttitle{Linear Theory of Tearing Instability}
\shortauthor{T. Shimizu and K. Kondoh}

\title{A New Approach of Linear Theory of \\
Tearing Instability in Uniform Resistivity}

\author{T. Shimizu \aff{1}
  \corresp{\email{shimizu@cosmos.ehime-u.ac.jp}},
 \and K. Kondoh \aff{1}}

\affiliation{\aff{1}
Research Center for Space and Cosmic Evolution, 
Ehime University, Bunkyo Town 2, 
Matsuyama City, Ehime Prefecture, 790-8577, Japan}

\begin{document}

\maketitle

\begin{abstract}
The linear perturbation equation of the tearing instability derived in LSC theory (Loureiro, Schekochihin, and Cowley \citep{Lour2007}) is numerically examined as an initial value problem, where the inner and outer regions are seamlessly solved under uniform resistivity. Hence, all regions are solved as the resistive MHD (magnetohydrodynamics). To comprehensively study physically acceptable perturbation solutions, the behaviors of the local maximum points required for physically acceptable solutions and zero-crossing points, at which $\phi=0$ and $\psi=0$, are examined. Eventually, the uniform resistivity assumed in the outer region is shown to play an important role in improving some conclusions derived from the theory. 
In conclusion, the upper limit $\lambda_{up}$ of the growth rate obtained in the improved (modified) LSC theory is shown to be regulated by the Alfven speed measured in the outer region. It is also shown to be partially consistent with the growth rate in the linear developing stage of the impulsive tearing instability observed in the compressible MHD simulation of the plasmoid instability (PI) based on uniform resistivity. 
\end{abstract}

\section{Introduction}

% Fast magnetic reconnection

The magnetic reconnection process is an energy conversion mechanism from  
magnetic energy to plasma kinetic energy. 
To explain the explosive energy conversion 
observed in the solar flares and geomagnetic substorms, 
the magnetic reconnection process must be fast. 
The fast magnetic reconnection process has been studied 
during the past exceeding 60 years.

% Sweet-Parker model vs Petschek model

 Some theories of the magnetic reconnection process were 
proposed in the 1960s \citep{vasy1975}, 
some of which have survived until today. 
In particular, two famous theoretical MHD (magnetohydrodynamic) models, i.e., the Sweet-Parker (SP) model and the Petschek (PK) model, are well known steady-state models \citep{spr1957,pt1964}. 
In general, the PK model is much more efficient than the SP model \citep{vasy1975}; hence, the PK model is believed to be a candidate 
for explaining the fast magnetic reconnection process required for solar flares and geomagnetic substorms.

% FKR Tearing Instability

FKR (Furth, Killeen, and Rosenbluth \citep{fkr1963}) theory was also proposed in the 1960s to explore the tearing instability caused by the magnetic reconnection process. 
Since FKR theory is a linear theory, 
the magnetic reconnection process based on FKR theory 
may be developed to some nonlinear instability. 
If the nonlinear instability finally settles to a steady state, 
either the SP model or PK model will be applicable. 
Meanwhile, if the nonlinear instability maintains 
any non-steady state, the resulting magnetic reconnection process 
may be characterized by neither the SP model nor the PK model. 
Recently, plasmoid instability (PI) has been actively studied as such a non-steady state model 
\citep{Lour2007, Baal2012,  Lour2009a, Lour2009b, Bhat2009, cask2009,huang2012, huang2013, baty2014, Shiba2015, veli2015a, veli2015b, LandiApj2015, loure2016, Huang2017}.

% Numerical Simulations: spontaneous vs externally driven models

In the 1970s, two theoretical models, i.e., the spontaneous and externally driven models, were proposed to numerically reproduce the PK model by means of MHD simulations. In the spontaneous model \citep{mu1977,mu1984,mu2005,ts2003,shi2009a,shi2009b,shi2013,shi2016} proposed by Ugai, the locally enhanced nonuniform resistivity at X-point results in the PK model. Such nonuniform resistivity is called anomalous resistivity. Ugai proposed current-driven anomalous resistivity, which is self-consistently enhanced by a positive feedback mechanism between the plasma inflow and outflow to and from the magnetic diffusion region, leading to the appearance of the PK model. The spontaneous model does not require any additional mechanism to establish the PK model, with the exception of the current-driven anomalous resistivity. Moreover, the externally driven model \citep{sato1978} proposed by Sato also requires current-driven anomalous resistivity, also resulting in the PK model. In contrast to the spontaneous model, some externally driven mechanism in the upstream is required to maintain the PK model. If the externally driven mechanism is removed, the fast magnetic reconnection process terminates. Hence, the assistance of the externally driven mechanism is essential to maintain the PK model in the externally driven model, and whether the spontaneous model or externally driven model causes the fast magnetic reconnection process has become a controversial topic.

% Uniform vs Anomalous Resistivity Models

The magnetic reconnection process requires some form of resistivity to reconnect magnetic field lines at X-point. As shown by many numerical MHD studies, nonuniform resistivity, such as anomalous resistivity, results in the PK model. Moreover, when the resistivity is assumed to be uniform in time and space rather than nonuniform, whether the PK model can be generated or not is a delicate problem \citep{Shiba2015,bskp1986,kuls2001,kuls2005,baty2006,kuls2011}. If the uniform resistivity can reproduce the PK model, nonuniform resistivity is not necessarily required for the fast magnetic reconnection process. If so, one may say that the fast magnetic reconnection needs some form of resistivity but is independent of the resistivity. Then, the uniform versus nonuniform resistivity for the PK model is also currently a controversial topic.

% Plasmoid Instability (PI) as a new paradigm shift \citep{loure2016}

Separate from steady-state models, such as the PK and SP models, non-steady state models may also be explored. The plasmoid instability (PI) is a non-steady state and multiple-tearing instability model developed from FKR and LSC theories. In fact, LSC theory \citep{Lour2007} predicts that when the uniform resistivity is extremely weak, the linear growth rate can be extremely high. When the Lundquist number exceeds a critical value $S_c$, such a high-speed non-steady and multiple-tearing instability is caused, which is called PI. In fact, some recent highly sophisticated numerical MHD studies reported that, once PI occurs, the reconnection rate exceeds the value predicted by the steady-state SP model \citep{Lour2007, Baal2012, Lour2009b, Bhat2009, huang2012, huang2013, Shiba2015, loure2016}, suggesting that PI is another potential cause of the fast magnetic reconnection process of solar flares and substorms.

In contrast, recent numerical MHD studies \citep{Ng2010,shi2017} suggest that when numerical dissipations are sufficiently removed, the tearing instability tends to be less active; hence, it suggests that PI under uniform resistivity cannot cause the fast magnetic reconnection process, at least in the spontaneous model. In other words, it suggests that $S_c$ does not exist, except in the externally driven model. From another perspective, PI may be associated with the non-steady state PK model rather than the non-steady state SP model. If so, the controversial topic of the uniform versus nonuniform resistivity for the PK model \citep{Shiba2015} may be extended to the non-steady state PK model. 

The main theme of this paper is to propose the modified LSC theory as a new linear theory of tearing instability, improving the original LSC theory \citep{Lour2007}. In Section 2, LSC theory is numerically examined, where the inner and outer regions are seamlessly solved under uniform resistivity as an initial value problem \citep{shi2018a,shi2018b}. Then, the upper limit $\lambda_{up}$ of the linear growth rate is obtained for physically acceptable perturbation solutions. In addition, a candidate of the physically acceptable perturbation solutions, which is proposed in this paper as zero-crossing solution, is studied. In Section 3, the MHD simulation of PI under uniform resistivity is shown, where how the modified LSC theory can be applied to the tearing instability in PI is demonstrated. The modified LSC theory is then shown to be partially consistent with the MHD simulation of PI. In Section 4, the following topics are discussed. First, it is shown that the existence of $S_c$ is not directly supported by the modified LSC theories. Second, the original and modified LSC theories are compared for PI application. Third, the viscosity effect, which is not considered in LSC theory but is employed in the MHD simulation for numerical stabilization, is briefly discussed. Fourth, FKR theory viewed from the perspective of the modified LSC theory is discussed. Section 5 provides a summary of this study.

% (see \S\ref{sec:filetypes} 
% \citep{Rogallo81} name and year
% \citet{Rogallo81} only year

\section{LSC theory}\label{sec:LSC theory}

\subsection{Basic equation}

In this section, LSC theory \citep{Lour2007} is numerically examined as an initial value problem \citep{shi2018a,shi2018b}, where the inner and outer regions are seamlessly solved on the assumption of uniform resistivity. LSC theory is based on incompressible one-component MHD equations. The largest difference between the FKR and LSC theories is that the zero-order flow field is, respectively, null and nonzero. By assuming a nonzero flow field, the thickness of the current sheet in LSC theory can be finite, thus maintaining the rigorous equilibrium. LSC theory can examine how the tearing instability grows in the nonzero flow field generated by the steady-state SP current sheet. In other words, FKR theory is not related to the SP model, because of the null flow field. At this point, LSC theory is better and more realistic than FKR theory. The basic equations derived in LSC theory are as follows. 

\vspace{0.25cm}

\begin{equation}
\phi''-\kappa^2\epsilon^2\phi=-\kappa f(\xi) \psi +\kappa f^2(\xi) \phi/\lambda +f''(\xi)\psi/\lambda, 
\label{phi-eq}
\end{equation}

\begin{equation}
\psi''-\kappa^2\epsilon^2\psi=\kappa \lambda \psi-\kappa f(\xi)\phi.
\label{psi-eq}
\end{equation}

\vspace{0.25cm}

Eqs.(\ref{phi-eq}) and (\ref{psi-eq}) basically follow LSC's notations, 
where $\phi$ and $\psi$ are the scalar potential functions of the 
2D incompressible velocity and 2D magnetic fields, respectively. 
The prime of $\phi$ and $\psi$ indicates the derivative with respect to $\xi$, which is normalized by the characteristic width $\delta_{cs}$ of the current sheet, as $\xi=y/\delta_{cs}$. In contrast to the LSC's notations, $x$ and $y$ are exchanged with each other, to compare LSC theory with the MHD simulation examined in Section 3. $\delta_{cs}$ is close to the half thickness of the current sheet but is not exactly the distance between the center (i.e. origin $\xi=0$) and edge of the current sheet. The edge is slightly separated from $\xi=1$, and rather, is located at $\xi=1.307$, i.e., $y=1.307\delta_{cs}$.

Following the LSC's notation, $\lambda$ is the growth rate 
normalized by $l_{cs}/V_A$, where  $l_{cs}$ is the wave length of the plasmoid chain and $V_A$ is Alfven speed measured in the upstream magnetic field region. 
$\kappa=\pi L_{cs}/l_{cs}$ is the wave number along the current sheet, 
where $L_{cs}$ is the total length of the steady state SP sheet along the sheet. In Section 3, $L_{cs}$ is calculated from the spatial gradient of the outflow velocity $u_{x0}$ measured at X-point, as $du_{x0}/dx=2V_A/L_{cs}$, 
where $u_{x0}$ is defined, as below. 
Following LSC's procedures, the zero-order nonzero flow field and antiparallel magnetic field in the inner region of the current sheet, i.e., $y<1.307\delta_{cs}$, are defined as described below.

\begin{eqnarray}
B_{y0}=0,\\
B_{x0}(\xi)=V_A f(\xi), \\
u_{y0}=-\Gamma_0 y, \\
u_{x0}=+\Gamma_0 x, 
\end{eqnarray}

Here, $\Gamma_0=2 V_A /L_{cs}$ is set. In addition, $\epsilon=2 \delta_{cs}/L_{cs}=2/\sqrt{S}$ corresponds to the uniform resistivity, where $S$ is the Lundquist number and the steady state SP model is assumed. Then, $f(\xi)$ is defined as shown below.

\begin{equation}
f(\xi)=\xi_0 e^{-\xi^2/2} \int^\xi_0 dz e^{z^2/2},  
\end{equation}

where $\xi_0 = 1.307$. Next, the field in the outside of the current sheet, i.e., $y>1.307 \delta_{cs}=y_0$, is defined as described below. 

\begin{eqnarray}
B_{y0}=0,\\
B_{x0}(\xi)=V_A, \\
u_{y0}=-\Gamma_0 y_0, \\
u_{x0}=0. 
\end{eqnarray}

Accordingly, these four fields are constant in $\xi$ space, where $f(\xi)=1$ is set instead of Eq.(2.7). Eqs.(2.3)-(2.11) are the zero-order equilibrium, by which the resistive MHD equations are rigorously satisfied by the connection of $dB_{x0}/d\xi=0$ at $\xi=1.307$. 

It is important that Eqs.(\ref{phi-eq}) and (\ref{psi-eq}) can be solved as an initial value problem when $\phi(0)$, $\psi(0)$, $\phi'(0)$ and $\psi'(0)$ are given as the initial values for a set of $\kappa$, $\epsilon$, and $\lambda$. Here, $V_A=1$ is assumed, and $\phi(0)=0$, $\psi(0)=1$, and $\psi'(0)=0$ are set without the lack of generality of solutions, which are based  on the symmetricity at $\xi=0$. 
Then, $\phi'(0)$ is a control parameter to uniquely determine a solution, in addition to $\kappa$, $\epsilon$, and $\lambda$. 
Then, $\psi'(0)=0$ means when $\Delta'$ index employed in FKR theory is rigorously zero. 
The tearing instability in FKR theory is caused in $\Delta'>0$. 
At this point, $\Delta'$ index defined in LSC theory 
is different from that of FKR theory 
but a discontinuity of $\Delta'>0$ defined in FKR theory 
is assumed at the origin. 
Meanwhile, that in this paper is caused in $\psi''(0)>0$, 
as explained in the third paragraph of Section 2.2. It means that 
any discontinuity at $\xi=0$ is not assumed in this paper. 

To numerically solve $\phi$ and $\psi$,  the forward Euler method is employed. The numerical resolution is set in 
$0.001>\Delta \xi>0.00025$ to suppress numerical errors. 
For a much higher numerical resolution, i.e., $0.00025>\Delta \xi$, 
the convergence test of the numerical results was done in some typical cases. 
The following results are extremely sensitive for 
changing some control parameters but basically not changed, 
evenwhen the Runge-Kutta method is employed instead of the forward Euler method.

\subsection{Physically acceptable $\phi$ and $\psi$}

Depending on the control parameters of 
$\phi'(0)$, $\kappa$, $\epsilon$, and $\lambda$, a lot of 
various $\phi$ and $\psi$ solutions can be obtained by numerically solving 
the initial value problem. 
We must select "physically acceptable" sets of $\phi$ and $\psi$ 
from those various solutions. 
Such selection means to specify the physically acceptable upstream condition 
of $\phi$ and $\psi$. 
Simply, $\phi$ and $\psi$, which converge to zero at $\xi=+\infty$, will be the most preferable candidate of physically acceptable solutions. Let us call it the zero-converging solution. In addition, $\phi$ and $\psi$, which reach zero in 
$0<\xi \leq +\infty$, will be another preferable candidate. 
Let us focus on when $\phi=0$ and $\psi=0$ are simultaneously attained at 
a finite $\xi$ value. Let us call it the zero-crossing solution as 
we define $\xi=\xi_c$ at the zero-crossing point of $\phi=\psi=0$.

In general, there will be many other physically acceptable solutions. 
In other words, the zero-converging and zero-crossing solutions are not 
the whole physically acceptable solutions. 
Rather, if $\phi$ and $\psi$ do not diverge to infinity in the finite $\xi$ range, they also may be a candidate for physically acceptable solutions in the range. It should be noted that they include extreme cases, in which 
$\phi$ and $\psi$ diverge to infinity at $\xi=+\infty$. 
In such extreme cases, the linear growth rate may exceed a unity, leading to 
super-Alfvenic plasmoid (tearing) instability \citep{HuangPoP2013}. 
 It will be fairly difficult to numerically explore the total behaviors of 
such physically acceptable solutions. 
Hence, in this paper, let us focus only on 
the zero-converging solutions and zero-crossing solutions introduced above, 
where the linear growth rate dose not exceed a unity, resulting in 
sub-Alfvenic tearing instability, as shown from Figs.6 to 11(b).

In the zero-converging and zero-crossing solutions, 
it is important that $\phi$ and $\psi$ must have local maximum points 
in $0<\xi<+\infty$, 
because, $\phi$ must start with $\phi(0)=0$ and $\phi'(0)>0$ and 
must reach zero in $0<\xi \le+\infty$. 
Note that $\phi'(0)>0$ means the plasma inflow toward X-point 
and the outflow from O-point. 
Additionally, $\psi$ must start with $\psi(0)=1$ and $\psi'(0)=0$ and 
must reach zero in $0<\xi \le +\infty$. 
At this point, $\psi''(0)>0$ is additionally required to 
cause the tearing instability 
because $\psi''(0)>0$ means when the current sheet becomes thin 
in the vicinity of X-point, 
and hence, corresponds to when $\Delta'$-index is positive. 
However, $\Delta'=0$ is resolved through this paper, because of $\psi'(0)=0$. 
In other words, the unstable mode of tearing instability requires that $\psi$ always has a downward convex feature at origin, and hence, 
a region of $\psi'>0$, at least, in the vicinity of origin. 
This is why $\psi$ must have a local maximum point in $0<\xi<+\infty$. 

As a result, we must find $\phi$ and $\psi$, which have local maximum points in $0<\xi<+\infty$. However, in numerical studies, such as the initial value problem started from $\xi=0$, it is impossible to precisely explore the behaviors of $\phi$ and $\psi$ at $\xi=+\infty$. 
Instead, as the first step of the exploration, 
let us observe the behaviors of 
$\phi$ and $\psi$ observed in the finite region, i.e., $0<\xi<\xi_b$, 
where $\xi_b$ is simply the end of the numerical calculation. 
In other words, $\xi_b$ is an upper limiter of $\xi$ to explore the 
local maximum points of $\phi$ and $\psi$ in the range of $0<\xi<\xi_b$. 
Increasing $\xi_b$, the behaviors at $\xi=+\infty$ can be deduced, 
followed by the upper limit of the growth rate $\lambda$. 
Let us call the upper limit $\lambda_{up}$ which is not necessarily equal to 
$\lambda$ but will be useful, as shown below. 
Next, as the second step of the exploration, 
let us focus on the zero-crossing solution. 
Increasing $\xi_c$, i.e., the crossing point, 
let us explore how the growth rate $\lambda$ of the 
zero-crossing solutions depends on $\xi_c$. 
Finally, the behaviors of zero-crossing solutions of $\xi_c=+\infty$ 
can be deduced, which will coincide with the zero-converging solution.

In addition, as will be discussed in Section 2.3.6, 
the zero-converging solutions and zero-crossing solutions 
observed in this paper 
may be classified into the "inner-triggered" and "outer-triggered" 
tearing instability. 
Let us define the former as when both local maximum points of 
$\phi$ and $\psi$ are located in the inner region of the current sheet. 
Then, let us define the latter as when either of the local maximum points 
is at least located in the outer region of the current sheet. 
The latter is different from the externally-driven reconnection process 
proposed by Sato, et al. \citep{sato1978}. 
At this point, since $\phi$ and $\psi$ are perturbation terms 
and hence have much weaker intensity than 
the zero-order equilibrium defined as Eqs.(2.3)-(2.11), 
the latter does not mean that the tearing instability is caused 
by some apparently-strong externally-driven force in the upstream region. 
Note that tearing instability is essentially 
driven by the Alfven waves propagating 
in the inner or outer region of current sheet. 
Thus, the inner-triggered tearing instability is triggered 
by the weak Alfven waves in the inner region 
and the outer-triggered tearing instability is all other cases. 
At this point, we may assume that the main part of the trigger, 
i.e., the maximum amplitude of the weak Alfven wave, 
is located at the local maximum point of $\phi$ and $\psi$.

\subsection{Numerical study}

\subsubsection{Basic features}

Figure 1(a) shows the behaviors of $\phi$ and $\psi$ numerically obtained from Eqs.(2.1) and (2.2) for $\phi'(0)=2.5$ and $\kappa=\epsilon=0.2$ as $\lambda$ varies between $0.1$ and $2.0$. This case is typical for the higher $\phi'(0)$ range and is limited in the lower $\lambda$ range. In the upper panel of Figure 1(a), as $\lambda$ increases from label a to c, the local maximum point of $\phi$ shifts away from $\xi=0$. Then, since $\phi$ of labels d-f exceeds the vertical axis scale range, the existence of a local maximum point is unclear. At this point, Figure 3(a) confirms the lack of local maximum point of $\phi$ in $0<\xi<5$. In the middle panel of Figure 1(a), the behavior of $\psi$ tends to be inverted with respect to that of $\phi$. In fact, $\psi$ of labels a-b does not have a local maximum point except at $\xi=0$, while $\psi$ of labels c-f does have a local maximum point slightly separated from the origin. Of note, $\psi$ of labels c-d appears to be a flat-top function but has a local maximum point slightly separated from the origin, as shown in $5 \psi'(\xi \ne 0)=0$ in the middle panel of Figure 5. It means $\psi''(0)>0$ for every $\psi$ shown in this figure. Eventually, for $0.2<\lambda<0.5$, i.e., between labels b and d, $\phi$ and $\psi$ may simultaneously have local maximum points in $0<\xi<5$. Such $\phi$ and $\psi$ may be physically acceptable.

Figure 1(b) shows the behavior of $\phi$ and $\psi$ for $\phi'(0)=2.5$ and $\kappa=\epsilon=0.2$ while $\lambda$ varies between $2$ and $100$. This is typical for the higher $\lambda$ range than Figure 1(a), where label f in Figure 1(a) is the same as label a in figure 1(b), excluding the vertical axis scale. As $\lambda$ increases, the local maximum point of $\phi$ goes to $\xi=0$, while that of $\psi$ shifts away from $\xi=0$. As a result, Figure 1(b) does not have $\phi$ and $\psi$ that simultaneously have local maximum points in $0<\xi<5$. At this point, labels a and b are unclear for the existence of the local maximum point of $\phi$, which is confirmed in Figure 3(a). Also, that of $\psi$ is unclear for labels c-f, which is confirmed in Figure 3(a). Eventually, no physically acceptable solution exists in Figure 1(b).

\renewcommand{\thefigure}{\arabic{figure}(\textbf{a})}

\hspace{1mm}
\vspace{10mm}
\begin{figure}
\begin{center}
\includegraphics[bb=0.00 0.00 512.00 512.00,width=0.50\hsize]
{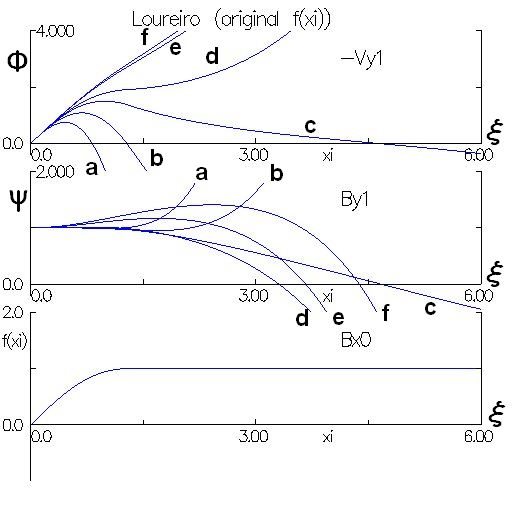}
   \caption{ $\phi$ (upper panel) and $\psi$ (middle panel) numerically obtained for $\kappa=\epsilon=0.2$,  $\phi'(0)=2.5$ , and $\lambda=$  (a) $0.1$, (b) $0.2$, (c) $0.349$, (d) $0.5$, (e) $1.5$, (f) $2$. The lower panel is $B_{x0}=f(\xi)$. 
 }
   \label{fig01a}
   \end{center}
   \end{figure}

\setcounter{figure}{0}
\renewcommand{\thefigure}{\arabic{figure}(\textbf{b})}

\hspace{1mm}
\begin{figure}
\begin{center}
\includegraphics[bb=0.00 0.00 512.00 512.00,width=0.5\hsize]
{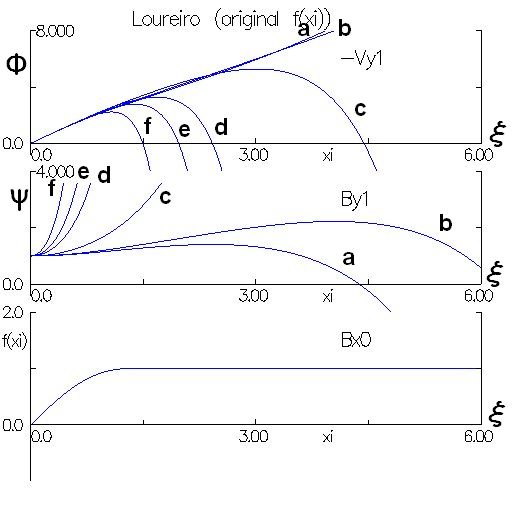}
   \caption{  $\phi$ (upper panel) and $\psi$ (middle panel) numerically obtained for  $\kappa=\epsilon=0.2$ ,  $\phi'(0)=2.5$ , and $\lambda=$  (a) $2$, (b) $2.5$, (c) $7$, (d) $30$, (e) $50$, (f) $100$ .  The lower panel is the same as that in Fig.1(a). Label a is the same as label f in Fig.1(a) but the vertical axis scale is different. 
 }
   \label{fig01b}
   \end{center}
   \end{figure}

Figire 2 shows the behaviors of $\phi$ and $\psi$ for $\phi'(0)=1.0$ and $\kappa=\epsilon=0.2$ while $\lambda$ is varied between $0.1$ and $100$. This is typical for the lower $\phi'(0)$ range than Figures 1(a) and (b). In Figure 2, it seems that $\phi$ and $\psi$ do not simultaneously have local maximum points in $0<\xi<5$.  At this point, the existence of the local maximum point is confirmed in Figure 3(a). Eventually, no physically acceptable solution exists in Figure 2.

\setcounter{figure}{1}
\renewcommand{\thefigure}{\arabic{figure}}

\begin{figure}
\begin{center}
\includegraphics[bb=0.00 0.00 512.00 512.00,width=0.5\hsize]
{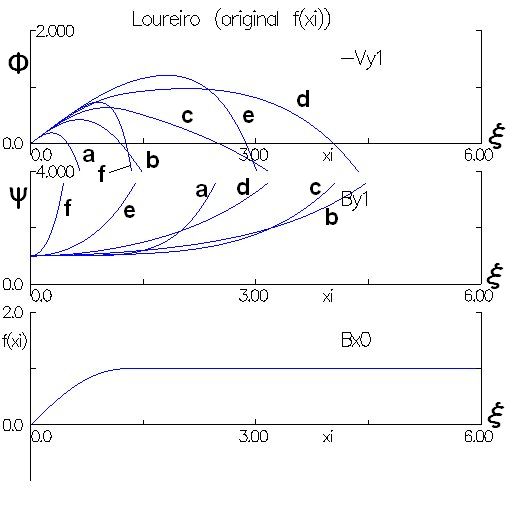}
   \caption{ $\phi$ (upper panel) and $\psi$ (middle panel) numerically obtained for  $\kappa=\epsilon=0.2$ ,  $\phi'(0)=1$ ,  $\lambda=$  (a) $0.1$, (b) $0.5$, (c) $1.1$, (d) $2.3$, (e) $10$, (f) $100$. The lower panel is the same as that in Fig.1(a). 
 }
   \label{fig02}
   \end{center}
   \end{figure}

\subsubsection{Existence of the local maximum points}

Figure 3(a) shows the existence of the local maximum points of $\phi$ and $\psi$ for $0.0001<\lambda<7$ and $0.0001<\phi'(0)<5$ for $\kappa=\epsilon=0.2$ and $0<\xi<\xi_b=5$, where $\xi_b$ is defined as the end point of finding the local maximum point. In the upper panel of Figure 3(a), the pink region indicates where the local maximum point of $\phi$ exists in $0<\xi<\xi_b$, and the white region indicates where it does not exist. In the lower panel, the pink and blue regions indicate where the local maximum point of $\psi$ exists in $0<\xi<\xi_b$, and the white and yellow regions indicate where it does not exist. Comparing with the upper and lower panels, the blue region in the lower panel represents where the local maximum points of $\phi$ and $\psi$ simultaneously exist in $0<\xi<\xi_b$, and the yellow region represents where both do not exist there. Hence, the blue region may include physically acceptable $\phi$ and $\psi$ in term of $0<\xi<\xi_b$.

The pink regions in the upper and lower panels are completely separated by the blue or yellow region in the lower panel. In fact, the blue and yellow regions are contacted around $\lambda=0.9$ and $\phi'(0)=1.52$. The contact point indicates the location of the upper limit of $\lambda$ for the physically acceptable $\phi$ and $\psi$. Let us define the upper limit as $\lambda_{up}$. The location of the contact point is difficult to specify exactly because the numerically obtained $\phi$ and $\psi$ are extremely sensitive to $\lambda$ and $\phi'(0)$. However, by carefully changing $\lambda$ and $\phi'(0)$, we can specify $\lambda_{up}$, as shown below.

Figure 3(b) shows the existence of the local maximum points of $\phi$ and $\psi$ for $0.90<\lambda<0.92$ and $1.52<\phi'(0)<1.54$ for $\kappa=\epsilon=0.2$ and $0<\xi<\xi_b=5$. Accordingly, Figure 3(b) shows only a limited region of Figure 3(a). The contact point between the blue and yellow regions is clearly observed between $0.9116<\lambda<0.9120$ and $1.5278<\phi'(0)=1.5281$. Hence, $\lambda_{up}=0.9116$ is roughly the upper limit of the growth rate for $\kappa=\epsilon=0.2$ and $0<\xi<\xi_b=5$.

\setcounter{figure}{2}
\renewcommand{\thefigure}{\arabic{figure}(\textbf{a})}

\begin{figure}
%\begin{center}
\vspace{60mm}
\includegraphics[bb=0 0 512 512,width=0.18\hsize]
{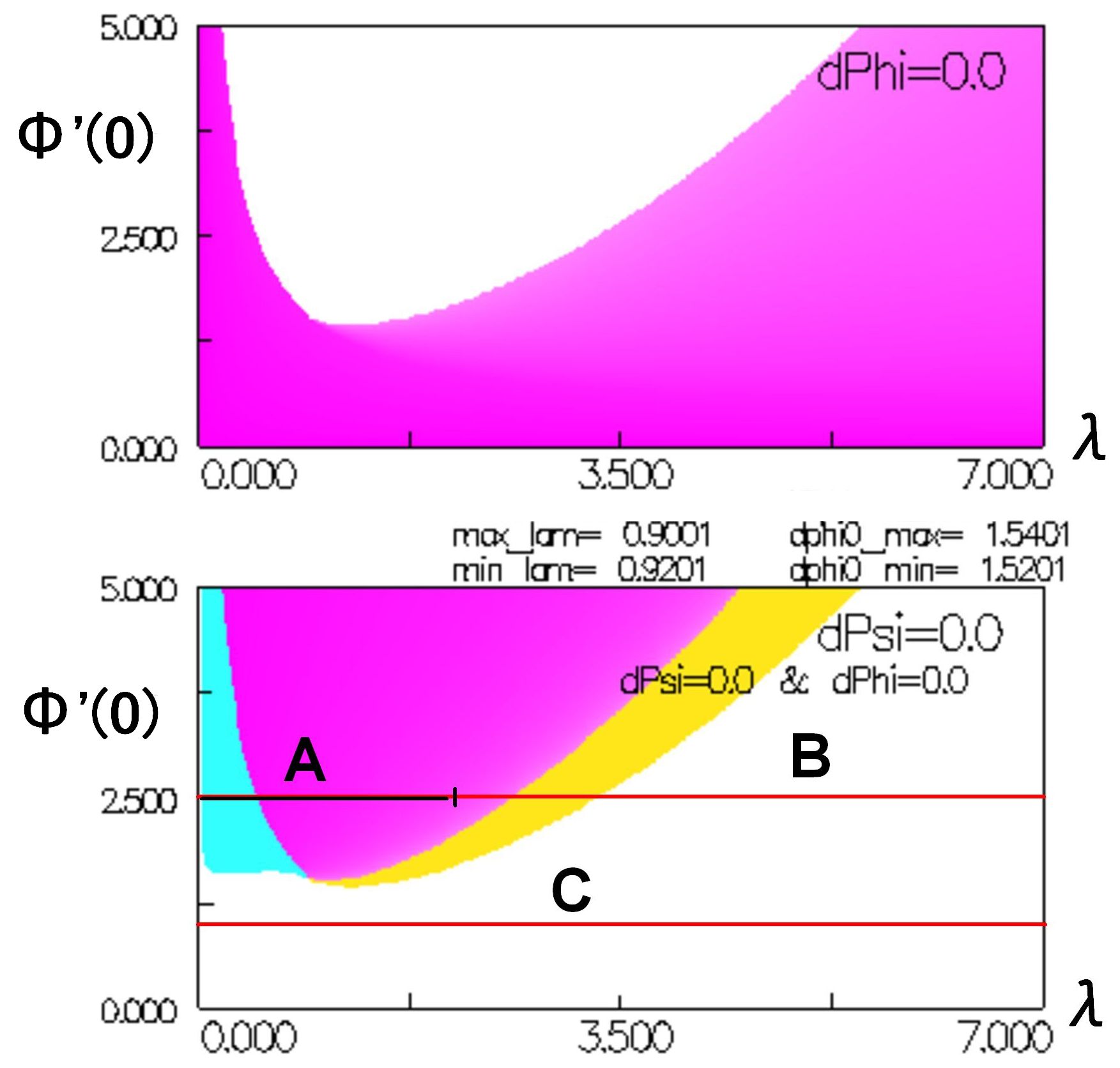}
   \caption{  The existence map of the local maximum points of  $\phi$ (upper panel) and  $\psi$ (lower panel) for $\kappa=\epsilon=0.2$ and $\xi_b=5$. The horizontal axis is shown for $0.0001<\lambda<7$, and the vertical axis is for $0.0001<\phi'(0)<5$. Labels A, B, and C, respectively, indicate the parameter range examined in Figs.1(a), (b), and 2. The slight color gradation shows how far the local maximum point is separated from the origin. 
 }
\label{fig03a}
%\end{center}
\end{figure}

\setcounter{figure}{2}
\renewcommand{\thefigure}{\arabic{figure}(\textbf{b})}

\begin{figure}
%\begin{center}
\vspace{60mm}
\includegraphics[bb=0 0 512 512,width=0.18\hsize]
{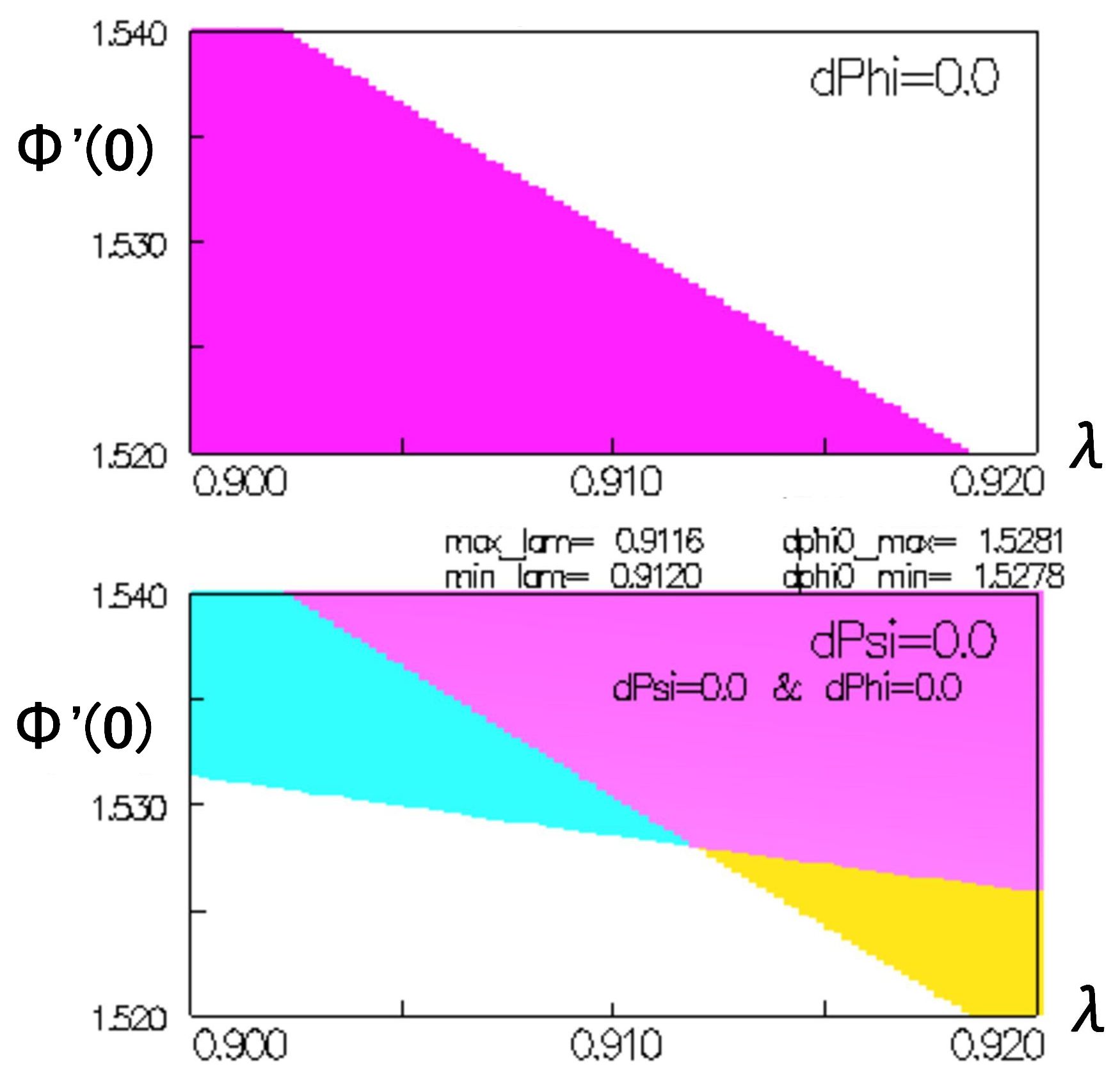}
   \caption{  The existence map of the local maximum points of $\phi$ (upper panel) and  $\psi$ (lower panel) for $\kappa=\epsilon=0.2$ and $\xi_b=5$. The horizontal axis is shown for $0.9<\lambda<0.92$ and the vertical axis is for $1.52<\phi'(0)<1.54$. The right edge of the blue region is detected at $\lambda=0.9116$ and the left edge of the yellow region is detected at $\lambda=0.9120$. 
Hence, the contact point between the blue and yellow regions is located in $0.9116<\lambda<0.9120$ and $1.5278<\phi'(0)<1.5281$. 
 }
\label{fig03b}
%\end{center}
\end{figure}

Figure 4 shows the existence of the local maximum points of $\phi$ and $\psi$ for the same parameter range as in Figure 3(a), with the exception that $0<\xi<\xi_b=1.307$, which corresponds to the case of the inner-triggered tearing instability. In other words, this figure shows when the local maximum points of $\phi$ and $\psi$ are found inside of the current sheet. In the same manner as discussed in Figures 3(a) and (b), $0.7001<\lambda_{up}<0.7101$ is obtained around $\phi'(0)=1.84$. In comparison to Figure 3(a), $\lambda_{up}$ tends to decrease as $\xi_b$ decreases from $5.0$ to $1.307$.

\setcounter{figure}{3}
\renewcommand{\thefigure}{\arabic{figure}}

\hspace{1mm}
\begin{figure}
%\begin{center}
\vspace{60mm}
\includegraphics[bb=0 0 512 512,width=0.18\hsize]
{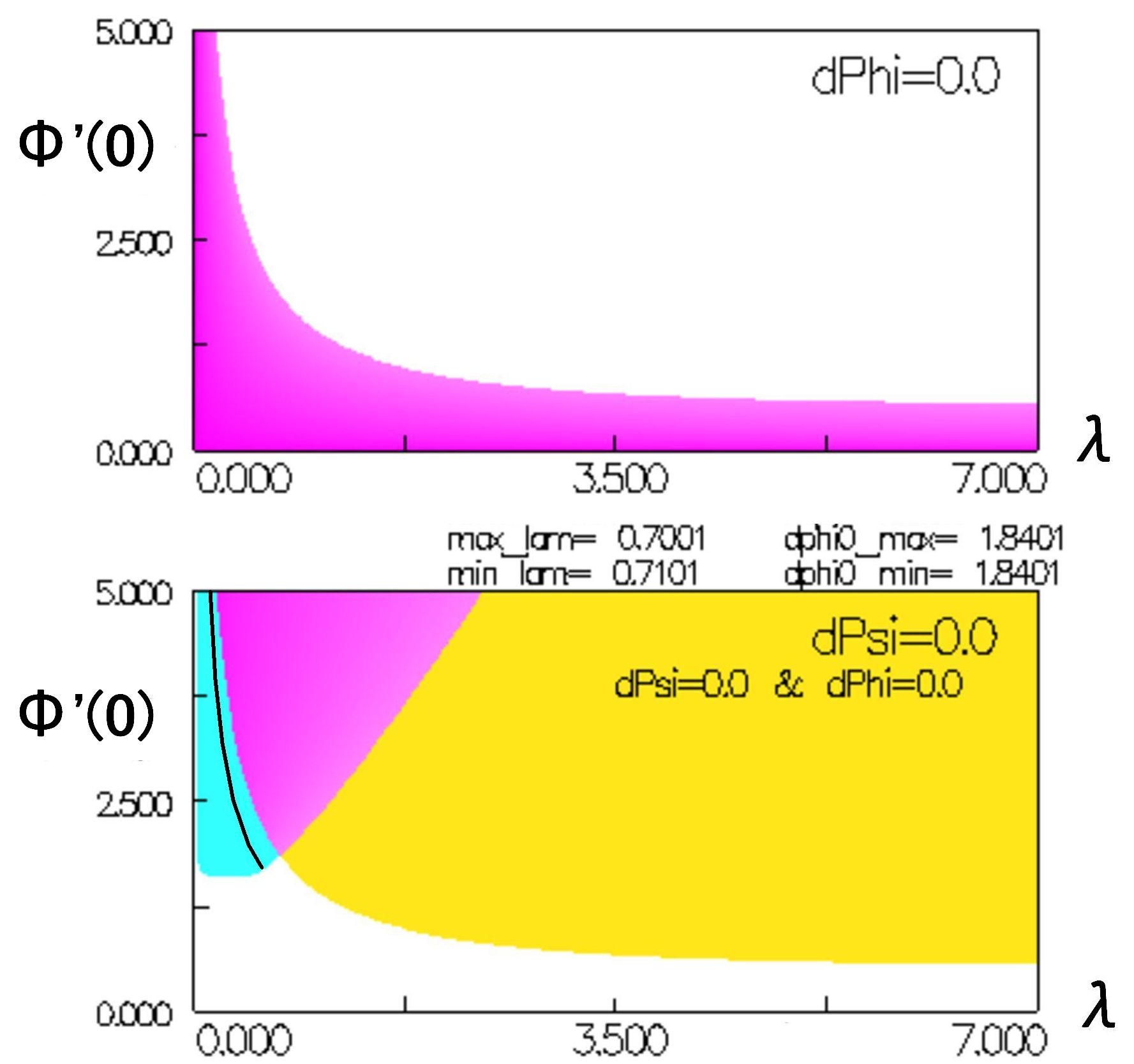}
   \caption{  The existence map of the local maximum points of $\phi$ (upper panel) and  $\psi$ (lower panel) for $\kappa=\epsilon=0.2$ and $\xi_b=1.307$. The horizontal axis is shown for $0.0001<\lambda<7$ , and the vertical axis is for $0.0001<\phi'(0)<5$ . The solid line drawn in the blue region indicates the variation of zero-crossing solutions, e.g. Fig.5 and label c in Fig.1(a), which are plotted between $(\lambda, \phi'(0))=(0.127,5.0)$ and $(0.629,1.749)$. 
 }
   \label{fig04}
%\end{center}
   \end{figure}

\subsubsection{Zero-crossing solutions}

We would ideally like to find $\phi$ and $\psi$ that converge to zero at $\xi=+\infty$, but as long as we rely on only a numerical analysis, such values are impossible to determine. Instead, by modifying the zero-order equilibrium, such a zero-converging solution can be found, as shown in Appendix A. However, such zero-converging solutions are not necessarily required for physically acceptable $\phi$ and $\psi$. Rather, if both $\phi$ and $\psi$ reach zero in a limited finite $\xi$ range, they may also represent a physically acceptable solution in the finite $\xi$ range. In this section, 
let us define such solutions as "zero-crossing solution" 
and try to find it through the parameter survey executed in Figures 1-3. 
In fact, such zero-crossing solutions are shown 
as label c in Figures 1(a) and 5.

Figure 5 shows a zero-crossing solution in which $\phi=0$ and $\psi=0$ are simultaneously established at $\xi=(\xi_c=)7.36525$, which is located outside of the current sheet, i.e., $\xi>1.307$. This solution is obtained for $\kappa=\epsilon=0.2$, $\lambda=0.540043$, and $\phi'(0)=1.9$. To precisely determine this zero-crossing solution, a primitive pinching method is employed, as shown in Appendix B. $\psi'$ is also plotted in the middle panel to clearly illustrate the local maximum point of $\psi$, which is separated from the origin. The local maximum points of $\phi$ and $\psi$ are, respectively, $\xi=1.071$ and $1.035$, which are located inside of the current sheet. Hence, as will be discussed in Section 2,3,6, this is the case for the inner-triggered tearing instability. It may be noted that label c of Figures 1(a) and 5 are obtained for the same $\kappa$ and $\epsilon$ but, respectively, for $\lambda=0.349$ and $0.540$. However, those crossing points $\xi_c$, i.e. the locations of upstream boundary, are different. 
It suggests that the growth rate significantly depends on the upstream boundary condition.

\setcounter{figure}{4}
\renewcommand{\thefigure}{\arabic{figure}}

\hspace{1mm}
\begin{figure}
\begin{center}
\includegraphics[bb=0.00 0.00 512.00 512.00,width=0.5\hsize]
{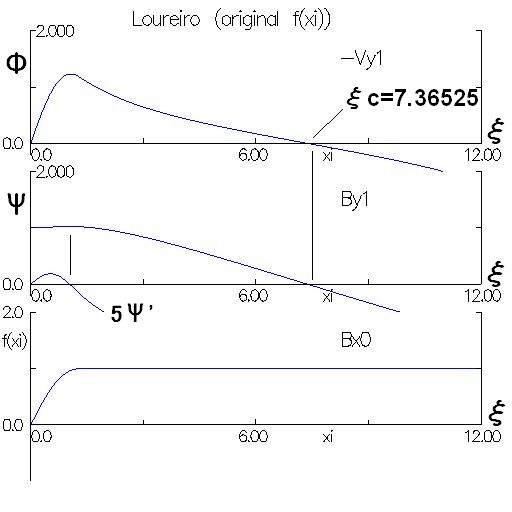}
   \caption{   The zero-crossing solution for  $\kappa=\epsilon=0.2$ ,  $\lambda=0.540043$, and  $\phi'(0)=1.9$ . The horizontal axis is scaled as $\xi$. The crossing points of $\phi=0$ and $\psi=0$ coincide at $\xi=7.36525$. To show the existence of the local maximum point of $\psi$, the middle panel shows $\psi'$, in addition to $\psi$. These $\phi$ and $\psi$ are the physically acceptable solution in  $0<\xi<7.36525$.
 }
   \label{fig05}
   \end{center}
   \end{figure}

\subsubsection{$\kappa$ and $\epsilon$ dependences of $\lambda_{up}$}

Figure 6 shows the $\xi_b$ dependence of the upper limit $\lambda_{up}$ for varying $\kappa$ and $\epsilon$, where $0<\xi<\xi_b$ is the parameter survey range of when the local maximum points of $\phi$ and $\psi$ are simultaneously found. Hence, $\xi_b$ may be considered to be another control parameter to explore the local maximum points, and then, measure $\lambda_{up}$. As $\xi_b$ increases, $\lambda_{up}$ monotonically increases. Hence, this figure shows that either of the local maximum points of $\phi$ and $\psi$, at least, shifts to a larger $\xi$ value for a larger growth rate $\lambda$. Additionally, Figure 6 suggests that $\lambda_{up}$ does not exceed unity, i.e., the Alfvenic rate $V_A/l_{cs}$ ; in other words, the tearing instability always proceeds at sub-Alfvenic speed. In addition, even for $\epsilon=0$, i.e., the zero resistivity limit, $\phi$ and $\psi$ can be obtained, and $\lambda_{up}$ appears to converge to unity as $\xi_b$ increases, which may be strange but not surprising, as will be discussed below.

\setcounter{figure}{5}
\renewcommand{\thefigure}{\arabic{figure}}

\begin{figure}
%\begin{center}
\vspace{10mm}
\includegraphics[bb=0.00 0.00 256.00 256.00,width=0.35\hsize]{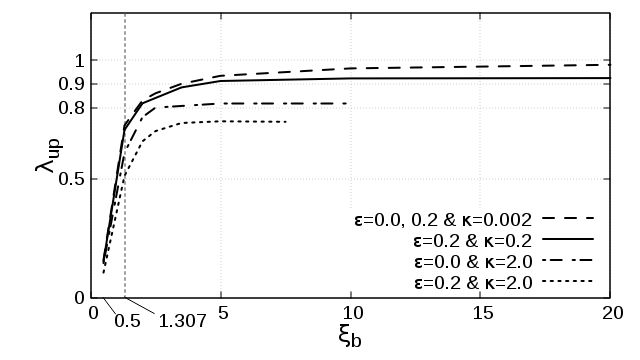}
   \caption{  The $\xi_b$ dependence of $\lambda_{up}$ for some cases of $\kappa$ and $\epsilon$.
 }
   \label{fig06}
%\end{center}
   \end{figure}

Figure 7 shows how $\lambda_{up}$ appears to converge to a value less than unity as $\epsilon$ is close to zero, where $\xi_b=1.307$ is set. 
As shown in figure 6, since $\lambda_{up}$ increases as $\xi_b$ increases, $\lambda_{up}$ for $\xi_b=1.307$ is considered to be the highest growth rate of the inner-triggered tearing instability. 
In addition, Figure 7 shows that, when $\kappa$ is close to zero,  
$\lambda_{up}$ is not sensitive to $\epsilon$ 
while $\lambda_{up}$ for larger $\kappa$ rapidly decreases as $\epsilon$ increases from zero.

Next, let us discuss why $\lambda_{up}$ can be obtained even for $\epsilon=0$, i.e., the zero resistivity limit. As simply predicted, the reconnection process, i.e., tearing instability, must stop in $\epsilon=0$, i.e. ideal MHD. However, the limit of $\epsilon=0$ examined in this paper does not directly mean $\epsilon=0$. In fact, the spatial scale $\xi$ is normalized by the thickness $\delta_{cs}$ of the current sheet, where $\delta_{cs}$ is defined for the steady-state SP sheet. Let us consider when $L_{cs}$ is constant. As $\epsilon=2 \delta_{cs}/L_{cs}$ approaches zero, $\delta_{cs}$ is unlimitedly close to zero. At this time, since the current density unlimitedly increases, the instability will be promoted, i.e., not stop. In other words, evenwhen the current sheet unlimitedly becomes thinner in the limit of $\epsilon=0$, tearing instability can rapidly occur in the thin current sheet. In that case, $\lambda_{up}$ normalized by $l_{cs}/V_A$ can become a value less than unity, which means the tearing instability can proceed at sub-Alfvenic speed in the zero resistivity limit, i.e., the infinite Lundquist number $S=+\infty$.

In addition, Figure 7 suggests that the tearing instability cannot rapidly proceed beyond Alfven speed when local maximum points of $\phi$ and $\psi$ exist, 
i.e. for zero-converging and zero-crossing solutions. 
It may be noted that $\lambda_{up}$ and $\lambda$ studied in this paper is 
different from the growth rate $\gamma/\Gamma_0$ 
studied in the original LSC theory 
(e.g. FIG.4 \citep{Lour2007}), as will be discussed in Section 2.3.7.

It may be noted that $\epsilon=2\delta_{cs}/L_{cs}=0$ can be established by either of $\delta_{cs}=0$ or $L_{cs}=+\infty$. However, to establish the $\epsilon=0$ limit,  $\delta_{cs}=0$ is not equivalent to $L_{cs}=+\infty$. Because, the $\delta_{cs}=0$ limit directly means that current sheet is unlimitedly thin but the $L_{cs}=+\infty$ limit changes $\kappa=\pi L_{cs}/l_{cs}$. Then, to keep a constant $\kappa$ value, $l_{cs}=+\infty$ is required, leading to the change of the $\lambda$ scale, i.e., $l_{cs}/V_A$.

\setcounter{figure}{6}
\renewcommand{\thefigure}{\arabic{figure}}

\begin{figure}
%\begin{center}
\vspace{10mm}
\includegraphics[bb=0.00 0.00 256.00 256.00,width=0.35\hsize]{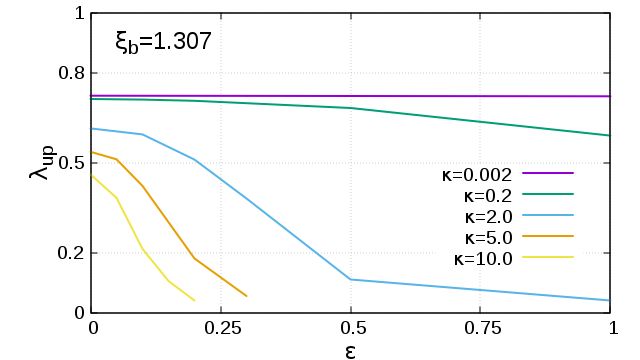}
   \caption{ The $\epsilon$ dependence of $\lambda_{up}$ for some cases of $\kappa$ for $\xi_b=1.307$. As $\kappa$ becomes close to zero, $\lambda_{up}$ appears to take a maximum around $0.724$.  }
   \label{fig07}
%\end{center}
   \end{figure}

\subsubsection{ $\kappa$, $\epsilon$ and $\xi_c$ 
dependences of $\lambda$ for zero-crossing solutions}

Figure 8 shows the $\kappa$ dependence of $\lambda_{up}$ for $\xi_b=5.0$, $1.307$, and $0.5$ for $\epsilon=0$. Regardless of $\xi_b$, as $\kappa$ increases from $0.002$ to $20$, three lines of $\lambda_{up}$ plotted for $\xi_b=0.5$, $1.307$ and $5$ tend to monotonically decrease. This result of $\lambda_{up}$ is applied to MHD simulation in the next section. 
Moreover, three lines of dot chain show the growth rate $\lambda$ of zero-crossing solutions for $\epsilon=0$, in which the crossing point is respectively fixed at $\xi_c=3.6$, $7.2$ and $10.8$. In contrast to $\lambda_{up}$, 
these three lines of the dot chain have a maximum point around $\kappa=1$. 
More exactly, as $\xi_c$ increases from $3.6$ to $10.8$, 
the maximum point gradually shifts to a smaller $\kappa$. 
Thus, the growth rate $\lambda$ of zero-crossing solutions 
depends on the crossing point $\xi_c$, i.e., the upstream boundary condition, 
as mentioned in end of Section 2.3.3. The case of the $\xi_c=+\infty$ limit will correspond to the zero-converging solutions which may coincide with $\lambda_{up}$ in the $\xi_b=+\infty$ limit. As shown in Figure 6, $\lambda_{up}$ in the $\xi_b=+\infty$ limit is predicted to be unity. 
 
Note that Figures 3 and 4 obtained for $\epsilon=0.2$ is directly inapplicable for Figure 8 for $\epsilon=0.0$. In addition, note that these three lines of dot chain are the growth rate $\lambda$, itself, but not the upper limit $\lambda_{up}$. However, as $\xi_c$ increases, $\lambda$ shown in Figure 8 tends to increase. 
This tendency is consistent with the $\lambda_{up}$ increase 
by increasing $\xi_b$, which is shown in Figure 6.

\setcounter{figure}{7}
\renewcommand{\thefigure}{\arabic{figure}}

\begin{figure}
\vspace{10mm}
%\begin{center}
\includegraphics[bb=0.00 0.00 256.00 256.00,width=0.35\hsize]{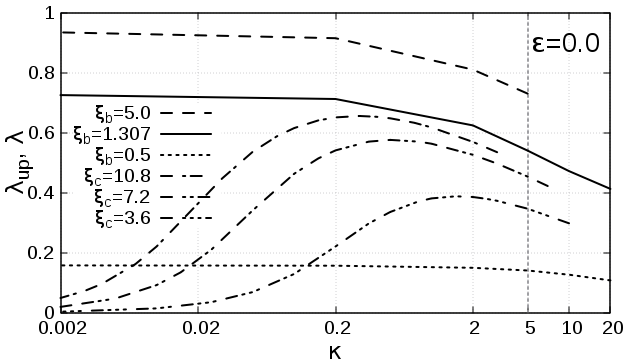}
   \caption{   The $\kappa$ dependence of $\lambda_{up}$ for $\epsilon=0$ and $\xi_b=0.5, 1.307$, and $5$. In addition, the one-dot, two-dot, and three-dot chain lines show the $\kappa$ dependence of $\lambda$ for the zero-crossing solution of $\epsilon=0$, in which the crossing point for each line is respectively fixed at $\xi_c=3.6$, $7.2$ and $10.8$. }
   \label{fig08}
%\end{center}
   \end{figure}

Figure 9 shows how $\lambda$ of zero-crossing solutions depend on $\xi_c$ 
for $\kappa=0.001,$ $ 0.01,$ $ 0.05,$ $ 0.1,$ $ 0.2,$ $ 1,$ $ 3$, and $5$ of 
$\epsilon=0$. 
In every $\kappa$ line, as $\xi_c$ increases, 
$\lambda$  monotonically increases.  Inversely observing, 
as $\xi_c$ decreases toward $1.307$, $\lambda$ seems to converge to zero. 
Due to the lack of numerical precision, 
the detail of $\lambda$ behaviors around $\xi_c=1.307$ cannot be observed. 
Hence, it is unclear whether $\lambda=0$ is attained exactly at $\xi_c=1.307$. 
As shown in Figure 8, 
Figure 9 also shows how the maximum point of $\lambda$ shifts 
in $\kappa$ space, depending on $\xi_c$. 
In fact, as $\xi_c$ increases, the maximum point tends to shift 
from higher $\kappa$ line to lower $\kappa$ line. 
E.g., the highest $\lambda$ value at $\xi_c=5$ is on the $\kappa=1$ line, 
while the highest $\lambda$ value at $\xi_c=10$ is on the $\kappa=0.2$ line.

\setcounter{figure}{8}
\renewcommand{\thefigure}{\arabic{figure}}

\begin{figure}
%\begin{center}
\includegraphics[bb=0 0 256 256 ,width=0.50\hsize]{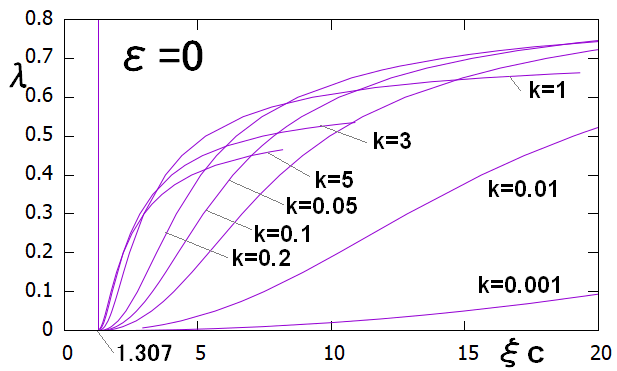}
   \caption{   
The $\xi_c$ dependence of $\lambda$ for zero-crossing solutions for 
 $\epsilon=0$ and various $k$ values, where $\xi_c$ is the location of the zero-crossing 
point.  }
   \label{fig09}
%\end{center}
   \end{figure}

Figure 10(a) shows how $\lambda$ of zero-crossing solutions depends on 
$\xi_c$ for $\kappa=0.1$ of $\epsilon=0, 0.2, 1, 2,$ and $5$. 
Evenwhen $\epsilon$ is non-zero, 
the monotonical increase in $\lambda$ for the larger $\xi_c$ 
observed in Figure 9 seems to be retained. 
As $\epsilon$ increases from zero, 
$\lambda$ tends to monotonically decrease. 
However, since $\lambda$ lines in $\epsilon=0$ and $0.2$ almost coincides, 
$\lambda$ is not sensitive to $\epsilon$ for $0<\epsilon<0.2$. 
For this reason, in Section 3, the growth rate measured at $\epsilon=0$ 
is applied to the MHD simulation of the tearing instability 
which is not the case of $\epsilon=0$ but close to $\epsilon=0$. 

Figure 10(b) is similar to Figure 10(a) but shows the case of 
higher $\kappa$ values, i.e., $\kappa=1$ for $\epsilon=0, 0.2$ and $0.5$ and 
the cases of $\kappa=3$ for $\epsilon=0$ and $0.2$. 
Evenwhen $\epsilon$ is non-zero, 
the monotonical increase in $\lambda$ for larger $\xi_c$ 
observed in Figures 9 and 10(a) seems to be retained. 
However, in contrast to Figure 10(a), Figure 10(b) shows that 
$\lambda$ for $\kappa>1$ rapidly decreases for increasing $\epsilon$.

\setcounter{figure}{9}
\renewcommand{\thefigure}{\arabic{figure}(\textbf{a})}

\begin{figure}
%\begin{center}
\includegraphics[bb=0.00 0.00 256.00 256.00,width=0.50\hsize]
{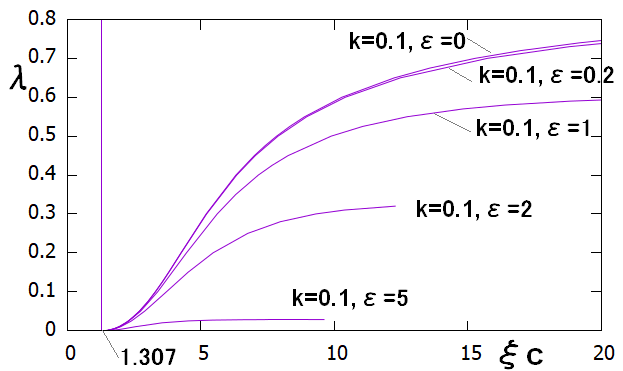}
   \caption{   
The $\xi_c$ dependence of $\lambda$ for zero-crossing solutions for 
 various $\epsilon$ values for $k=0.1$, 
where $\xi_c$ is the location of the zero-crossing point. 
}
   \label{fig10a}
%\end{center}
   \end{figure}

\setcounter{figure}{9}
\renewcommand{\thefigure}{\arabic{figure}(\textbf{b})}

\begin{figure}
%\begin{center}
\includegraphics[bb=0.00 0.00 256.00 256.00,width=0.50\hsize]
{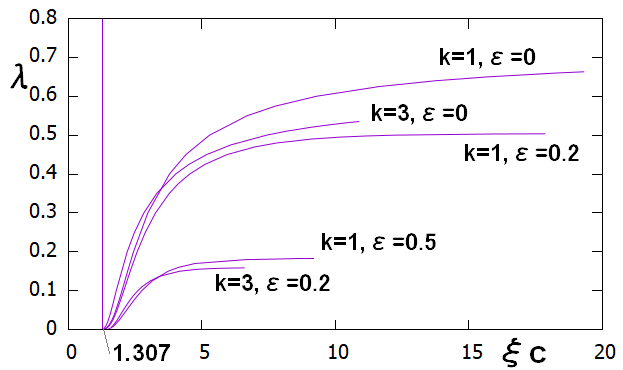}
   \caption{  
The $\xi_c$ dependence of $\lambda$ for zero-crossing solutions for 
 various $\epsilon$ values of $k=1$ and $3$, 
where $\xi_c$ is the location of the zero-crossing point. }
   \label{fig10b}
%\end{center}
   \end{figure}

\subsubsection{Inner and outer-triggered tearing instability}

As discussed in the end of Section 2.2, 
these numerical results shown in this section may be classified 
into the inner-triggered and outer-triggered tearing instability. 
The inner-triggered tearing instability has the local maximum points 
of $\phi$ and $\psi$ located in the inner region of the current sheet. 
Since such a tearing instability is initiated inside the current sheet, 
this will be called spontaneous tearing instability. 
Meanwhile, the outer-triggered tearing instability is all other cases, 
i.e., when either of the local maximum points of $\phi$ and $\psi$ 
is located in the outer region.

Figure 11(a) shows how the growth rate $\lambda$ of zero-crossing solutions 
observed in Figures 8 and 9 changes 
for the movements of local maximum point of $\phi$ in $\xi$ space, 
where the horizontal axis is 
the location $\xi$ of the local maximum point of $\phi$. 
Due to the lack of numerical precision, 
Figure 11(a) is unclear for the vicinity of $\lambda=0$ and larger $\lambda$. 
The unclear range (the end of each line) largely depends on $\kappa$. 
As the location of the local maximum point of $\phi$ is separated from zero, 
$\lambda$ increases. This feature is consistent with Figure 6 
in which, as $\xi_b$ increases, $\lambda_{up}$ monotonically increases. 
In addition, as $\kappa$ is close to zero, 
those curves tend to get lower. 
However, for all $\kappa$ cases, it seems that the locations 
do not exceed $\xi=1.307$, i.e., the boundary point of 
the inner and outer regions of the current sheet.

Figure 11(b) is similar to Figure 11(a) but shows 
the movements of the local maximum point of $\psi$, 
where the horizontal axis is the location $\xi$ of the local maximum point of $\psi$. As the location of the local maximum point of $\psi$ is separated from zero, 
$\lambda$ increases. This feature is consistent with Figures 6 and 11(a). 
In contrast to Figure 11(a), 
the local maximum point of $\psi$ can shift beyond $\xi=1.307$. 
Hence, when the location of the local maximum point of $\psi$ is located in $\xi<1.307$ is the case of the inner-triggered tearing instability. 
Figure 11(b) shows that 
the growth rate $\lambda$ in the outer-triggered tearing instability 
is higher than that of the inner-triggered tearing instability.

\setcounter{figure}{10}
\renewcommand{\thefigure}{\arabic{figure}(\textbf{a})}

\begin{figure}
%\begin{center}
\includegraphics[bb=0.00 0.00 256.00 256.00,width=0.50\hsize]
{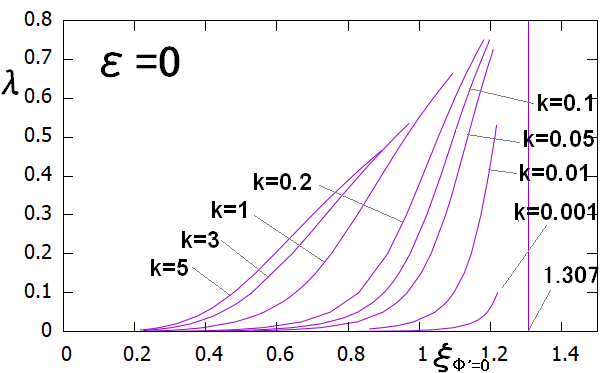}
   \caption{  
The $\lambda$ change for the movement of local maximum point 
$\xi_{\phi'=0}$ for $\phi$. }
   \label{fig11a}
%\end{center}
   \end{figure}

\setcounter{figure}{10}
\renewcommand{\thefigure}{\arabic{figure}(\textbf{b})}

\begin{figure}
%\begin{center}
\includegraphics[bb=0.00 0.00 256.00 256.00,width=0.50\hsize]
{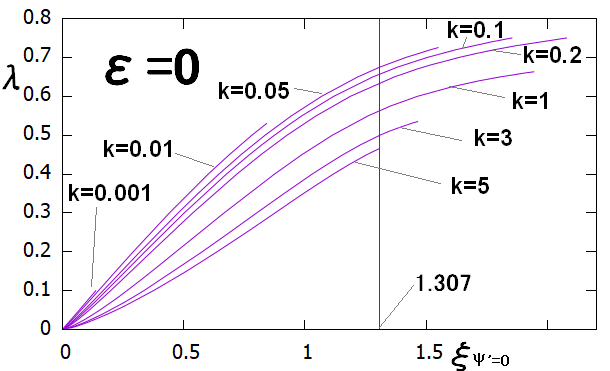}
   \caption{  
The $\lambda$ change for the movement of local maximum point 
$\xi_{\psi'=0}$ for $\psi$. }
   \label{fig11b}
%\end{center}
   \end{figure}

\subsubsection{Comparison of the growth rate with original LSC theory.}

As mentioned in Section 2.1, 
the growth rate in this paper is normalized by $l_{cs}/V_A$. 
When we compare the growth rate $\lambda$ and upper limit 
$\lambda_{up}$ studied in this paper with the growth rate 
in the original LSC theory \citep{Lour2007},  
note that their growth rate is measured to be 
$\kappa \lambda=\gamma/\Gamma_0$, which is not $\lambda$, itself. 
In addition, their growth rate is obtained by separating 
inner and outer regions with the $\Delta'$-index. 
For example, FIG.4 in Loureiro,et al. shows 
that $\kappa \lambda$ increases as $\kappa^{2/3}$ for a lower $\kappa$ regime and decreases as $\kappa^{-2/5}$ for higher $\kappa$ regime. 
The finding means that $\lambda$ decreases as $\kappa^{-1/3}$ for a lower $\kappa$ regime and decreases as $\kappa^{-7/5}$ for a higher $\kappa$ regime. 
Hence, $\lambda$ monotonically decreases as $\kappa$ increases. 
This feature is qualitatively consistent with Figure 8 
for $\lambda$ of $\kappa>1$ and $\lambda_{up}$ of the entire $\kappa$ range. 
However, those behaviors of $\lambda$ are difficult to quantitatively 
compare, because, $\lambda$ and $\lambda_{up}$ in Figure 8 
is very difficult to explore in $\kappa>5$ 
due to the lack of numerical precision. 
Moreover, FIG.4 in Loureiro,et al. \citep{Lour2007} appears to be unclear in 
$\kappa<<1$, where $\lambda$ may be higher because 
it varies as $\kappa^{-1/3}$. 
To explore extremely high $\kappa$ range of Figure 8, 
the $\lambda=0$ limit solutions of Eqs.(\ref{phi-eq}) and 
(\ref{psi-eq}) will be helpful, 
which are extensively discussed in Appendix C.

Figure 8 is similar to what was reported 
in some papers \citep{ LandiApj2015, PucciApjL2014, TeneraniApj2015, ZannaJoP2016, PapiniJoP2018}, where Eqs.(\ref{phi-eq}) and (\ref{psi-eq}) were solved in the upstream boundary conditions which are different from this paper. 
At this point, this paper shows 
that the growth rate significantly depends on the upstream boundary condition, 
i.e., the locations of the zero-crossing point $\xi_c$. 
In addition, their equilibrium seems to be different 
from that of this paper. 
At this point, most of them took the $\Gamma_0=0$ limit 
in the equilibrium of the LSC theory but this paper does not. 
In other words, they did not directly apply Eqs.(2.3)-(2.11) 
introduced by Loureiro. In fact, instead of Eqs.(2.4) and (2.7), 
the Harris type of current sheet as the magnetic field equilibrium 
was employed, and, instead of Eqs.(2.5), (2.6) and (2.10), 
the null-flow field was assumed. In the $\Gamma_0=0$ limit, 
since $L_{cs}=2V_A/\Gamma_0$ becomes infinity, 
their results cannot be applied to the MHD simulation in the next section.

More recently, Shi,et.al. also numerically studied perturbation equations similar to Eqs.(2.1) and (2.2) (\cite{ShiApj2018}), 
where the inner and outer regions were seamlessly solved under 
uniform resistivity. 
Hence, their approach may be similar to what is shown in this section. 
However, there are some differences between their results and 
what is shown in this section. For example, they reported that 
tearing instability is stabilized by the background flow, i.e.,  
Eq.(2.6), when Lundquist number $S$ is less than $70$. 
Meanwhile, as shown in Appendix C, this paper suggests that 
the growth rate $\lambda$ can keep to be positive 
even when resistivity $\epsilon$ is extremely large, 
as long as the criterion of $\kappa \epsilon <1.0915$ is satified. 
Since low $S$ corresponds to large $\epsilon$, 
the criterion suggests that tearing instability can occur at small $\kappa$ 
even in extremely low $S$. 

It is unclear why the difference is caused but 
let us put an intuitive explanation for 
the tearing instability in small $\kappa$ and low $S$, as below. 
First, note that tearing instability can occur 
when the current sheet at X-point becomes thin. 
Such a thinning steadily occurs if the outflow from the X-point 
is stronger than the inflow to the X-point. 
Because, at the time, the X-point becomes close to vacuum. 
The equilibrium in linear theory takes 
the balance between the outflow and inflow even in extremely low $S$. 
Hence, if the perturbed outflow is stronger than the perturbed inflow, 
i.e., $\phi'(0)$ is large, the tearing instability will start. 
Next, let us consider about meanings of small $\kappa$. 
As defined in Fig.17, when $dVx/dx$ measured at the X-point is large, 
$L_{cs}=2V_A/(dVx/dx)$ is small, leading to small $\kappa=\pi L_{cs}/l_{cs}$. 
At the time, resistivity will prevent the thinning at the diffusion speed. 
However, if we focus only on the sufficiently vicinity of the X-point, 
the thinning will be able to overcome the diffusion even in extremely low $S$. 
Hence, we cannot say that sufficiently low $S$ 
steadily stabilizes the tearing instability.

However, it may be noted 
that the WKB approximation employed in LSC theory \citep{Lour2007} 
fails for the small $\kappa$. 
Hence, the intuitive explanation mentioned above 
must be carefully re-examined in the future, improving the WKB approximation.

\section{MHD simulation}\label{MHD simulation}

\subsection{Spontaneous plasmoid instability}

In this section, we apply the modified LSC theory, which resulted in Figure 8, to the MHD simulation of the spontaneous PI. At the end of this section, the modified LSC theory is shown to be partially consistent with the MHD simulation results.

The word "spontaneous" means that the subsequent tearing instability is spontaneously driven by the preceding tearing instability after the first tearing instability is externally initiated by an initial disturbance. Let us call such multiple tearing instabilities spontaneous PI. The spontaneous PI is enhanced and developed by a kind of positive feedback mechanism, including the PI itself. In addition to the spontaneous PI model, an externally driven PI model may be possible, where the tearing instability is maintained by an externally driven mechanism which may be a weak noise \citep{Ng2010}. Hence, if the driven mechanism is removed, the PI stops. However, the existence of such an externally driven mechanism in solar flares and substorms is unclear. In this paper, we focus on the spontaneous PI model.

In our previous paper \citep{shi2017}, we found that when the current sheet becomes extremely thin, the numerical error may fatally affect the simulation results of the spontaneous PI model. Similar result has also been reported by Ng. et al.\citep{Ng2010}. In fact, the active PI observed at a lower numerical resolution tends to be less active at a higher numerical resolution. At a higher numerical resolution, the magnetic reconnection rate may momentarily exceed the value predicted by the steady state Sweet-Parker (SP) theory but may not constantly reach the level independent of the resistivity, which suggests that the critical Lundquist number $S_c$ does not exist\citep{shi2017}. In this paper, to avoid the numerical error due to lower numerical resolution, the case of relatively large resistivity, i.e. a relatively low Lundquist number $S$, is examined. 
Then, on the basis of modified LSC theory, the $S=+\infty$ limit is discussed in Sections 4.1.

\subsection{Simulation procedures}

The one-component compressible 2D MHD eqs. are numerically solved. The reconnection process studied herein is essentially the same as that of our previous studies \citep{mu1984,ts2003,shi2017}. In this paper, 2-step Lax-Wendroff scheme \citep{mu1977} is employed, but the results shown in this section are independent of the numerical scheme, because the numerical resolution is kept to be sufficiently high. The first tearing instability is initiated by a small resistive disturbance induced around the origin of the 1D current sheet. After the resistive disturbance is removed at $t=4$, uniform resistivity is maintained for $t>4$. Then, the thinning of the current sheet spontaneously starts around the most intensive region, i.e. the origin.  Accordingly, we can examine whether the 1D current sheet is spontaneously destabilized by the tearing instability. In fact, after the first tearing instability is terminated by nonlinear saturation, subsequent tearing instabilities intermittently start in the current sheet. At $t>4$, since no externally driven mechanism is applied to compress the current sheet to maintain the reconnection process, this situation represents the spontaneous PI model under uniform resistivity.

The simulation box size is limited to $0<x<L_x$ and $0<y<L_y$, where $(L_{x},L_{y})=(400,200)$. All boundary conditions of the simulation box are set to symmetric boundary conditions. The initial current sheet has a 1D structure, i.e., the magnetic field ${\bf B}=[B_{x}(y),0,0]$ is assumed to be $B_{x}(y)=B_{x0} \tanh y$ for $0<y<L_y$. In contrast to previous studies\citep{shi2017}, an inversed current sheet does not exist in the upstream region, but instead, $L_y=200$ is much larger than the initial thickness of the current sheet. Accordingly, the upstream boundary at $y=L_y$ will have a minimal numerical effect on the tearing instability in the current sheet near $y=0$.

The plasma static pressure $P$ initially satisfies the pressure-balance condition, i.e., $2P+B_{x}^{2}=1+\beta_{0}$, where $\beta_{0}$ is the ratio of the plasma pressure to the magnetic pressure in the magnetic field region of $B_x=1.0$. In this paper, $\beta_{0}=0.15$. The initial fluid velocity is ${\bf u}=(0,0,0)$ throughout the simulation box. Hence, the initial state is not equilibrium for uniform resistivity, leading to the magnetic annihilation. However, the magnetic reconnection, i.e., tearing instability, overcomes the magnetic annihilation and grows. The initial plasma density $\rho = (P/2)/(1+\beta_{0})$, where the initial plasma temperature is uniform. The Alfven speed $V_A=B_x/\sqrt{\rho}$ measured in the upstream region of $B_x=1.0$ is approximately $5.5$ with $\rho=0.033$. The local speed of sound $C_s=\sqrt{\gamma P/\rho}$ in the center of the current sheet ($y=0$) is approximately $2.0$ with $P=0.575$, $\gamma=5/3$, and $\rho=0.25$.

The initial resistive disturbance for $0<t<4$ is as follows. 

\begin{equation}
\eta_a= \eta_{a0} e^{- (x^2+y^2)}
\label{eta-a}
\end{equation}

\noindent
where $\eta_{a0}=0.005$. This resistive disturbance initiates the first tearing instability around the origin, i.e., $(x,y)=(0,0)$. The corresponding Lundquist number $S$ for this resistive disturbance in $0<t<4$ is estimated to be $S=L_x V_A / \eta_{a0} = 440000$. 

After $t=4$, $\eta_a$, which is defined by Eq.(\ref{eta-a}), is removed and, instead, uniform resistivity $\eta_b=0.016$ in time and space is assumed, which drives the first tearing instability, leading to PI. The corresponding Lundquist number $S$ in $t>4$ is estimated to be $S=L_x V_A / \eta_{b} = 137500$. Since the first tearing instability initially occurs in a much smaller region than $L_x$, a realistic $S$ will be much smaller than $137500$ and, hence, may be smaller than the critical Lundquist number $S_c=10^5 \sim 10^6$ predicted previously \citep{Lour2007}. Because of the relatively low $S$, we can sufficiently suppress the numerical errors that may fatally affect the tearing instability. Hence, the numerical errors are not discussed in this paper, in contrast to our previous paper \citep{shi2017}. The $S=+\infty$ limit is then discussed in Section 4.1, using Figure 8. The simulation box $(L_{x},L_{y})$ is divided by numerical grids of $\Delta x(=\Delta y)=0.02$ that are constant in time and space, where the time step $\Delta t=0.004$ is set to maintain the CFL conditions. Accordingly, the grid size is 
$(N_x,N_y)=(20000,10000)$.

Unfortunately, the 2-step Lax-Wendroff scheme requires artificial viscosity to maintain numerical stability. In this paper, artificial viscosity that is uniform in time and space is applied to the mass, momentum, and energy conservation equations in the MHD eqs. but not for the magnetic flux conservation equation, i.e., Faraday's law, because uniform resistivity $\eta_b$ has been applied. The intensity of the artificial viscosity corresponds to $0.5$ when translated as the magnetic Prandtl number. By contrast, LSC theory does not include any viscosity, i.e., the magnetic Prandtl number is zero. This discrepancy in viscosity between the theories and MHD simulations is briefly discussed in Section 4.3.

\subsection{ Overview of the numerical simulations}

In this section, the first and second tearing instabilities are examined. 
After them, the third and subsequent tearing instabilities 
are impulsively repeated. After the third tearing instability ends, the numerical result starts to be gradually affected by numerical error, because of thinning of the current sheet. Hence, the numerical results after the third tearing instability is unreliable, which are not shown in this paper.

Figure 12(a) shows the magnetic field lines and current density $J_z$ at $t=200$. 
 The initial resistive disturbance $\eta_a$ has been removed until this time, and the uniform resistivity $\eta_b$ is assumed. Red and yellow represent negative $J_z$, and blue, which does not appear in this figure and is observed in Figures 12(b) and (c), represents positive. The $J_z$ intensity is indicated by the darkness of each color. More exactly, as shown in the color scale bar, the color intensities between red, white, and blue are respectively normalized by the maximum, zero, and minimum of $J_z$ in this figure. This figure shows the beginning of the first tearing instability directly resulting from the initial resistive disturbance, i.e., Eq.(3.1). The red colored current sheet is strongly localized around the origin. The red current sheet extending toward the surroundings of the plasmoid located at $20<x<50$ is similar to the slow shock layer often observed in PK model. Hence, the first tearing instability may be expected to develop to the PK model but is immediately changed to SP-like sheet, as will be shown later. In addition, the color of the initial 1D current sheet observed in $100<x<400$ is yellow and hence the current intensity is weakened because of magnetic annihilation, by which the current sheet is gradually diffused by uniform resistivity $\eta_b$. Note that the initial 1D current sheet is not exactly the equilibrium under uniform resistivity. Eventually, the first tearing instability grows, overcoming the magnetic annihilation. The details of the first tearing instability are briefly presented in Appendix D.

Figure 12(b) shows the magnetic field lines and the current density $J_z$ at $t=308$. The SP-like sheet, which is shown in red, is formed around $0<x<75$. Additionally, a large-scale plasmoid, generated by the first tearing instability, grows around $75<x<200$. In addition, the second tearing instability is starting, and a new X-point appears around $x=75$. However, the new X-point is still invisible at this time.

Figure 12(c) shows the magnetic field lines and the current density $J_z$ at $t=360$. At this time, the second tearing instability develops around $x=140$, and the red current sheet is strongly localized around this point. As the second tearing instability proceeds, a new plasmoid appears around $80<x<130$. Since the second tearing instability is caused by the first tearing instability, it is a spontaneous tearing instability. In comparison to Figure 12(a), the second tearing instability develops more rapidly than the first tearing instability because the first tearing instability develops slowly during $4<t<300$, whereas the second tearing instability develops rapidly during $308<t<380$. After Figure 12(c), the third and fourth tearing instability start, leading to fully developed PI. 

\setcounter{figure}{11}
\renewcommand{\thefigure}{\arabic{figure}(\textbf{a})}

\begin{figure}
%\begin{center}
\vspace{20mm}
\includegraphics[bb=0 0 300 300,width=0.25\hsize]
{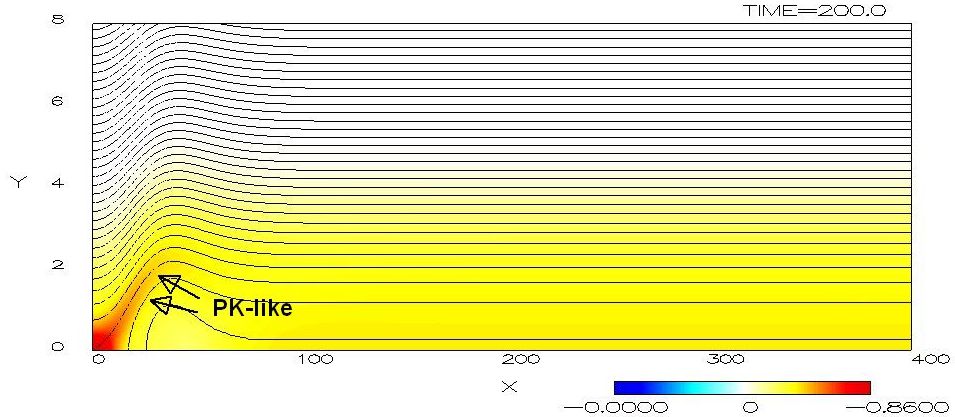}
   \caption{ Magnetic field lines and current density $J_z$ color contour map at $t=200$. This figure is shown for $0<x<L_x=400$ and $0<y<8(<<L_y=200)$, which is a limited part of the simulation box. Accordingly, note that the aspect ratio of this figure is extremely distorted to show the vicinity of the neutral sheet. The first tearing instability has been slowly developing around the origin. As in the PK-like current sheet, the red current sheet slightly splits from the neutral sheet, i.e., $y=0$.  }
   \label{fig12a}
%\end{center}
   \end{figure}

\setcounter{figure}{11}
\renewcommand{\thefigure}{\arabic{figure}(\textbf{b})}

\begin{figure}
%\begin{center}
\vspace{20mm}
\includegraphics[bb=0 0 300 300,width=0.25\hsize]
{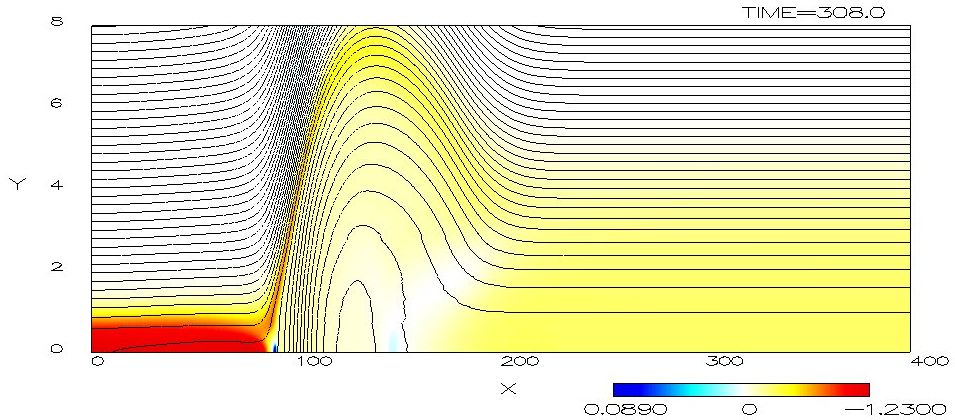}
   \caption{  Magnetic field lines and current density $J_z$ color contour map at $t=308$. The first tearing instability has been saturated, and an SP-like sheet is formed around $0<x<80$. At this time, the second tearing instability has started near $x=75$ but is still invisible. 
 }
   \label{fig12b}
%\end{center}
   \end{figure}

\setcounter{figure}{11}
\renewcommand{\thefigure}{\arabic{figure}(\textbf{c})}

\begin{figure}
%\begin{center}
\vspace{20mm}
\includegraphics[bb=0 0 300 300,width=0.25\hsize]
{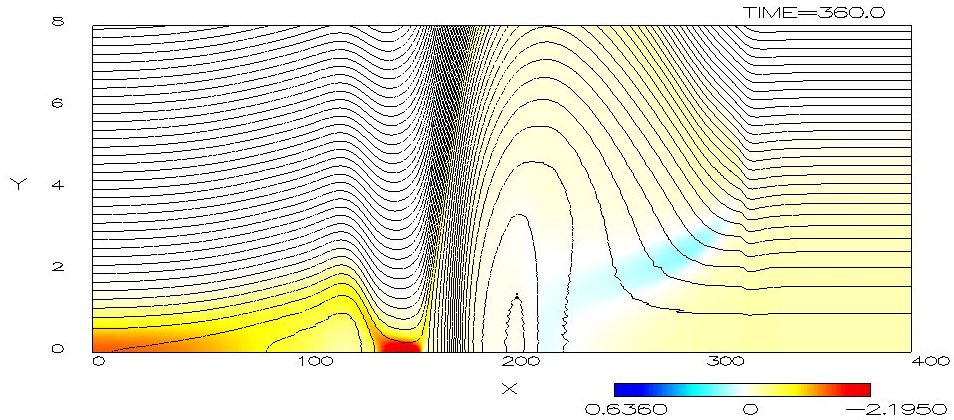}
   \caption{Magnetic field lines and current density $J_z$ color contour map at $t=360$. The second tearing instability is developing around $x=140$. 
 }
   \label{fig12c}
%\end{center}
   \end{figure}

Figure 13(a) shows the magnetic field lines and $B_y$ contour map, which is the reconnected field intensity at $t=308$. The red region represents $B_y>0$, and the blue region represents $B_y<0$. The $B_y$ intensity is indicated by the darkness of each color. In fact, the left-side region of the large-scale plasmoid formed around $75<x<150$ is deep red, and the right side is deep blue. These deep colors are associated with steady growth of the large-scale plasmoid. At this moment, the second tearing instability is starting, but the associated X-point and plasmoid are still invisible. However, a small light-blue region appears in the SP-like sheet, as indicated by a black arrow. The appearance of the light-blue region indicates the appearance of the second plasmoid and, hence, the beginning of the second tearing instability.

Figure 13(b) shows the magnetic field lines and $B_y$ contour map at $t=360$. In addition to the first (large-scale) plasmoid observed at $150<x<300$, the second plasmoid generated by the second tearing instability grows in $80<x<120$. Accordingly, the small light-blue region observed in Figure 13(a) develops into a vertically-wider deep-blue region in $120<x<140$ in Figure 13(b). Note that the aspect ratio of this figure is extremely distorted to show the vicinity of the neutral sheet.

\setcounter{figure}{12}
\renewcommand{\thefigure}{\arabic{figure}(\textbf{a})}

\begin{figure}
%\begin{center}
\vspace{20mm}
\includegraphics[bb=0 0 300 300,width=0.25\hsize]
{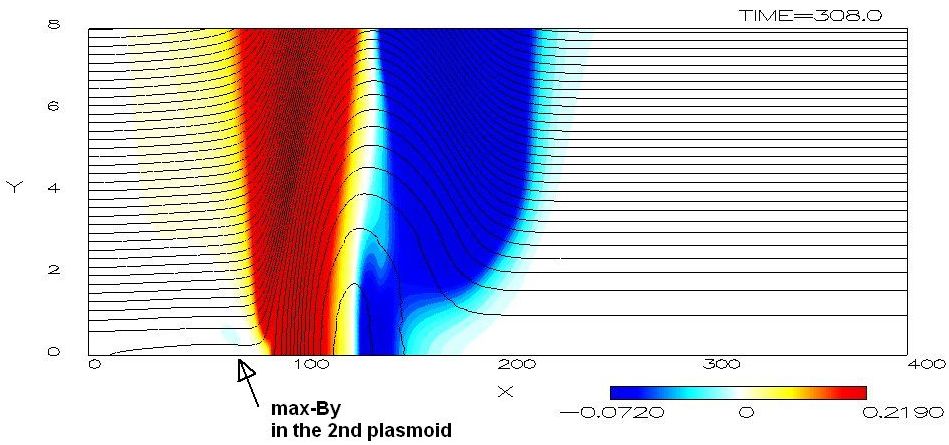}
   \caption{ Magnetic field lines and $B_y$ color contour map at $t=308$. The local maximum point of $B_y$ starts to appear around $x=75$ and is shown by the light-blue region indicated by a black arrow. This point is located in the second plasmoid generated by the second tearing instability. 
 }
   \label{fig13a}
%\end{center}
   \end{figure}

\setcounter{figure}{12}
\renewcommand{\thefigure}{\arabic{figure}(\textbf{b})}

\begin{figure}
%\begin{center}
\vspace{20mm}
\includegraphics[bb=0 0 300 300,width=0.25\hsize]
{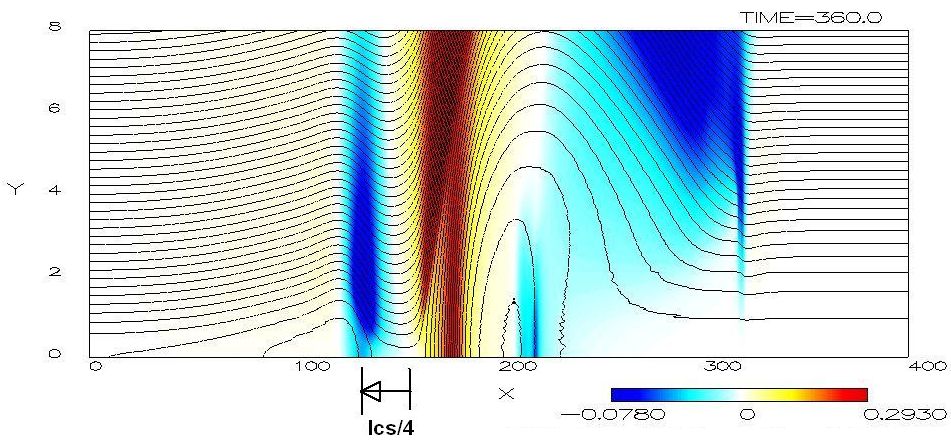}
   \caption{Magnetic field lines and $B_y$ color contour map at $t=360$. The local maximum point of $B_y(<0)$ grows around $x=120$ and shifts with the propagation of the second plasmoid around $0<x<150$. 
 }
   \label{fig13b}
%\end{center}
   \end{figure}

Figure 14(a) shows the magnetic field lines and $V_y$ contour map at $t=308$. The blue region, i.e., $V_y<0$, represents the inflow toward the current sheet, and the red region, i.e., $V_y>0$, represents the outflow from the current sheet. A negative high-intensity region of $V_y$ indicated by deep blue (i.e., labels A and B) is observed in the surroundings of the plasmoid at $75<x<150$. Meanwhile, the color around the X-point at the origin, i.e., $x=0$, is light blue or no color. This result is inconsistent with LSC theory because the highest negative $V_y$ (i.e., deep-blue) region assumed in the theory must be located around the X-point rather than around the plasmoid. This inconsistency will be associated with the compressibility and nonlinearity of this MHD simulation, while the theory is based on an incompressible and linearized MHD.

Figure 14(b) shows the magnetic field lines and $V_y$ contour map at $t=360$. As in Figure 14(a), a negative high-intensity region of $V_y$ (i.e., deep-blue region) is observed in the surroundings of the large scale plasmoid rather than the two X-points, i.e., at $x=0$ and $x=140$. This result suggests that the strong inflow to the SP-like sheet is driven by the movement of the plasmoids. 
In addition, it suggests that the plasma inflow around the X-point 
weakens as the plasmoid moves away from the X-point. 
To study how the inflow speed $V_y$ weakens, 
let us observe the time dependence of $V_y$. 

\setcounter{figure}{13}
\renewcommand{\thefigure}{\arabic{figure}(\textbf{a})}

\begin{figure}
%\begin{center}
\vspace{20mm}
\includegraphics[bb=0 0 300 300,width=0.25\hsize]
{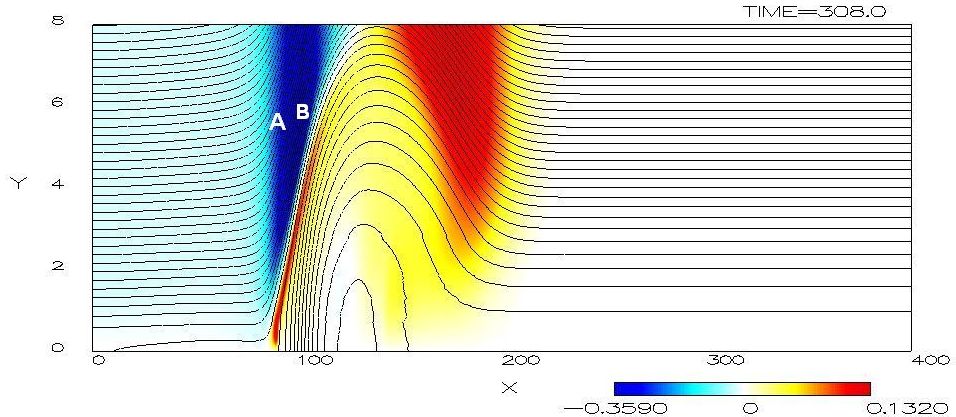}
   \caption{ Magnetic field lines and $V_y$ color contour map at $t=308$. Labels A and B show the $V_y<0$ region colored in blue, which represents the inflow region toward the SP-like sheet and plasmoid. 
 }
   \label{fig14a}
%\end{center}
   \end{figure}

\setcounter{figure}{13}
\renewcommand{\thefigure}{\arabic{figure}(\textbf{b})}

\begin{figure}
%\begin{center}
\vspace{20mm}
\includegraphics[bb=0 0 300 300,width=0.25\hsize]
{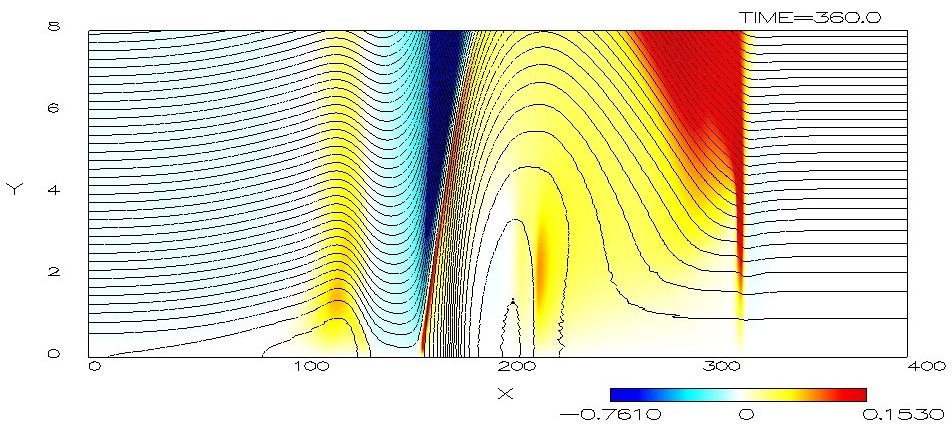}
   \caption{ Magnetic field lines and $V_y$ color contour map at $t=360$.}
   \label{fig14b}
%\end{center}
   \end{figure}

Figure 15 shows the magnetic field lines and $\Delta V_y$ contour map at $t=308$, where $\Delta V_y=V_y(t,x,y)-V_y(t-4,x,y)$ is defined as a velocity increment. The red region ($\Delta V_y>0$) indicates either deceleration of the inflow speed ($V_y<0$) or acceleration of the outflow speed ($V_y>0$). Inversely, the blue region indicates either acceleration of the inflow speed or deceleration of the outflow speed. Note that the locations of label A in Figure 14(a) and 15 are exactly the same. Since label A represents $V_y<0$ in Figure 14(a) and $\Delta V_y>0$ in Figure 15, the plasma inflow toward the SP-like sheet is decelerated around label A. Furthermore, the locations of label B in Figures 14(a) and 15 are also exactly the same. Since label B is $V_y<0$ in Figure 14(a) and $\Delta V_y<0$ in Figure 15, the plasma inflow toward the SP-like sheet is accelerated around label B. Hence, as the SP-like sheet elongates and the plasmoid moves away from the X-point, the plasma inflow around the plasmoid accelerates but the plasma inflow around the X-point decelerates. This deceleration suggests that the linear growth of the tearing instability is terminated as the SP-like sheet elongates, as will be discussed in Section 3.3.7.

\setcounter{figure}{14}
\renewcommand{\thefigure}{\arabic{figure}}

\begin{figure}
%\begin{center}
\vspace{20mm}
\includegraphics[bb=0 0 300 300,width=0.25\hsize]
{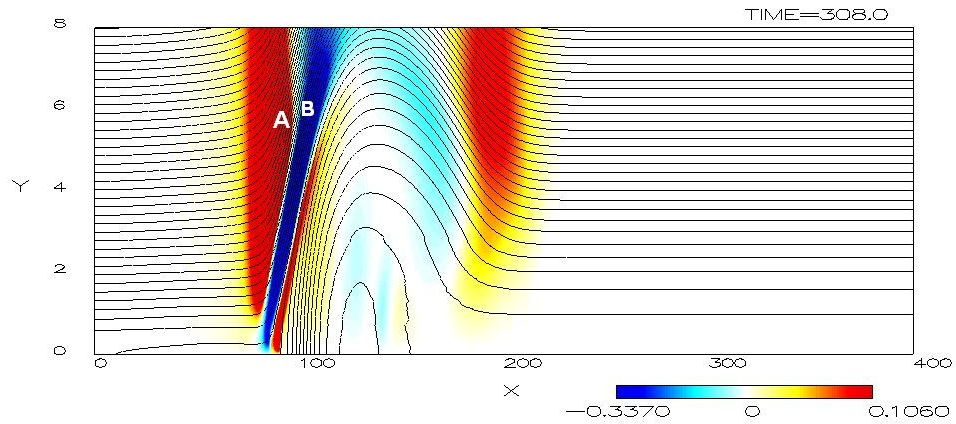}
\caption{ Magnetic field lines and $\Delta V_y$ color contour map at $t=308$. 
Labels A and B are the same locations as those in Fig.14(a). 
 }
   \label{fig15}
%\end{center}
   \end{figure}

\subsubsection{ $x$-directional profiles of $J_z$ and $V_x$}

In this section, we focus mainly on the $x$-directional behaviors of the second tearing instability observed during $308<t<380$. Figure 16 shows the $-J_z$ profile in the $x$-direction at $y=0$ at $t=308, 320, 340$, and $360$. As shown in Figures 12(a) to (c), keeping the first X-point at the origin, the second X-point has appeared at $t=360$. The $-J_z$ peak observed at the origin in Figure 16 corresponds to the first X-point, and another $-J_z$ peak corresponding to the second X-point appears around $x=80$ at $t=308$. Figure 16 shows that the second peak gradually grows, moving in the $+x$ direction. In FKR and LSC theories, $\psi''(0)>0$, which corresponds to $\Delta'>0$, predicts that this $-J_z$ peak appears at X-point. At this point, this MHD simulation is consistent with those theories.

\setcounter{figure}{15}
\renewcommand{\thefigure}{\arabic{figure}}

\begin{figure}
%\begin{center}
\vspace{40mm}
\includegraphics[bb=0 0 256 256,width=0.25\hsize]{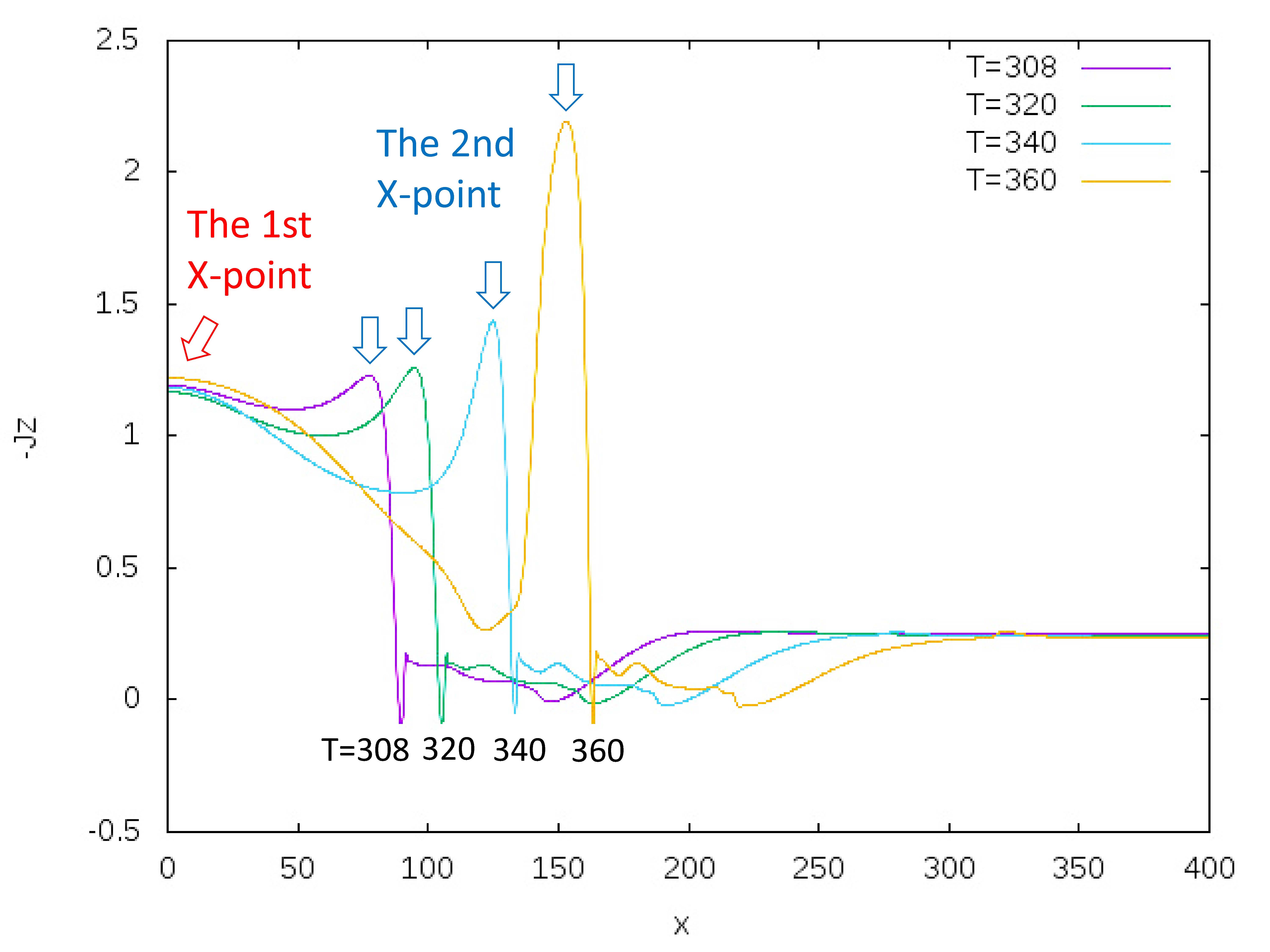}
   \caption{ The $x$-directional profile of the current density $-J_z$ along the neutral sheet ($y=0$) at $t=308, 320, 340$, and $360$. The first X-point is located at the $-J_z$ peak of $x=0$. The second X-point corresponds to another $-J_z$ peak moving from $x=75$ to $150$. }
   \label{fig16}
%\end{center}
   \end{figure}

Figure 17 shows the reconnection outflow speed $V_x$ profile in the $x$-direction at $y=0$ at $t=308, 320, 340$, and $360$. The $V_x$ peak, moving from $x=90$ at $t=308$ to $x=160$ at $t=360$, corresponds to where the reconnection jet collides with the plasmoid. Accordingly, the SP-like sheet is formed between the origin and the $V_x$ peak. This figure shows that $V_x$ does not yet reach the Alfven speed at these times, which is initially $V_A=5.5$ and almost unchanged over time. Hence, the first and second tearing instabilities do not yet reach the steady-state SP model. In addition, at $t=308$ and $320$, as $x$ increases, $V_x$ monotonically increases in the SP-like sheet, while at $t=340$ and $360$, the second domed $V_x$ peak newly appears in the SP-like sheet, which are, respectively, located at $x=75$ and $90$. The appearance of the second domed $V_x$ peak is associated with the interaction between the first and second tearing instabilities. In addition, since $V_x$ at the second X-point is not zero and takes a positive value, the second X-point itself is moving in the $+x$ direction. 
Because of the movement of X-point, the tearing instability in this MHD simulation cannot be directly applied to LSC theory, as will be discussed later in Section 3.3.4. 

Additionally, Figure 17 explains how to measure $L_{cs}$ in the modified LSC theory. Once $\partial V_x/\partial x$ is measured at X-point, $L_{cs}$ is calculated from $L_{cs}=2V_A/(\partial V_x/\partial x)$, where $V_A$ is measured in the upstream magnetic field. In the same manner, $L_{cs}$ can be measured in each tearing instability, such as first, second, third and so on.

\setcounter{figure}{16}
\renewcommand{\thefigure}{\arabic{figure}}

\begin{figure}
%\begin{center}
\vspace{40mm}
\includegraphics[bb=0 0 256 256,width=0.25\hsize]{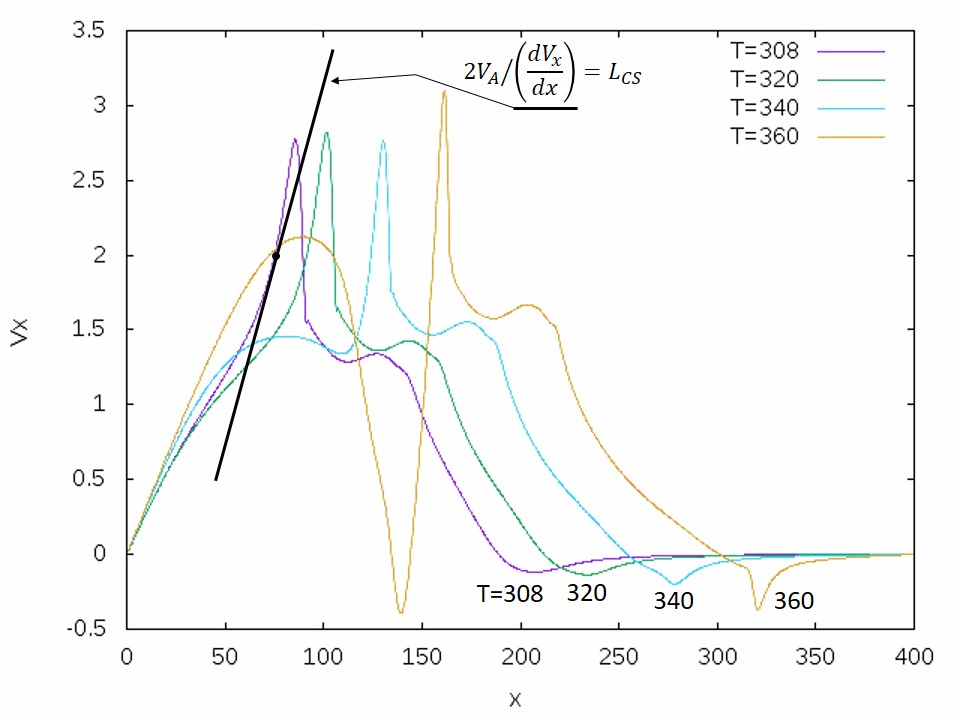}
   \caption{ The $x$-directional profile of the outflow speed $V_x$ along the neutral sheet ($y=0$) at $t=308, 320, 340$, and $360$. As shown, $L_{cs}$ is measured from the Alfven speed $V_A$ and the spatial gradient $\partial Vx/\partial x$ at the X-point, which is located around $x=75$ at $t=308$. 
 }
   \label{fig17}
%\end{center}
   \end{figure}

\subsubsection{ $y$ directional profiles of $J_z$ at the X-point and the plasmoid}

Figure 18(a) shows the $-J_z$ profile in the $y$-direction at $t=308, 320, 340$, and $360$ at the second X-point. Note that since the symmetry boundary condition is set at $y=0$, $-J_z$ profile for $y<0$ is mirrored by this $-J_z$ profile for $y>0$. In $308<t<340$, the height and width of the $-J_z$ peak located at $y=0$ are almost unchanged, but the $-J_z$ peak rapidly becomes higher during $340<t<360$. Thus, the rapid increase of the $-J_z$ peak is delayed by the growth of the second tearing instability started at $t=308$. In addition, Figure 18(a) shows that the $-J_z$ peak has a double current sheet structure during $308<t<340$. In other words, the thin current sheet of $0<y<0.5$ is embedded in the thick current sheet of $0<y<2$. These two sheets are separated by the dent indicated by label A in Figure 18(a). This double current sheet structure is similar to what was observed in a previous study \citep{PapiniApj2019}. This double current sheet structure cannot be directly applied to LSC theory, as will be discussed later in Section 3.3.6.

Figure 18(b) shows the $-J_z$ profile in the $y$-direction at $t=308, 320, 340$, and $360$ in the second plasmoid, where these profiles are plotted at the local maximum point of $B_y$ at each time, e.g., which is located in the blue contour region around $x=75$ of Figure 13(a) and $x=120$ of Figure 13(b). 
The $-J_z$ peak at $t=360$, which is indicated by label A, is separated from $y=0$. This separation is caused by the plasmoid formation, where label A corresponds to the outer edge of the plasmoid. Hence, as the tearing instability is developed, the current sheet around the plasmoid shown in Figure 18(b) gradually thickens due to the growth of the plasmoid, while the thickness around the X-point shown in Figure 18(a) gradually thins. As a result, the current sheet thickness gradually becomes nonuniform along the sheet. This nonuniformity means that the linear theory, such as the modified LSC theory, is inapplicable at this moment. It has entered into a nonlinear phase until $t=360$.

\setcounter{figure}{17}
\renewcommand{\thefigure}{\arabic{figure}(\textbf{a})}

\begin{figure}
%\begin{center}
\vspace{40mm}
\includegraphics[bb=0 0 256 256,width=0.25\hsize]{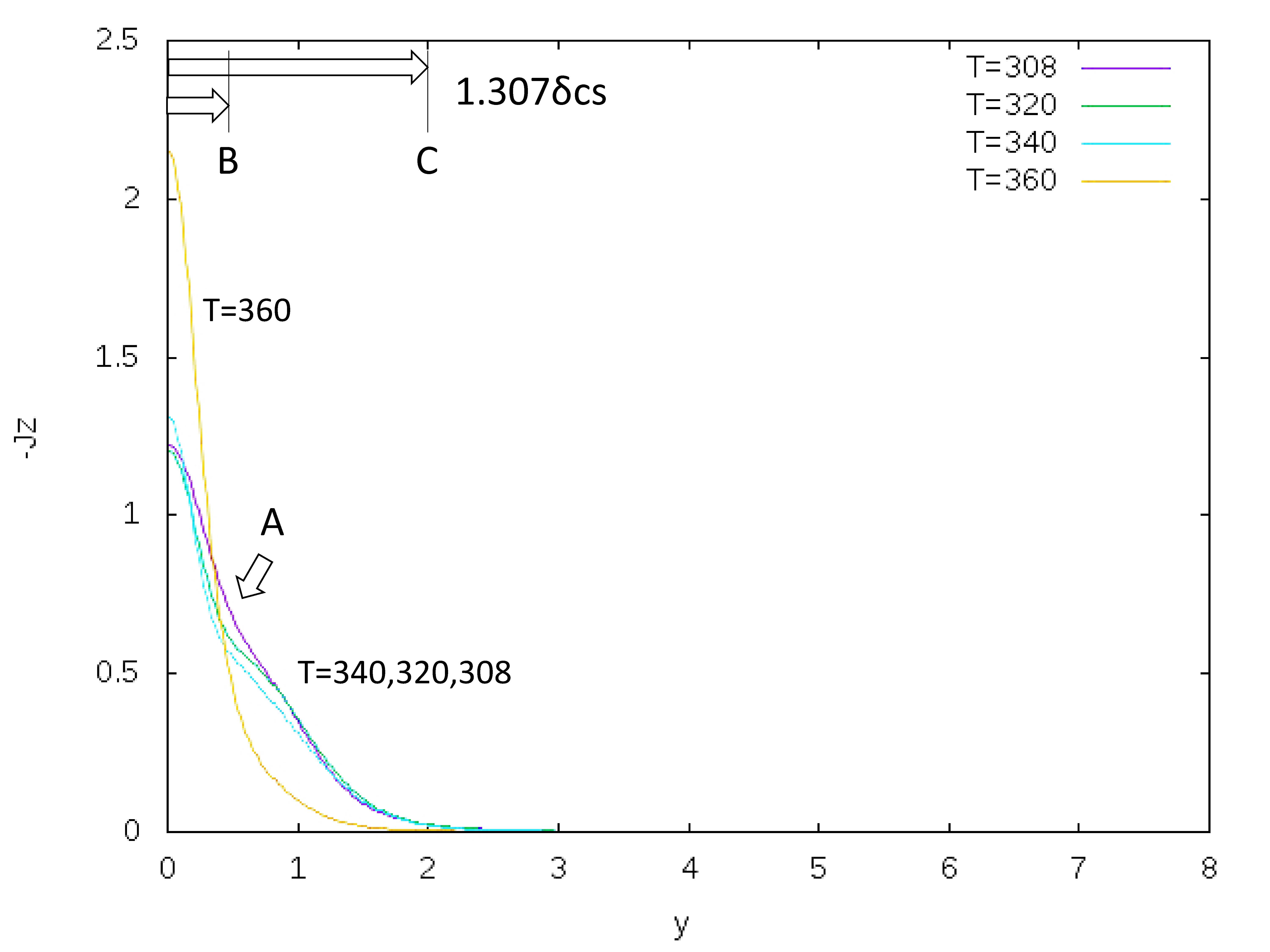}
   \caption{  The $y$-directional profile of the current density $-J_z$ at the second X-point shown in Fig.12(c). The X-points at $t=308, 320, 340$, and $360$, respectively, are located at $x=78.9, 97.5, 127.8$, and $155.1$. The dent in the $-J_z$ profile observed at $t=320$ and $340$ is indicated by label A, which is the boundary between the thin (inner) and thick (outer) current sheets. 
 }
   \label{fig18a}
%\end{center}
   \end{figure}

\setcounter{figure}{17}
\renewcommand{\thefigure}{\arabic{figure}(\textbf{b})}

\begin{figure}
%\begin{center}
\vspace{40mm}
\includegraphics[bb=0 0 256 256,width=0.25\hsize]{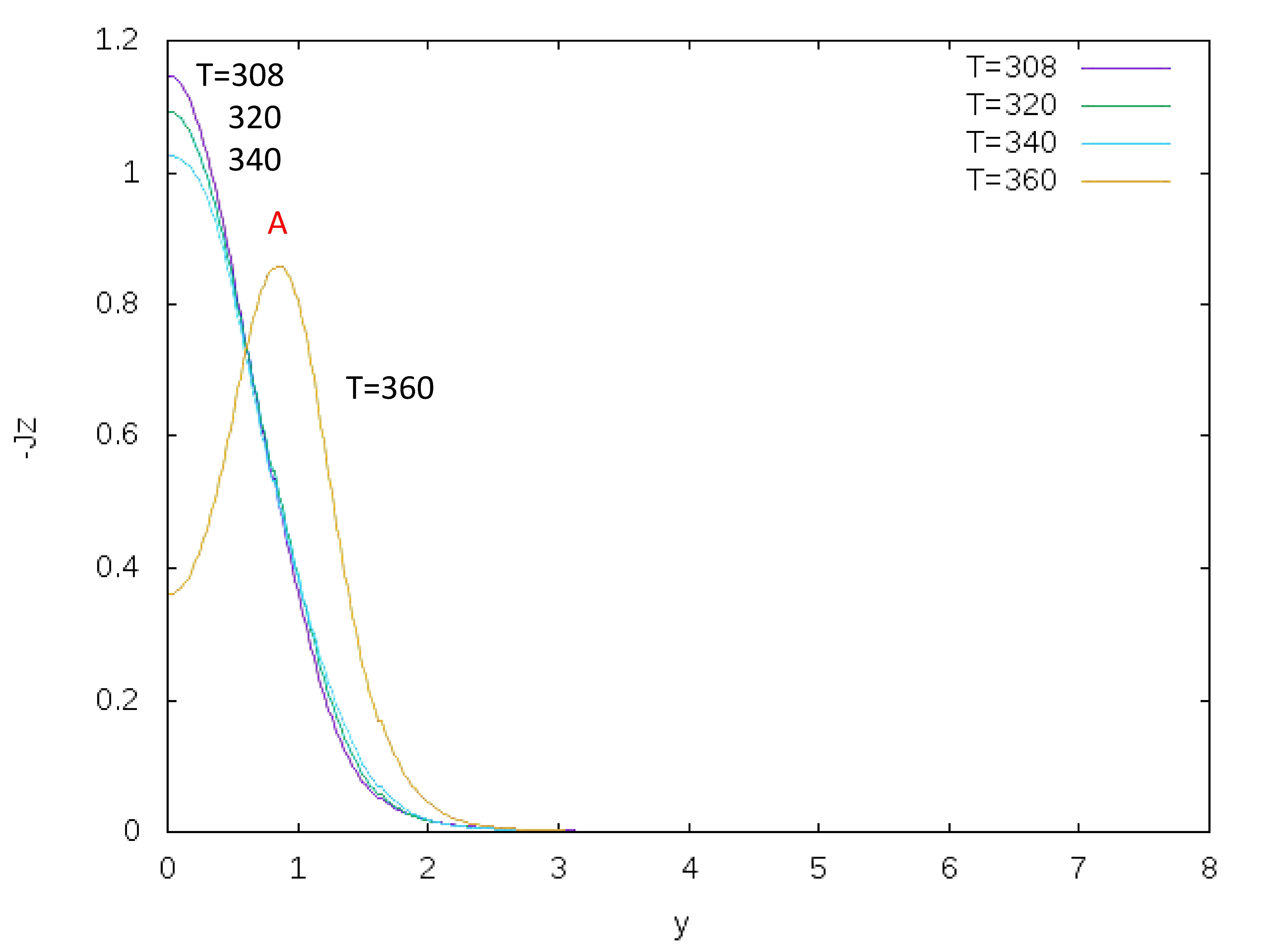}
   \caption{ The $y$-directional profile of the current density $-J_z$ in the second plasmoid. These profiles are plotted at the local maximum points of $B_y$ at $t=308, 320, 340$, and $360$, which are, respectively, located at $x=67.5, 84.6, 114.0$, and $134.1$. 
 }
   \label{fig18b}
%\end{center}
   \end{figure}

\subsubsection{The local maximum point of $B_y$ in the plasmoid}

Figure 19(a) shows the $-B_y$ profile in the $y$-direction at $t=308, 320, 340$, and $360$ in the second plasmoid.  This $B_y$ profile corresponds to the $\psi$ in LSC theory. In fact, as shown in Figure 19(a), these profiles have a local maximum point indicated by thick blue arrows. These profiles are plotted at the local maximum point of $B_y$ in the second plasmoid at each time. Hence, the $x$ locations plotted at each time are exactly the same as those in Figure 18(b). 
First, the local maximum point of $B_y$ appears around $y=0.6$ at $t=308$, 
which corresponds to the small light-blue region in Figure 13(a). 
Then, the local maximum point gradually shifts to a larger $y$ value from around $y=0.6$. Finally, it grows and reaches $y=1.56$ at $t=360$, which is located around the outer edge of the growing plasmoid, i.e., the outer edge of the current sheet of $t=360$ of Figure 18(b). Since LSC theory assumes that the local maximum point does not move, this MHD simulation is inconsistent with the theory at this point.

Figure 19(b) shows how the local maximum point of $-B_y$ observed in Figure 19(a) moves in the $y$-direction. The purple solid line shows the $y$-directional movement with respect to time. The green and blue solid lines, respectively, show the movements of the locations of the one-half and one-quarter value of the maximum value of the $-J_z$ peaks measured in Figure 18(b). The dashed line shown only for $304<t<308$ is the prediction line extrapolated from the purple solid line by which the generation point of the local maximum point of $B_y$ can be deduced. Since the beginning of the second tearing instability is during $304<t<308$, the local maximum point of $B_y$ is deduced to be generated at $y=0.42$ at $t=304$. Hence, at the beginning of the tearing instability, the local maximum point of $B_y$ appears to be separated from the origin, i.e. $y=0$.

\setcounter{figure}{18}
\renewcommand{\thefigure}{\arabic{figure}(\textbf{a})}

\begin{figure}
%\begin{center}
\vspace{40mm}
\includegraphics[bb=0 0 256 256,width=0.25\hsize]{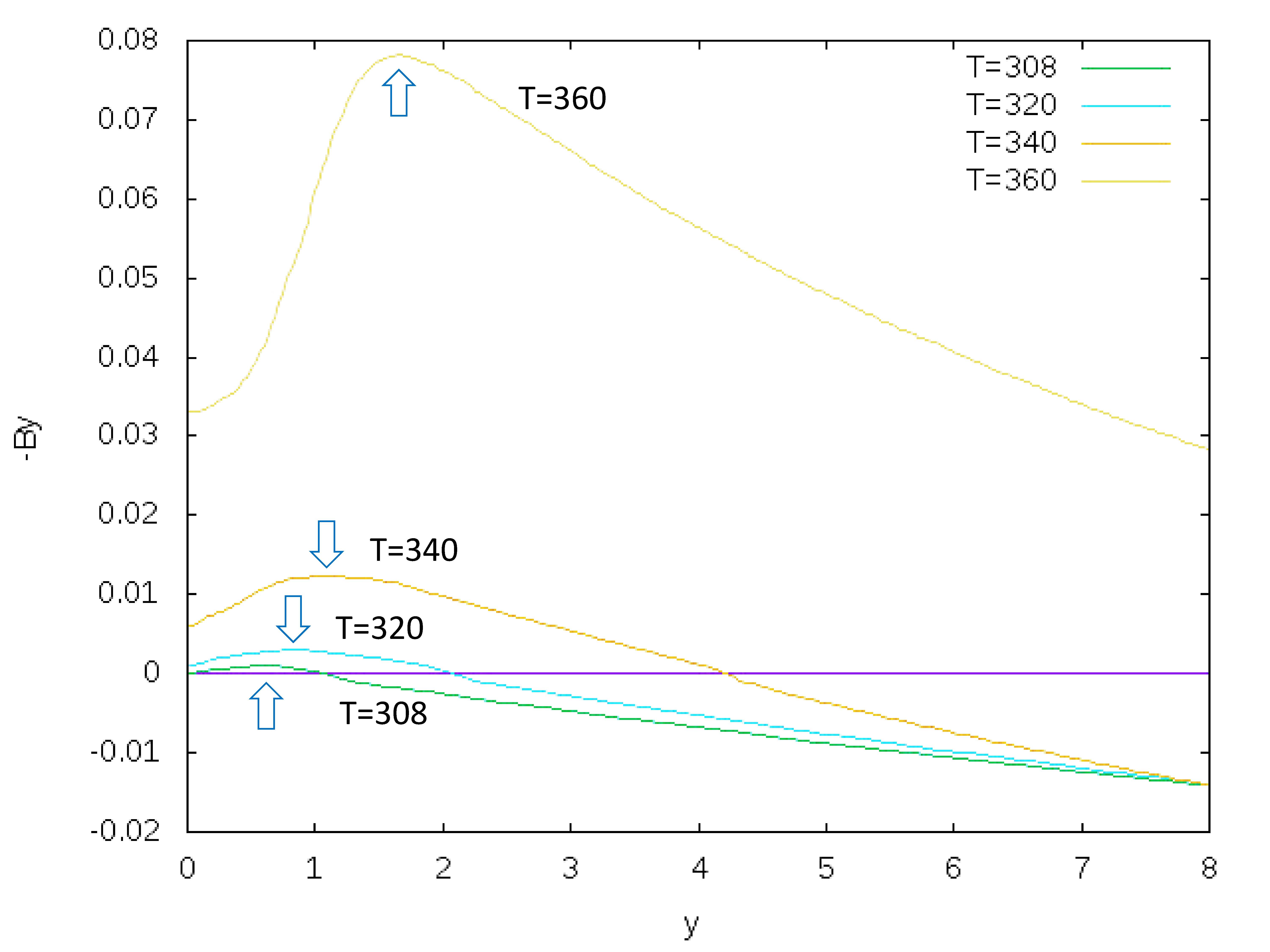}
   \caption{ The $y$-directional profiles of $-B_y$ plotted at the same location as that in Fig.18(b). The local maximum points of $B_y$ in the second plasmoid are indicated by thick blue arrows in each profile. The local maximum point of $B_y$ at $t=308$ is barely observed as the small light-blue region in Fig.13(a). 
 }
   \label{fig19a}
%\end{center}
   \end{figure}

\setcounter{figure}{18}
\renewcommand{\thefigure}{\arabic{figure}(\textbf{b})}

\begin{figure}
%\begin{center}
\vspace{20mm}
\includegraphics[bb=0 0 256 256,width=0.3\hsize]
{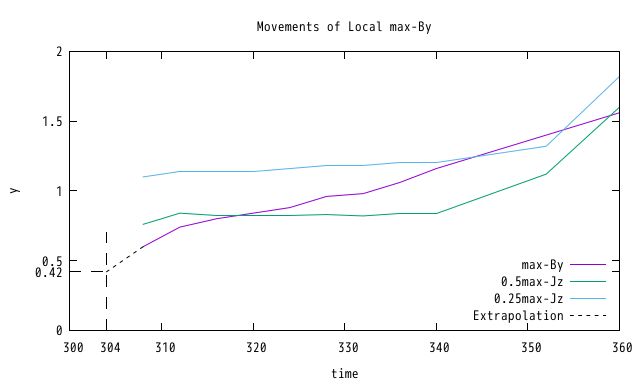}
   \caption{ The $y$-directional movements of the local maximum point of $B_y$. The movements of the half (green solid line) and quarter (blue solid line) thickness locations of the current sheet are indicated at the same $x$ locations as measured in Fig.18(b). The local maximum point appears somewhere in $304<t<308$ because it is not yet observed at $t=304$ but is observed at $t=308$, as shown in Figs.13(a) and 19(a). Hence, the generation point of the local maximum point predicted from the movement in $t>308$ is around $y=0.42$ at $t=304$. 
 }
   \label{fig19b}
%\end{center}
   \end{figure}

\subsubsection{The local maximum point of $\Delta V_y$ around the X-point}

Figure 20(a) shows the $-V_y$ profile in the $y$-direction at $t=308, 320, 340$, and $360$ at the second X-point. As the tearing instability grows, the plasma inflow toward the X-point is accelerated in $308<t<340$. In fact, since the inflow speed $V_y$ takes a negative value, i.e., $V_y<0$, the $-V_y$ profile shown in Figure 20(a) tends to rise over time until $t=340$. Then, the overall $-V_y$ profile for $y>1.5$ starts to fall between $t=340$ and $360$. This fall indicates that the second tearing instability starts to terminate. Nevertheless, the second plasmoid still continues to grow beyond $t=340$, as shown in Figure 19(a). Thus, the reconnection process is maintained in the SP-like sheet beyond $t=340$. In addition, the $-V_y$ profiles in $308<t<340$ are slightly distorted around $0.4<y<1.0$. This distortion is associated with the double current sheet structure observed in Figure 18(a). Then, the distortion disappears until $t=360$; this disappearance appears to be associated with the disappearance of the double current sheet structure in Figure 18(a).

To adapt the $V_y$ profile observed in Figure 20(a) to $\phi$ in the modified LSC theory, at least, three problems must be considered. First, the largest problem is that $V_y$ does not have only the perturbed component $\phi$ but also the zero-order component defined by Eqs.(2.5) and (2.10). The perturbed component $\phi$ is difficult to rigorously extract from $V_y$ because at the beginning of the second tearing instability, i.e., $t=304 \sim 308$, the current sheet has already been in an unsteady state that is not the equilibrium. However, the growth of $\phi$ can be approximately measured by the increment of $V_y$, i.e., $\Delta V_y=V_y(t,x,y)-V_y(t-\Delta T,x,y)$, where $\Delta T=4$ is the data sampling time in the MHD simulation. Then, the growth rate $\Lambda$ normalized by the MHD simulation time scale is calculated from $\Delta V_y(t)/\Delta V_y(t-\Delta T)=e^{\Lambda \Delta T}$, and then, translated to the growth rate $\lambda_{\Delta V_y}$ normalized in the time scale of the modified LSC theory, as will be shown in Table 1.

The second and third problems, respectively, originate in the compressibility and moving of the second X-point in the MHD simulation. Because of the compressibility, the intensive region of $-\Delta V_y(\propto \partial V_y/\partial t)$ in Figure 15 is located around the plasmoid rather than around the X-point. This result is inconsistent with the assumption in incompressible LSC theory that the local maximum point of $\phi$ is located around the X-point. Adapting to the theory, we measure the local maximum value of $\Delta V_y(t)$ around the X-point by using the $-V_y$ profiles shown in Figure 20(a). At this point, $-\Delta V_y$ around the X-point will not be seriously affected by the compressibility because the measurement point is close to the X-point. In addition, the third problem is that the X-point gradually moves from $x=78.9$ to $x=155.1$ during $308<t<360$. At this point, $\Delta V_y$ must be measured by an observer attached to the moving X-point. As a result, the $\Delta V_y$ value is different from $\Delta V_y$ in Figure 15, which was measured by an observer standing on the ground.

Figure 20(b) shows the $y$-directional profiles of $-\Delta V_y$, which is redefined as $\Delta V_y=V_y(t,x_1(t),y)-V_y(t-\Delta T,x_1(t-\Delta T),y)$, where $x_1$ is the $x$ location of the moving X-point and $\Delta T=4$. Hence, this $-\Delta V_y$ is measured by an observer attached to the moving X-point. In Figure 20(b), the local maximum point of the redefined $-\Delta V_y$ is barely observed at $t=308$. Then, the point for $312<t<360$ moves in $0.3<y<0.6$ and is always in the inner region of the current sheet shown in Figure 18(a). Then, during $340<t<360$, the peak height rapidly decreases, indicating that the growth of the second tearing instability is terminated around $t=340$.

Figure 20(c) shows how the local maximum point of $-\Delta V_y$ redefined in Figure 20(b)  moves in the $y$-direction with respect to the thickness of the current sheet observed in Figure 18(a). 
The dashed line at $304<t<312$ predicts the generation location of the local maximum point, which is extrapolated from the purple solid line observed at $312<t$. The local maximum point of $-\Delta V_y$ is predicted to be generated around $x=0.3$ at $t=304$. Hence, the local maximum point stagnates near the outer edge of the inner current sheet of the double sheet structure observed in Figure 18(a). At the beginning of the tearing instability, the local maximum point of $-\Delta V_y$  appears to be separated from the origin, i.e. $y=0$.

\setcounter{figure}{19}
\renewcommand{\thefigure}{\arabic{figure}(\textbf{a})}

\begin{figure}
%\begin{center}
\vspace{40mm}
\includegraphics[bb=0 0 256 256,width=0.25\hsize]{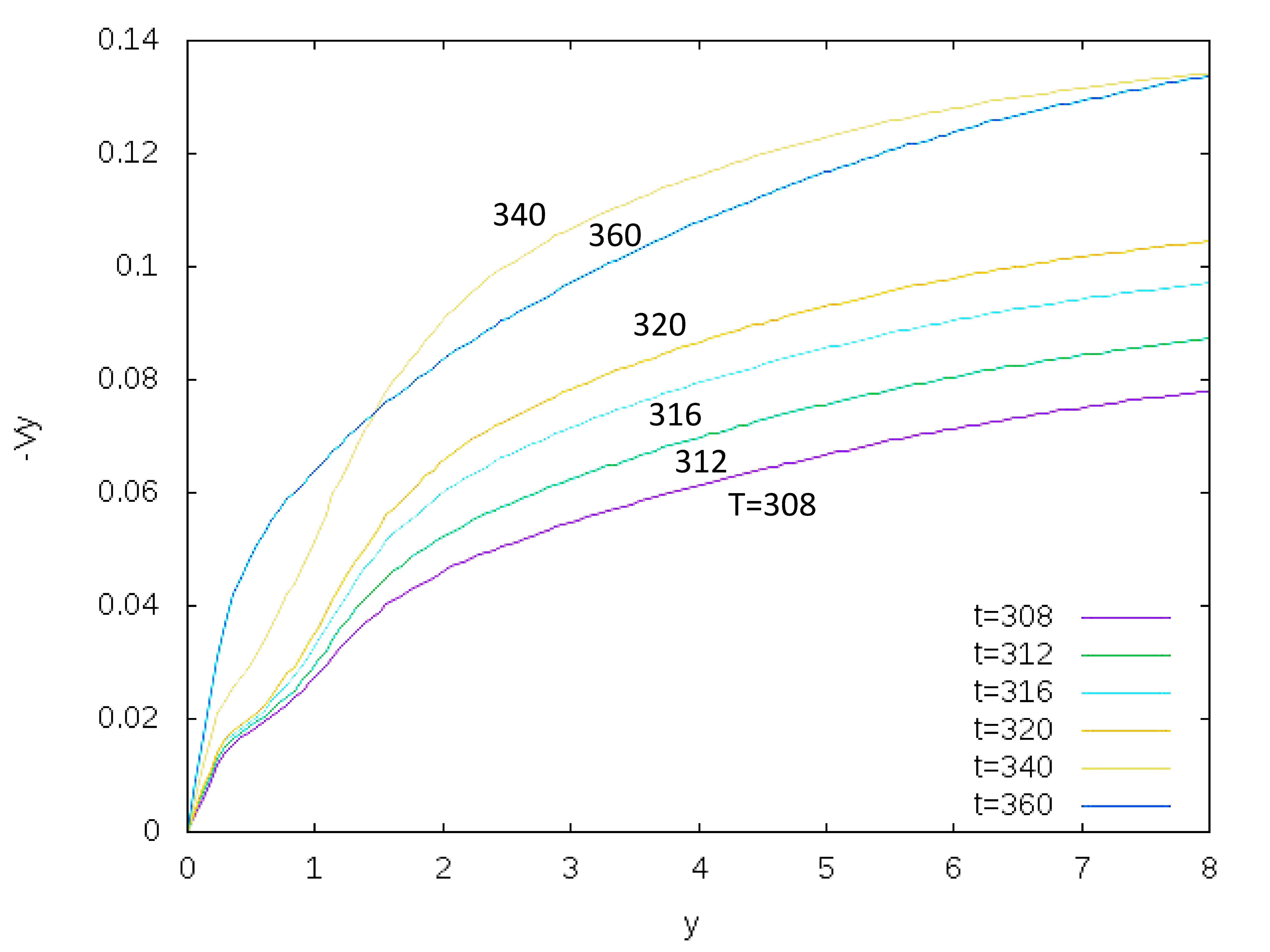}
   \caption{  The $y$-directional profile of $-V_y$ at the second X-point, where the $x$ location is the same as that in Fig.18(a). The dent in the profiles observed around $0.4<x<1$ and $308<t<340$ is associated with that of the double current sheet structure shown in Fig.18(a). 
 }
   \label{fig20a}
%\end{center}
   \end{figure}

\setcounter{figure}{19}
\renewcommand{\thefigure}{\arabic{figure}(\textbf{b})}

\begin{figure}
%\begin{center}
\vspace{40mm}
\includegraphics[bb=0 0 256 256,width=0.25\hsize]{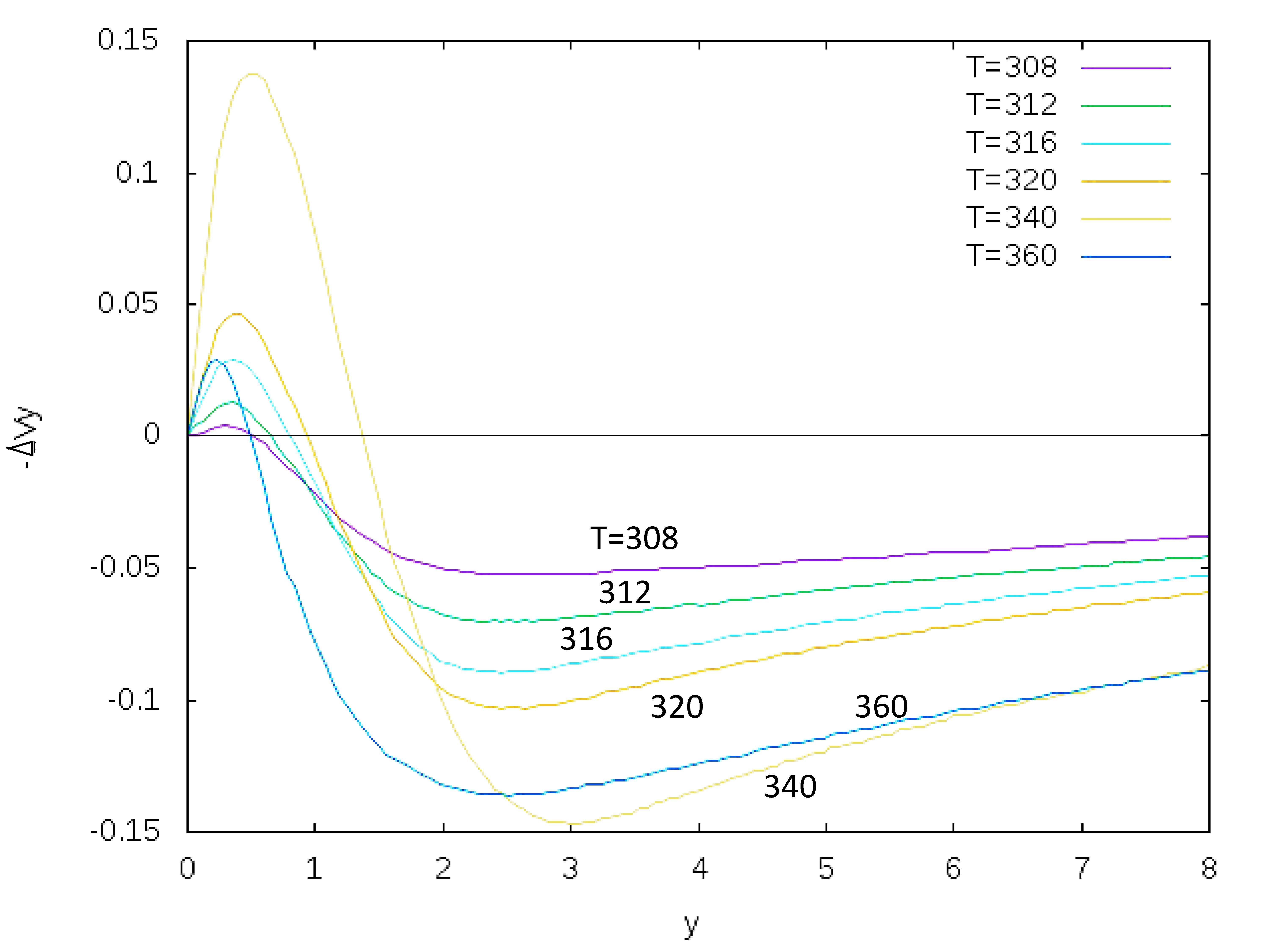}
   \caption{  The $y$-directional profile of $-\partial V_y/\partial t$ at the second X-point. This time differential is approximately measured as $\Delta V_y/\Delta T=(V_y(t,x_1(t))-V_y(t-4,x_1(t-4)))/4$ with time interval $\Delta T=4$, where $x_1(t)$ is the location of the X-point in Fig.18(a). Hence, this figure corresponds to $-\partial V_y/\partial t$ measured by an observer located at the moving X-point. In comparison to Fig.18(a), the local maximum points are always located inside the current sheet, i.e., $y<1$. 
 }
   \label{fig20b}
%\end{center}
   \end{figure}

\setcounter{figure}{19}
\renewcommand{\thefigure}{\arabic{figure}(\textbf{c})}

\begin{figure}
%\begin{center}
\vspace{40mm}
\includegraphics[bb=0 0 256 256,width=0.25\hsize]
{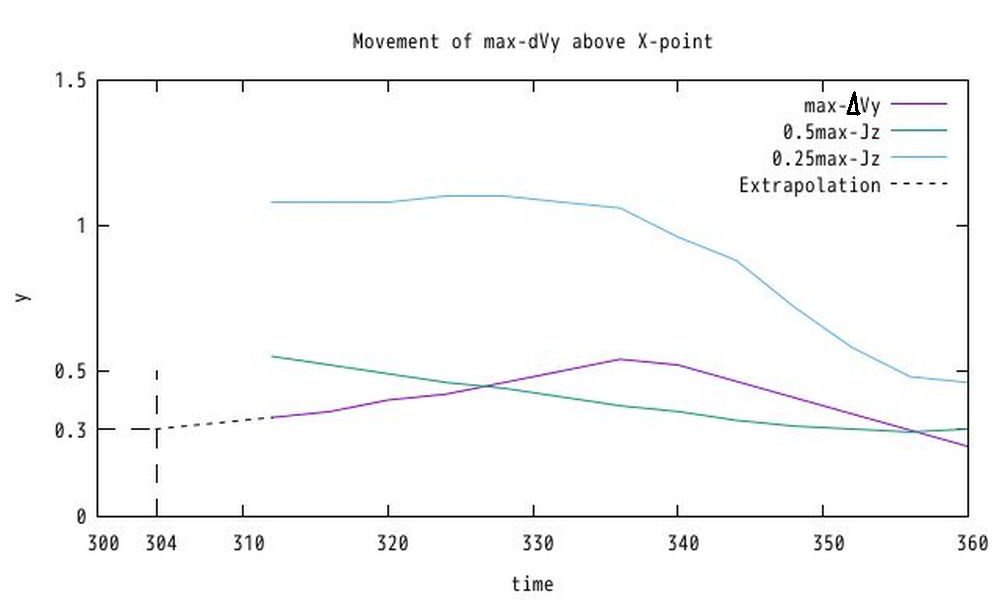}
   \caption{ The $y$-directional movements of the local maximum point of $-\partial V_y/\partial t$, i.e., $-\Delta V_y/\Delta T$, shown in Fig.20(b). The movements of the half and quarter thickness locations of the current sheet, which are measured in Fig.18(a), are also indicated. The local maximum point is not yet observed at $t=304$. The generation point of the local maximum point predicted from the movement observed at $t>312$ is around $y=0.3$ at $t=304$. 
 }
   \label{fig20c}
%\end{center}
   \end{figure}

\subsubsection{The growth rates, $\lambda_{B_y}$ and $\lambda_{\Delta V_y}$, 
measured in the MHD simulation}

Figure 21 shows the time variations of $B_{ymax}$ and $\Delta V_{ymax}$, which are defined as the local maximum values of $-B_y$ and $-\Delta V_y$, respectively, observed in Figures 19(a) and 20(b). Both $B_{ymax}$ and $\Delta V_{ymax}$ monotonically increase at the beginning of the tearing instability. Then, $B_{ymax}$ continues to increase until $t=360$, but $\Delta V_{ymax}$ is saturated around $t=340 \sim 350$. The former growth is still maintained by the reconnection process based on the SP model, but the latter saturation indicates the termination of the second tearing instability. In other words, around $t=340 \sim 350$, the second tearing instability switches from the linear phase to the nonlinear phase.

\setcounter{figure}{20}
\renewcommand{\thefigure}{\arabic{figure}}

\begin{figure}
%\begin{center}
\vspace{40mm}
\includegraphics[bb=0 0 256 256,width=0.25\hsize]
{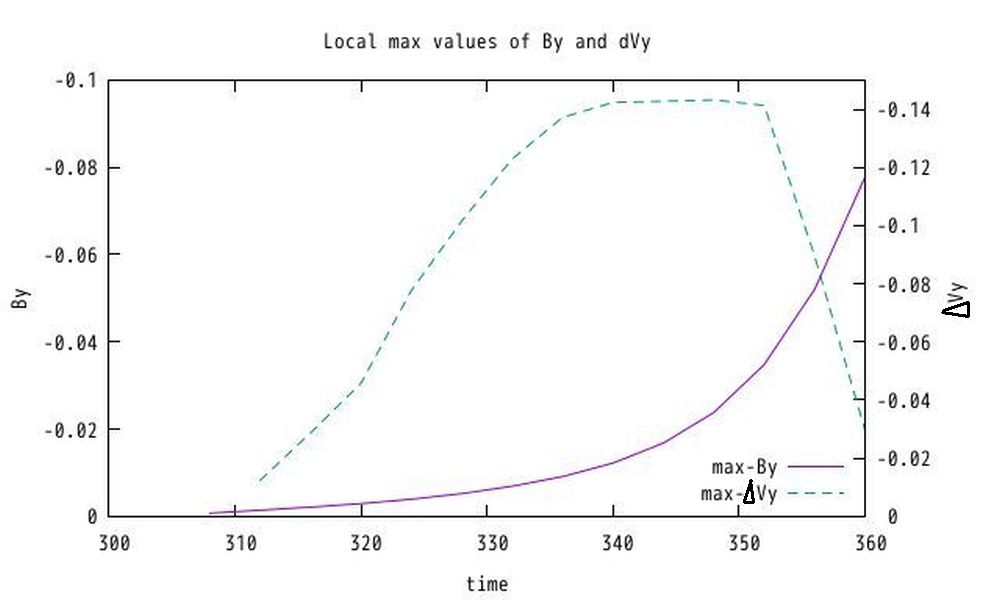}
   \caption{ The time variations of the local maximum values defined as $B_{ymax}$ and $\Delta V_{ymax}$, which are presented respectively by solid line and dashed line. The former is measured in Fig.19(a), and the latter is measured in Fig.20(b). The former (i.e., $B_{ymax}$) grows smoothly until the end of the second tearing instability, but the latter (i.e., $\Delta V_{ymax}$) is saturated at $t=340$, which is around the end of the linear phase of the instability.  }
   \label{fig21}
%\end{center}
   \end{figure}

Table 1 shows the numerical data plotted in Figure 21 and also the growth rates $\lambda_{B_y}$ and $\lambda_{\Delta V_y}$ calculated from the numerical data. For example, $\lambda_{B_y}=0.21$ shown at $t=312$ in Table 1 is calculated from $B_{ymax}(t=312)/B_{ymax}(t=308)=-0.0014/-0.0007=e^{2\pi \lambda_{B_y} (312-308)V_A/l{cs}}$, where $V_A=5.5$ is the Alfven speed measured in the upstream region and $l_{cs}$ is the average value between $t=308$ and $312$, i.e., $(l_{cs}(t=308)+l_{cs}(t=312))/2$ $=41.5$. As shown in Figure 13(b), $l_{cs}$ is measured as 4 times of the x-directional distance between the second X-point and the local maximum point of $B_y$ in the plasmoid left side of the second X-point, e.g., which is also indicated by the black arrow in Figure 13(a). 
At this point, since the plasmoid chain shown in Figure 13(b) is not exactly the sinusoidal shape assumed by linear theory, this $l_{cs}$ measurement is just an approximation, where $B_{ymax}$ is assumed to be located between X-point and O-point and the distance between the X-point and O-point is assumed to be $1/2$ of the wave length of the plasmoid chain. 
Table 1 shows that $l_{cs}$ almost monotonically increases from $41$ to $162$ during $312 \leq t \leq 380$, and $\lambda_{B_y}$ varies between $0.11$ and $0.23$ during $312 \leq t \leq 360$. In $364 \leq t \leq 380$, $-J_z$ at the X-point and $\lambda_{B_y}$ start to slowly decrease due to over-elongation of the SP-like sheet, which is characterized by simultaneous increases in $l_{cs}$ and $L_{cs}$. The $\lambda_{B_y}$ values can be compared with the upper limit $\lambda_{up}$ obtained in the modified LSC theory, as discussed below.

%\title{Table 1}
\begin{table}
\begin{center}

%\scalebox{0.8}{

\begin{tabular}{ccccccccccc}

\hline
$t$ & $J_z$ &
$B_{ymax}$ $(x,y)$ & 
$\lambda_{B_y}$ & 
$\Delta V_{ymax}$ $(x,y)$&  
$\lambda_{\Delta V_y}$ & $l_{cs}$& $L_{cs}$& 
$\kappa$ & 
$\lambda_{up}$ (in$\xi_b$) & 
$\lambda_{up}$ (out$\xi_b$) \\

\hline
308& -1.23 &
-0.0007 (68,0.54) & ---& 
-------- (78,--)& ---& 
 42& 150& 11.2&---&---\\
\hline
312& -1.22 &
-0.0014 (75,0.58) & \bf{0.21}& 
-0.0124 (85,0.34)& ---& 
 41& 127& 9.8& 0.4 (0.9)&$<$ 0.1 (0.2)\\
\hline
316& -1.21 &
-0.0021 (80,0.64) & 0.14& 
-0.0287 (92,0.34)& \bf{0.27}& 
 47& 114& 7.6& 0.4 (1.0)&$<$ 0.1  (0.2)\\
\hline
320& -1.21 &
-0.0029 (85,0.72) & 0.12& 
-0.0458 (98,0.36)& 0.17& 
 52& 106& 6.5& 0.4 (1.0)&$\sim$ 0.1  (0.3)\\
\hline
324& -1.19 &
-0.0039 (90,0.74)& 0.12& 
-0.0777 (104,0.42)& 0.20& 
 54& 100& 5.8& 0.4 (1.0)&$\sim$ 0.1  (0.3)\\
\hline
328& -1.19 &
-0.0052 (97,0.78) & 0.11& 
-0.1016 (110,0.46)& 0.10& 
 53& 97& 5.8& 0.5 (1.2)&$\sim$ 0.1  (0.3)\\
\hline
332& -1.20 &
-0.0069 (103,0.86) & 0.11& 
-0.1229 (116,0.50)& 0.07& 
 52& 92& 5.6& 0.5 (1.3)&$\sim$ 0.1  (0.3)\\
\hline
336& -1.23 &
-0.0091 (109,0.92)&0.11&
-0.1372 (122,0.52)& 0.04&
 53& 83& 4.9& \bf{0.6} (\bf{1.4})&0.1 (0.4)\\
\hline
340& -1.28 &
-0.0122 (114,1.02) & 0.12& 
-0.1423 (128,0.56)& 0.01& 
 55& 75& 4.3& 0.6 (1.3)&$\sim$ 0.1  (0.3)\\
\hline
344& -1.39 &
-0.0168 (119,1.14) & 0.13& 
-0.1427 (134,0.52)& 0.001& 
 56& 66& 3.7& 0.6 (1.3)&$\sim$ 0.1  (0.3)\\
\hline
348& -1.55 &
-0.0238 (124,1.26)&0.15&
-0.1432 (139,0.48)& 0.002&
 60& 58& 3.0& 0.5 (1.0)&$\sim$ 0.1  (0.3)\\
\hline
352& -1.74 &
-0.0348 (129,1.36) & 0.18& 
-0.1413 (145,0.44)& \bf{-0.001}& 
 65& 51& 2.5& 0.5 (0.9)&$<$ 0.1  (0.2)\\
\hline
356& -1.98 &
-0.0520 (131,1.46) & 0.22& 
-0.0895 (150,0.32)& -0.23& 
 74& \bf{50}& 2.1&0.4 (0.8)&$<$ 0.1  (0.2)\\
\hline
360& -2.16 &
-0.0778 (134,1.66)& 0.23& 
-0.0295 (155,0.24)& -0.63&
 84& 53& 2.0& 0.3 (0.7)&$<$ 0.1  (0.2)\\
\hline
364& \bf{-2.23} &
-0.1112 (136,1.76)& 0.23& 
-0.0004 (160,0.04)& ---&
 95& 60& 2.0& ---& ---\\
\hline
368& -2.22 &
-0.1477 (138,1.80)& 0.21&
-------- (165,--)& ---&
 107& 70& 2.1& ---& ---\\
\hline
372& -2.17 &
-0.1837 (140,2.00)& 0.18&
-------- (170,--)& ---&
 122& 82& 2.1& ---& ---\\
\hline
376& -2.09 &
-0.2183 (141,2.18)& 0.18&
-------- (176,--)& ---&
 139& 95& 2.1& ---& ---\\
\hline
380& -2.01 &
-0.2508 (142,2.36)& 0.18&
-------- (182,--)& ---&     
 162& 104& 2.0& ---& ---\\
\hline
\end{tabular}

\caption{
The growth rates, $\lambda_{B_y}$ and $\lambda_{\Delta V_y}$, respectively, are measured from the time variations in the local maximum values, $B_{ymax}$ and $\Delta V_{ymax}$, shown in Fig.21. $J_z$ is measured at X-point. How to measure $l_{cs}$ and $L_{cs}$ is explained in Figs.13(b) and 17, respectively. Then, $\kappa=\pi L_{cs}/l_{cs}$ is obtained. The upper limit $\lambda_{up}$ is obtained from Fig.8, for either of the inner and outer $\xi_b$, where $\epsilon=0$ is assumed. The inner $\xi_b$ ($=$in$\xi_b$) and outer $\xi_b$ ($=$out$\xi_b$) are measured for the inner and outer current sheet, respectively shown as labels B and C of Fig.18(a). 
}

% } %scalebox

\end{center}
\end{table}

Similarly, $\lambda_{\Delta V_y}$ shown in Table 1 is approximately calculated in the same manner as $\lambda_{B_y}$. For example, $\lambda_{\Delta V_y}=0.27$ shown at $t=316$ in Table 1 is calculated from $\Delta V_{ymax}(t=316)/\Delta V_{ymax}(t=312)=-0.0287/-0.0124=e^{2\pi \lambda_{\Delta V_y}(316-312)V_A/l{cs}}$. As shown in Table 1, $\lambda_{\Delta V_y}$ takes its highest value at $t=312$, i.e., the beginning of the tearing instability, and then monotonically decreases over time. Notably, $\lambda_{\Delta V_y}$ takes negative values after $t=352$; thus, the exponential growth of the tearing instability is terminated at this time. Since $\lambda_{B_y}>0$ is maintained even at $t>352$, the reconnection process still continues to follow the SP model, maintaining the growth of the plasmoid. The $\lambda_{\Delta V_y}$ values can be compared with $\lambda_{up}$ in the modified LSC theory, as shown next.

\subsubsection{The upper limit $\lambda_{up}$ of 
growth rate predicted by the modified LSC theory}

This section presents how to compare the growth rates obtained from the MHD simulation and modified LSC theory, where it is shown that both growth rates, to some extent, appear to be consistent. First, note that $\lambda_{up}$ obtained from Figure 8 is just the upper limit of the growth rate. In the theory, the exact growth rate is unknown but generally depends on the upstream condition, such as zero-converging, zero-crossing and so on. The upstream condition in the theory is not easy to be matched to that of MHD simulation. Rather, let us focus on the fact that the exact growth rate is always smaller than $\lambda_{up}$. As mentioned in Sections 2.3.4 and 5, $\lambda_{up}$ is determined from $\kappa$, $\epsilon$, and $\xi_b$, which can be measured in the MHD simulation result. In fact, $\kappa$, which is defined as the ratio of $L_{cs}$ and $l_{cs}$, is shown in Table 1. The measurement of $l_{cs}$ is shown in Figure 13(b). The assumption of $\epsilon=0$ and measurements of $L_{cs}$ and $\xi_b$ are presented below.

As shown in Figure 17, $L_{cs}$ is obtained from $\partial V_x/\partial x$ at the X-point and Alfven speed $V_A$. This definition of $L_{cs}$ is essentially the same as that in the original LSC theory but the measurement method is evidently different. Because, LSC theory assumes that the SP sheet is in the steady state, but the MHD simulation in this section has not yet reached a steady state. In fact, the $V_x$ profile shown in Figure 17 does not reach $V_A=5.5$ at the maximum peak. Even in such a unsteady state, $L_{cs}$ can be measured. According to our measurement, Table 1 shows that $L_{cs}$ decreases from $150$ to $50$ during $308 \leq t \leq 356$, which is associated with the fact that the magnetic diffusion region is localized by the tearing instability until $t=356$. After $t=356$, $L_{cs}$ starts to increase, which means that the magnetic diffusion region starts to elongate along the current sheet, resulting in the SP-like current sheet.

The current sheet thickness $\delta_{cs}$ must be measured to determine $\epsilon=2 \delta_{cs}/L_{cs}$. As shown in Figure 18(a), the thickness is measured at approximately $y=2$, that is, between $y=0$ and the outer edge of the current sheet, and it is much smaller than the $L_{cs}$ shown in Table 1, resulting in $\epsilon<0.1$. According to Figure 7, $\lambda_{up}$ is not sensitive to $\epsilon$ when $\epsilon<0.1$ and $\kappa<2$; hence, below we assume $\epsilon=0$. Figure 7 shows that $\lambda_{up}$ rapidly decreases in $\kappa>2$. However, since $\lambda_{up}$ tends to decrease for larger $\epsilon$, $\lambda_{up}$ obtained for $\epsilon=0$ will be available even for $\kappa>2$. Then, $\lambda_{up}$ can simply be obtained from Figure 8 when $\kappa$ and $\xi_b$ are known. By contrast, $\lambda_{up}$ is sensitive to $\xi_b$.

The measurement of $\xi_b$, where $\xi_b$ is equal to either the local maximum point's location of $-\Delta V_y$ or $B_y$, is associated with some controversial problems to be studied in the future. In this paper, we employ $-\Delta V_y$ to measure $\xi_b$. At this point, we can employ $B_y$. However, as shown in Figures 19(b) and 20(c), the $\xi_b$ values obtained from $B_y$ and $-\Delta V_y$ are not drastically different; thus, we employ $-\Delta V_y$ to measure $\xi_b$ in Table 1.

Another problem to be studied in the future is that the current sheet consists of double sheets, as shown in Figure 18(a). Therefore, two choices, i.e., the inner or outer current sheet, are available to measure the current sheet thickness. Both cases are examined below. 

First, we employ the thickness of the outer current sheet defined by label C in Figure 18(a), which is measured to be approximately $y=2$. In addition, the local maximum point of $-\Delta V_y$ is located near $y=0.34$ at $t=312$, as shown in Figure 20(c). The resulting $\xi_b$ is roughly estimated to be $0.2=1.307*0.34/2$. Note that the factor "$1.307$" originates from the fact that the outer edge of the current sheet in Eq.(2.4) is located at $\xi=1.307$ on the $\xi$ scale. The $\xi_b$ value listed as the outer $\xi_b$ in Table 1 varies between $0.2$ and $0.4$. Because $\xi_b<1.307$ at all times, this result represents the case of the inner-triggered tearing instability. Considering the $\kappa$ listed in Table 1 and $\epsilon=0$, $\lambda_{up}<0.1$ is roughly estimated from Figure 8. Because $\lambda_{up}$ is the upper limit, $\lambda_{up}<0.1$ is too small to be consistent with the $\lambda_{B_y}$ and $\lambda_{\Delta V_y}$ measured in the MHD simulation.

Second, we employ the thickness of the inner current sheet defined by label B in Figure 18(a), which is measured to be approximately $y=0.5$. In the same manner as for the outer $\xi_b$, the resulting $\xi_b$ varies between $0.7$ and $1.4$, as listed in the inner $\xi_b$ of Table 1. Since $\xi_b=1.4$ at $t=336$ is slightly larger than $1.307$, this scenario may be classified as the outer-triggered case. Considering that $\lambda_{up}$ is the upper limit, $\lambda_{up}$ for the inner $\xi_b$ is consistent with the $\lambda_{B_y}$ and $\lambda_{\Delta V_y}$ measured in the MHD simulation. Finally, none of the growth rates obtained in the MHD simulation or modified LSC theory exceed unity, i.e., they are sub-Alfvenic. This result is reasonable for tearing instabilities driven by Alfven waves.

\subsubsection{Relation of $t_s$ (linear phase) and $t_v$ (linear $+$ nonlinear phases).}

As shown in Figures 12(a)-(c), the tearing instability is impulsively repeated, where each tearing instability grows and then slows during elongation of the SP-like sheet. Then, the next tearing instability starts in the over-elongated SP-like sheet. Evidently, there is a time interval $t_v$ for repeating each tearing instability and a duration time $t_s(<t_v)$ for which the modified LSC theory is applicable, i.e., a linear phase. For the second tearing instability, $t_v$ is roughly measured to be $380-308=72$ because the second tearing instability starts at $t=308$ and the third tearing instability starts at $t=380$ which is not shown in this paper to reduce the number of Figures. Furthermore, $t_s$ is less than $348-308=40$ because $\lambda_{\Delta V_y}$ changes from positive to negative during $348<t<352$, as shown in Table 1. Thus, the linear growth of the tearing instability has been terminated until $t=348$. The factors that dominate $t_v$ are unclear, but $t_s$ will be regulated by two time scales, i.e., $l_{cs}/V_A$ and $l_{cs}/C_s$, where $C_s$ is the local speed of sound in the current sheet. In other words, as Alfven and sound waves spread over the wavelength of the plasmoid chain \citep{Shibata2001}, the tearing instability will shift from the linear phase to the nonlinear phase, where the linear phase is explained by the modified LSC theory and the nonlinear phase may be explained by the SP model. In this MHD simulation, the PK model is not observed in the nonlinear phase. According to Table 1, $t_{s}=l_{cs}/V_A$ varies between $41/5.5=7.5$ at $t=312$ and $84/5.5=15$ at $t=360$. In fact, as shown in the time variation of $\lambda_{\Delta V_y}$ in Table 1, the second tearing instability, which started around $t=308$, rapidly slows during $308+7.5<t<308+15$. Then, as the SP-like sheet elongates along the current sheet, $\lambda_{up}$ for the inner $\xi_b$ and $\lambda_{\Delta V_y}$ gradually separate. In contrast to $\lambda_{\Delta V_y}$, $\lambda_{B_y}$ does not drastically decrease because the reconnection process itself is maintained by SP-like sheet formation. This discussion is further continued in Section 4.1. Finally, beyond $t=364$, $\lambda_{B_y}$ and $-J_z$ start to decrease, leading to the start of the third tearing instability. Since the MHD simulation studied in this paper is for the low beta plasma, the discussion based on $t_{s}=l_{cs}/C_s$ is not substantially different from that of $t_{s}=l_{cs}/V_A$ mentioned above.

\subsubsection{The first and third tearing instabilities.}

Finally, let us briefly examine the first and third tearing instabilities observed in the MHD simulation. As shown in Figures 12(a)-(c), the growth of the first tearing instability at the origin in $4<t<280$ is much slower than that of the second tearing instability in $308<t<348$. According to the modified LSC theory proposed in this paper, this slow growth can be explained by the extremely large $L_{cs}$ measured at the origin. In fact, the $\partial V_x/\partial x$ at the origin is smaller than that of the second tearing instability. The resulting $\kappa$ is then larger, leading to the slower growth rate predicted from Figure 8. The details of the first tearing instability are summarized in Appendix D.

Conversely, the growth of the third tearing instability at $380<t<420$ is faster than that of the second tearing instability, which is not shown in this paper to suppress the number of figures. However, as reported in many MHD simulations of PI \citep{Lour2009a,Bhat2009,cask2009,LandiApj2015,shi2017,PapiniApj2019}, as the tearing instability is repeated, the wave length $l_{cs}$ of the plasmoid chain gradually becomes shorter and the growth becomes faster. According to the modified LSC theory, this rapid growth can be explained by the small $L_{cs}$. In addition, the relatively small $l_{cs}$ enhances the growth because the realistic growth rate $\Lambda$ defined as $B_{ymax}(t+\Delta T)/B_{ymax}(t)=e^{2\pi \lambda \Delta T V_A/l_{cs}}=e^{\Lambda \Delta T}$ tends to be larger for smaller $l_{cs}$ evenwhen $\lambda$, $V_A$ and $\Delta T$ are constant 
\citep{Tajima2002}. 
Then, as PI nonlinearly proceeds, $l_{cs}$ tends to be much smaller in the subsequently repeated tearing instabilities. Hence, the subsequent tearing instabilities tend to grow at a higher $\Lambda$ evenwhen $\lambda$ is constant. 
This is, as PI proceeds, why each single event of the tearing instability tends to be gradually faster in the realistic time scale of the MHD simulation.

\section{Discussions}\label{Discussions}

\subsection{Application of the modified LSC theory for PI}

This section shows how the modified LSC theory proposed in Section 2 can be applied to PI. At this point, the modified LSC theory cannot be directly applied to PI. Furthermore, the original LSC theory\citep{Lour2007} is also inapplicable; thus, the existence of the critical Lundquist number $S_c$ is not directly supported by those two theories. If PI actually can occur, it may have to be studied as a nonlinear process, which unfortunately is not considered in this paper.

Figure 22 schematically shows a basic image of PI based on LSC theory. This image shows how a plasmoid chain appears in the SP sheet of length $L_{cs}'/2$, which is not necessarily equal to $L_{cs}/2$ defined in this paper, i.e., Figure 17. At this point, note that $\Gamma$ in Figure 22 is not necessarily equal to $\Gamma_0$ defined in Eqs.(2.5) and (2.6) because $L_{cs}'$ defined in the macroscopic SP sheet may be different from $L_{cs}$ defined in each single event of tearing instability. As shown in Figure 22(a), the tiny plasmoid (magnetic island) generated around the origin gradually grows and propagates downstream, i.e., in the $+x$ direction. This gradual growth is accompanied with the plasma outflow, which is shown as the solid oblique straight line in Figure 22(b), where $\Gamma=2V_A/\L_{cs}'$.

\begin{figure}
\begin{center}
\vspace{40mm}
\includegraphics[bb=0 0 512 512,width=0.6\hsize]
{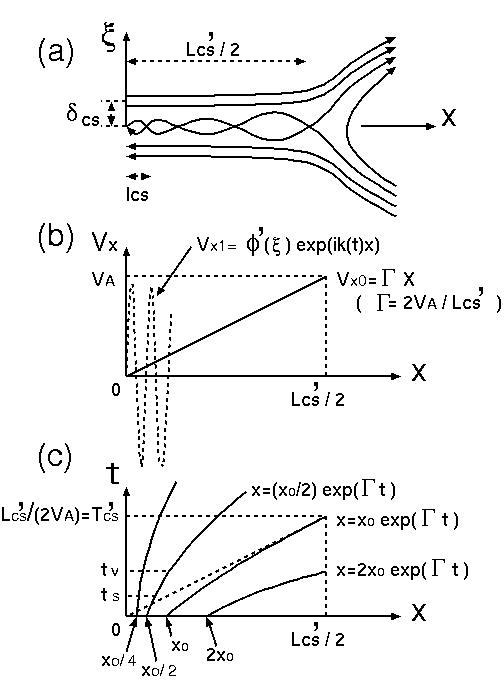}
   \caption{ Schematic images of Plasmoid Instability based on LSC theory.
 }
   \label{fig22}
\end{center}
   \end{figure}

In general, as long as the amplitude of the perturbation solution is kept to be sufficiently weaker than that of the zero-order equilibrium, each trajectory $x(t)$ of the X-points in the plasmoid chain will be traced as $dx(t)/dt=V_{x0}=\Gamma x(t)$. This process results in $x(t)=l_{cs} e^{\Gamma t}$, where $l_{cs}$ is defined as the wavelength of the plasmoid chain generated around the origin. For example, Figure 22(c) shows four trajectories, i.e., for $l_{cs}=x_0/4$, $x_0/2$, $x_0$, and $2x_0$, where $x_0$ is defined as $x_0=L_{cs}'/(2e^{\Gamma T_{cs}'})=L_{cs}'/(2e)=L_{cs}'/5.436$, where $T_{cs}'$ is defined as $T'_{cs}=L_{cs}'/(2 V_A)$. In other words, $T'_{cs}$ is defined as the travel time from the origin to $x=L'_{cs}/2$  
when a tiny plasmoid is assumed to move constantly at $V_A$, 
as shown in the dashed oblique straight line in Figure 22(c).

According to Loureiro's paper \citep{Lour2007}, $L_{cs}'=L_{cs}$, i.e. $\Gamma=\Gamma_0$, is assumed. Then, Eqs.(\ref{phi-eq}) and (\ref{psi-eq}) are established when $\Gamma=2V_A/L_{cs}'$ is sufficiently smaller than the growth time and $l_{cs}$ is almost constant in time. These conditions correspond to the case of $l_{cs}<<L_{cs}'$, e.g., $l_{cs}=x_0/4$ and $x_0/2$ shown in Figure 22(c), i.e., $l_{cs}< x_0=L'_{cs}/5.436$. As shown in the $x_0/4$ trajectory of Figure 22(c), since the movement of the X-point is very slow, the plasmoid remains in the SP sheet for a long time until it reaches $x=L_{cs}'/2$. In fact, as shown in Section 3, the X-point of every tearing instability remains almost still until each tearing instability is terminated by nonlinear saturation, i.e., the over-elongation of the SP sheet. Thus, $t_{s}=l_{cs}/V_A<<T_{cs}'$. Note that, as $l_{cs}$ becomes shorter, $t_{s}$ also tends to become shorter.

Two remarkable points are noted for the reality of the basic image of PI shown in Figure 22. First, as shown in Section 3, $L_{cs}$ and $l_{cs}$ strongly depend on each single event of tearing instability. For example, when the first tearing instability is terminated, the perturbed amplitude $V_{x1}$, which is shown as the dashed line in Figure 22(b) and corresponds to $\phi$, becomes almost as large as the zero-order amplitude of $V_{x0}$, which is shown as the solid oblique straight line in Figure 22(b). In fact, as will be shown in Appendix D, the $V_x$ profiles in Figure 17 are largely different from those in Figure 27. Hence, $L_{cs}$ for the first tearing instability may be equal to $L_{cs}'$ defined in Figure 22(a) but is evidently not applicable for the second tearing instability: $L_{cs}$ for the second and subsequent tearing instabilities must be individually measured. Furthermore, $l_{cs}$ must also be individually measured for each tearing instability. Hence, after the first tearing instability is terminated, the total length $L_{cs}'$ of the macroscopic SP sheet does not affect how each subsequent tearing instability occurs. Hence, $L_{cs}$ is more important than $L_{cs}'$. 

The second remarkable point is that, as mentioned in Section 2, $\lambda_{up}$ obtained for $\epsilon<0.1$ and $\kappa<5$ is almost independent of $\epsilon$, i.e., the resistivity. In addition, even for $\epsilon=0$, $\lambda_{up}$ takes a finite value less than unity. Note that the $\xi$ scale is normalized by the current sheet thickness; thus, the tearing instability in the real time and space of the MHD simulations follows the similarity law. In other words, the MHD simulation based on finite resistivity is directly applicable to the much smaller resistivity case, where $l_{cs}$ and $L_{cs}$ are much smaller for the smaller resistivity, and the time intervals of $t_{s}$ and $t_v$ will also become much smaller. Moreover, note that the Alfven speed $V_A$ in the upstream magnetic field region and the local speed of sound $C_s$ in the current sheet do not change, even in the limit of $\epsilon=0$. Hence, the case of much smaller resistivity, i.e., much higher Lundquist number, will not essentially change how PI occurs, at least on the basis of the modified LSC theory. Thus, the existence of $S_c$ is not directly supported by the modified LSC theory, i.e., linear theory.

\subsection{Comparison of the original and modified LSC theories for PI}

The growth rate $\kappa \lambda(=\gamma/\Gamma_0)$ of the original LSC theory\citep{Lour2007} is normalized by $L_{cs}'$ defined in Figure 22, as shown in FIG.4 in the paper, while $\lambda$ and $\lambda_{up}$ of the modified LSC theory, shown in Figure 8, are normalized by $l_{cs}$, which will be much smaller than $L_{cs}'$. Therefore, the growth rate derived in the original LSC theory may be assumed to be constant during $0<t<T_{cs}'$, i.e., until the plasmoid generated around the origin reaches the exit of the macroscopic SP sheet at $x=L_{cs}'/2$. However, as shown in Table 1, the growth rate measured in the MHD simulation immediately slows at $t>t_{s}$, where $t_{s}$ is much shorter than $T_{cs}'$. Thus, the growth rate $\kappa \lambda$ derived in the original LSC theory \citep{Lour2007} appears to be overestimated for the PI application, at least, in terms of the MHD simulation shown in this paper.

\subsection{ Viscosity effect}

Here, we discuss the viscosity effect, which is the largest difference between the theory and MHD simulation shown in this paper. Unfortunately, viscosity cannot be easily embedded in the modified LSC theory because Eqs.(\ref{phi-eq}) and (\ref{psi-eq}) are fairly complicated by viscosity. Moreover, the nonviscous MHD simulation results in a fatal numerical explosion due to the appearance of an extremely thin current sheet. Hence, rigorously, the modified LSC theory shown in this paper will be insufficient to be applied to the MHD simulations of PI. 

It may be additionally noted that the nonviscous tearing instability is similar to the nonviscous turbulence in normal fluid dynamics, in which the simulation numerically fails because of the appearance of unlimitedly short wave length. Note that, since the thickness of the current sheet in LSC theory is normalized by $\delta_{cs}$, it can have unlimitedly thin current sheet in real space of $(x,y)$ even in $\epsilon \ne 0$. In other words, the real thickness of the steady state SP sheet can be unlimitedly thin even in $\epsilon \ne 0$, i.e., the double limits of $\delta_{cs}=0$ and $L_{cs}=0$. 
In addition, Figures 7, 10(a) and (b) suggest that such a unlimitedly thin current sheet always has a unstable $\kappa$ range, in which $\lambda>0$ is established (also, see Appendix C for the stability criterion), with exception of that LSC theory fails at $\kappa =0$. Then, the finite viscosity can suppress the appearance of short wave length 
and will contribute to stabilize the tearing instability. 

Hence, the introduction of viscosity is very important for the realistic linear theory of the tearing instability, and also MHD simulations of PI. There are some studies  for the visco-resistive tearing instability \citep{TeneraniApj2015, BetarApj2020}, where the growth rate is observed to tend to decrease for a larger viscosity. Viscosity will be considered also in our next work.

\subsection{ FKR theory viewed from the perspective of modified LSC theory}

The upper limit $\lambda_{up}$ of the growth rate in FKR theory can be explored in consideration of the limit of $L_{cs}=+\infty$ in the modified LSC theory, because, $L_{cs}=+\infty$ results in $\partial V_x/\partial x=0$ at the X-point, which corresponds to the null-flow field of the equilibrium. Then, $L_{cs}=+\infty$ leads to $\kappa=\pi L_{cs}/l_{cs}=+\infty$. According to Figure 8, $\lambda_{up}$ tends to decrease as $\kappa$ increases; hence, $\lambda_{up}$ predicted in FKR theory will be close to zero. Thus, the tearing instability in FKR theory may be too slow in real cases. In other words, it may be important that the growth of the tearing instability is essentially dominated by the nonzero gradient $\partial V_x/\partial x$ at the X-point, i.e. the nonzero gradient of outflow speed along the neutral sheet. Further discussion of FKR theory is provided at the end of Appendices C and E.

\section*{Summary}\label{Summary}

In this paper, LSC theory \citep{Lour2007}  under rigorously uniform resistivity has been examined, where inner and outer regions were seamlessly solved as an initial value problem. In other words, we did not use the $\Delta'$ index which is widely used to connect the inner and outer regions (Appendix E). Then, it was shown that the upper limit $\lambda_{up}$ of the growth rate can be estimated given the wave number $\kappa$, resistivity $\epsilon$, and $\xi_b$, which is related to the local maximum point location of all physically acceptable solutions, including the zero-converging and zero-crossing solutions. In addition, it was shown that the growth rate $\lambda$ of the zero crossing solutions can be obtained from $\kappa$, $\epsilon$, and $\xi_c$, which represents the upstream condition, i.e.  the zero-crossing point. These results for $\lambda_{up}$ and $\lambda$ are summarized in Figure 8.

The MHD simulation presented in this paper is not new, i.e. it has been widely studied in many previous MHD studies. Rather, as shown in Table 1, here we showed that the modified LSC theory is applicable only to the beginning of each tearing instability, which exists in the linear phase. Then, when $t_s=l_{cs}/V_A$ is exceeded, the linear growth of the tearing instability terminates and enters the non-linear phase. In the MHD simulation presented in this paper, the nonlinear phase always resulted in the SP model, where the magnetic diffusion region was elongated along the current sheet. In other words, the nonlinear phase did not appear to result in the PK model, which is consistent with previous studies \citep{bskp1986}. Then, after an interval of time $t_v$ from the beginning, the next tearing instability started in the over-elongated SP-like sheet. How $t_v$ is determined is unclear, but the MHD simulation in this paper suggests $t_v>>t_s$. If $t_v>>t_s$ is always satisfied, the modified LSC theory is directly inapplicable for PI because PI should be fully nonlinear.

Most importantly, the modified LSC theory can predict tearing instabilities in the zero resistivity limit, i.e., $S=+\infty$, as shown in Figure 8. In the limit, the current sheet becomes unlimitedly thin, but the tearing instability can still occur at a sub-Alfvenic growth rate, i.e., $\lambda_{up}<1$. Apparently, the zero resistivity "limit" case is different from the "perfect" zero resistivity case, in which the reconnection process is completely stopped. Additionally, it should be noted that the tearing instability essentially follows the similarity law in realistic time and space. Hence, the tearing instability in the zero resistivity limit will be similar to that in the finite resistivity case shown in Section 3. In other words, evenwhen $S$ increases to $+\infty$, no drastic changes appear in the tearing instabilities of PI, at least, in term of linear theory. 

\appendix

\section{Zero-converging solutions in modified LSC theory}\label{appA}

Moreover, let us extend the modified LSC theory. For $\phi$ and $\psi$ to be physically acceptable, it is preferable that they converge to zero at $\xi=+\infty$. However, in Section 2, we avoided finding such $\phi$ and $\psi$. 
In this section, let us attempt to find such zero-converging solutions 
by modifying $f(\xi)$. In LSC theory,  $f(\xi)$ is assumed to be constant for $\xi>\xi_0$, but in this section, $f(\xi)$ is modified to be constant for $\xi>\xi_1$, where $\xi_1=\xi_0(=1.307)$ is not necessarily assumed. In this case, we can easily obtain the following analytical solutions for Eqs.(\ref{phi-eq}) and (\ref{psi-eq}) in $\xi>\xi_1$. 

\begin{eqnarray}
\phi(\xi)= C e^{\beta \xi} +D e^{-\beta \xi} \\
\psi(\xi)= E e^{\beta \xi} +F e^{-\beta \xi}
\end{eqnarray}

If $C=E=0$ and $\beta>0$, both $\phi$ and $\psi$ converge to zero at $\xi=+\infty$. The $C$ and $E$ values are determined by the values of $\phi(\xi_1)$, $\psi(\xi_1)$, $\phi(\xi_1)/d\xi$ and $\psi(\xi_1)/d\xi$. Then, inserting Eqs.(A1) and (A2) into Eqs.(\ref{phi-eq}) and (\ref{psi-eq}), $C=E=0$ requires that the following three conditions are simultaneously satisfied. 

Condition 1: 
\begin{eqnarray}
(d\phi(\xi_1)/d\xi)/\phi(\xi_1)=-\sqrt{G_1}, -\sqrt{G_2} 
\end{eqnarray}

Condition 2: 
\begin{eqnarray}
(d\psi(\xi_1)/d\xi)/\psi(\xi_1)=-\sqrt{G_1}, -\sqrt{G_2} 
\end{eqnarray}

Condition 3:
\begin{eqnarray}
\phi(\xi_1)/\psi(\xi_1)=f(\xi) \kappa \lambda/(f^2(\xi) \kappa
         +\lambda \kappa^2 \epsilon^2-\lambda G_2), \\ \nonumber
 f(\xi) \kappa \lambda/(f^2(\xi) \kappa
         +\lambda \kappa^2 \epsilon^2-\lambda G_1) 
\end{eqnarray}

where $G_1$ and $G_2$ are as follows. 

\begin{eqnarray}
G_1=\kappa^2 \epsilon^2+(\kappa/\lambda) (\lambda^2+f^2(\xi)), \\
G_2=\kappa^2 \epsilon^2
\end{eqnarray}

We recognize that there are two types of solutions as for $G_1$ and $G_2$. Hence, the general solutions will have the form of linear combinations of those types. However, let us only examine either type to confirm the existence of the zero-converging solutions. 
For $C=E=0$, either of $G_1$ or $G_2$ for Eqs.(A3), (A4), and (A5) must be satisfied by adjusting $\lambda$, $\xi_1$, and $\phi'(0)$ for a given set of $\kappa$ and $\epsilon$. Figure 23 shows an example for $G_2$. Hence, $\phi$ and $\psi$ shown in Figure 23 simultaneously converges to zero at $\xi=+\infty$, where $f(\xi)$ is set to be constant in $\xi>\xi_1=2.55625$, as shown at the bottom panel of Figure 23. Inevitably, $f'(\xi)$ is discontinuous at $\xi=2.55625$. Since Conditions 1, 2, and 3 can be exactly satisfied by adjusting $\lambda$, $\phi'(0)$ and $\xi_1$, the zero-converging perturbation solution actually exists.

Note that Figures 1-11 shown in Section 2 are inapplicable for this zero-converging solution because $f(\xi)$ has been modified in $\xi>\xi_0$. However, the growth rate $\lambda=0.868631$ in Figure 23 is less than $1.0$, i.e., it is sub-Alfvenic. In addition, the local maximum points of $\phi$ and $\psi$ are, respectively, located at $\xi=1.6005$ and $1.67275$, i.e., between $\xi_0$ and $\xi_1$, which is located in the reversed current sheet. This condition is the case for the outer-triggered tearing instability.

\setcounter{figure}{22}
\renewcommand{\thefigure}{\arabic{figure}}

\begin{figure}
\begin{center}
\includegraphics[bb=0 0 512 512,width=0.5\hsize]
{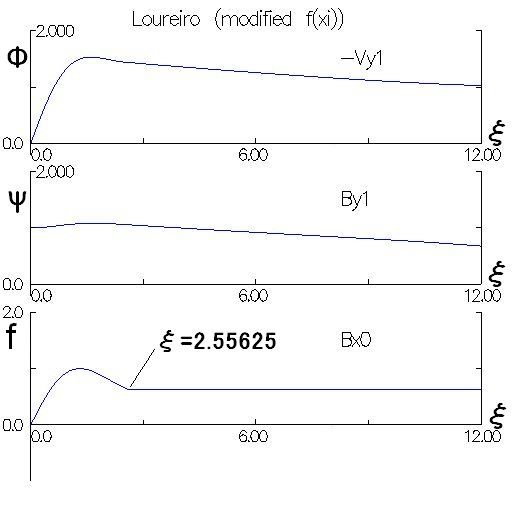}
   \caption{ $\phi$ (upper panel) and $\psi$ (middle panel) of the zero-converging solution for $\kappa=\epsilon=0.2$, $\lambda=0.868631$, $\phi'(0)=1.79892$, where $f(\xi)$ (lower panel) switches at  $\xi_1=2.55625$ from Eq.(2.4) to a constant value $f(\xi_1)$ that is not equal to Eq.(2.9). 
 }
   \label{fig23}
\end{center}
   \end{figure}

\section{Pinching method to find zero-crossing solution}

This is a very primitive and steady technique, as shown in Figure 24. 
The format of Figure 24 is similar to the upper and 
middle panels of Figure 1(a) 
but this figure schematically shows how to numerically 
find the zero-crossing solution. 
In each panel of Figure 24, three lines labeled as "1", "2", and "3" 
correspond to $\phi$ and $\psi$ solved for three different values of 
a control parameter, e.g., $\lambda_1$, $\lambda_2$, and $\lambda_3$. 
In this case, the zero-crossing solution is established 
at $\lambda=\lambda_2$, 
i.e., line "2", but it is impossible to exactly find line "2" 
by solving the numerical initial value problem 
because of inevitable numerical errors of the numerical study.

Rather, if lines "1" and "3" are found, and 
the transition from line "1" to "3" is assumed to be continuous, 
line "2" between lines  "1" and "3" should actually exist, 
Because, as $\lambda$ changes from $\lambda_1$ to $\lambda_3$, 
the zero-crossing point of $\phi$ shifts 
from $\xi_{\phi1}$ to $\xi_{\phi3}$ and 
the zero-crossing point of $\psi$ shifts 
from $\xi_{\psi1}$ to $\xi_{\psi3}$. 
It is important that those zero-crossing points of $\phi$ and $\psi$ 
shift in the opposite direction.

\setcounter{figure}{23}
\renewcommand{\thefigure}{\arabic{figure}}

\begin{figure}
\begin{center}
\includegraphics[bb=0 0 512 512,width=0.75\hsize]
{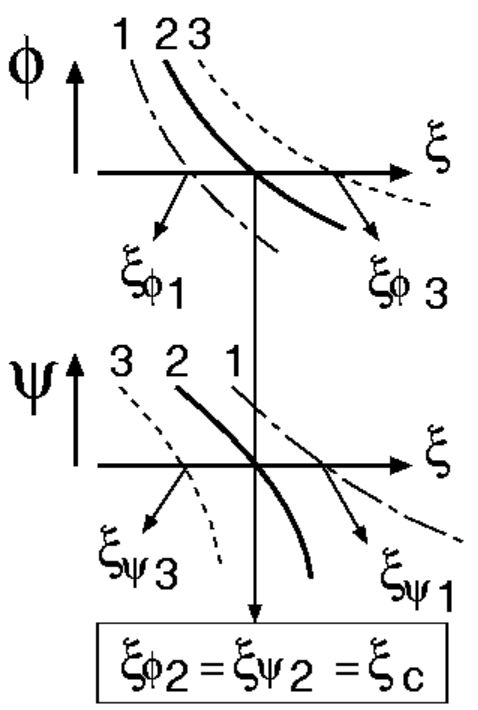}
   \caption{ Schematic image of the pinching method employed to find zero-crossing solutions, for which $\phi=\psi=0$ at $\xi=\xi_c$.}
   \label{fig24}
\end{center}
   \end{figure}

\section{$\lambda=0$ solutions of Eqs.(\ref{phi-eq}) and (\ref{psi-eq})}

If there is a critical condition between stable and unstable modes for tearing instability, it will be worth finding $\phi$ and $\psi$ solutions in $\lambda=0$ limit. Taking the limit of $\lambda=0$, Eqs.(\ref{phi-eq}) and (\ref{psi-eq}) are modified, 
as follows. 

\begin{eqnarray}
\psi''=(\kappa^2\epsilon^2+f''(\xi)/f(\xi))\psi, \\
\phi=-(f''(\xi)/\kappa f^2(\xi))\psi
\end{eqnarray}

Eq.(C1) can be numerically resolved as the initial value problem 
with $\psi(0)=1$ and $\psi'(0)=0$. 
Then, $\phi$ is directly obtained from Eq.(C2). 
Figure 25 shows the result for wave number $\kappa=5.4575$ and resistivity 
$\epsilon=0.2$, which is a zero-crossing solution, 
because $\psi=0$ at $\xi=5$ and $\phi=0$ is constantly kept in $\xi>1.307$. 
However, Eq.(C2) does not satisfy $\phi(0)=0$, 
because $\phi(0)$ diverges at the origin, 
as shown in the upper panel of Figure 25. 
In addition, $\phi$ generally has a discontinuity at $\xi=1.307$, 
because $f''(\xi)$ has a discontinuity, which disappears for $\kappa=3.215$ 
because $\xi=1.307$ becomes the zero-crossing point of $\phi$ and $\psi$. 
It is important that, as long as $\kappa \epsilon=1.0915$ is kept, 
the zero-crossing solution is kept for any $\kappa$ and $\epsilon$, 
because Eq.(C1) is not changed. Hence, in $\epsilon=0$ limit, 
 the zero-crossing solution shifts to $\kappa=+\infty$.

\setcounter{figure}{24}
\renewcommand{\thefigure}{\arabic{figure}}

\begin{figure}
\begin{center}
\vspace{20mm}
\includegraphics[bb=0 0 512 512,width=0.5\hsize]{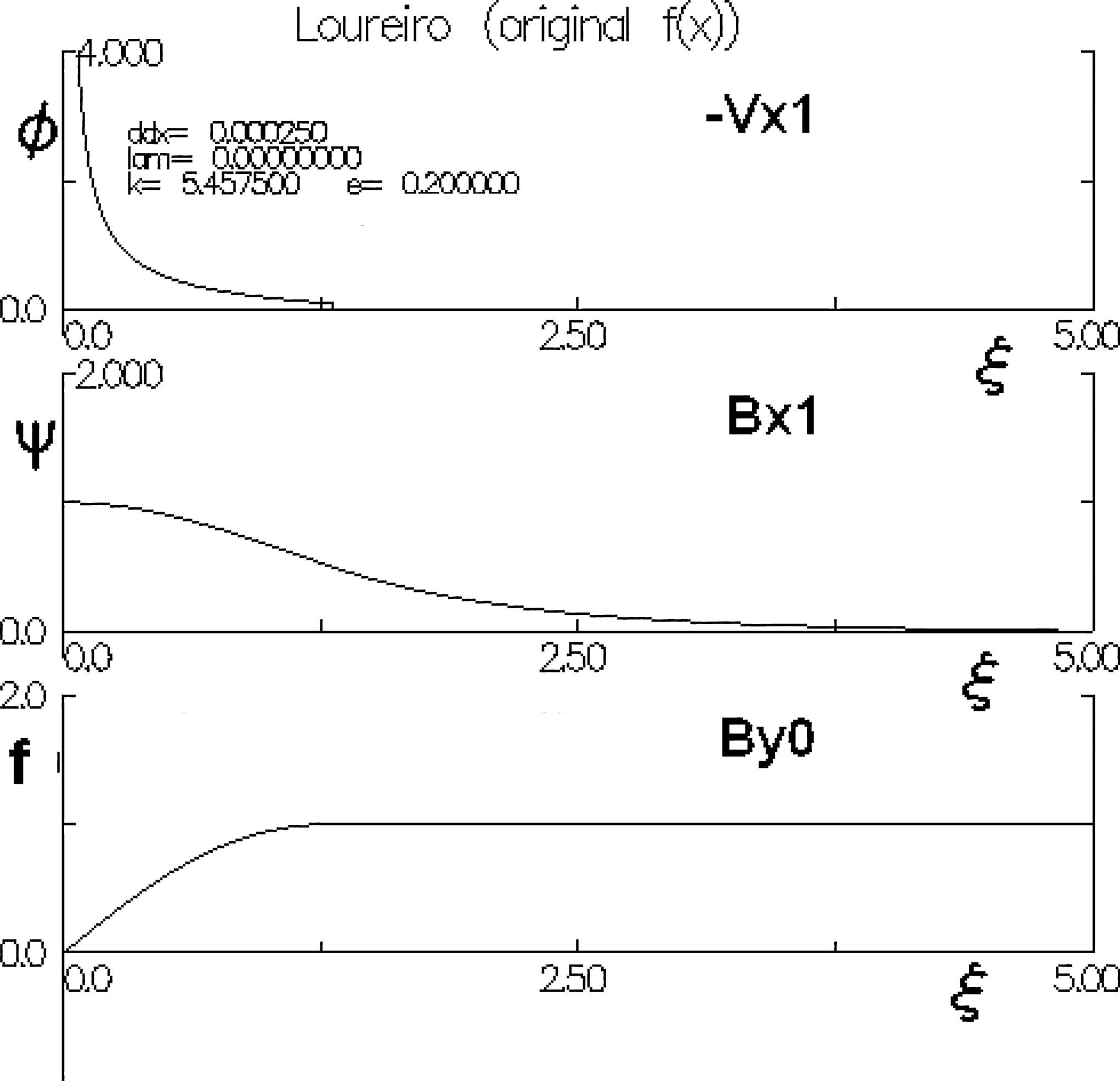}
   \caption{ $\phi$ (upper panel) and $\psi$ (middle panel) of $\lambda=0$ limit solution for $\kappa=5.4575$ and $\epsilon=0.2$, where $\psi$ has a zero-crossing point at $\xi=5$ and $\phi=0$ is kept in $\xi>1.3069$. Because of $\phi(0)=+\infty$ in the upper panel, this is not the physically acceptable solution 
that must be started from $\phi(0)=0$. In addition, $\phi$ has a discontinuity at $\xi=1.307$ but is almost invisible in this figure.  }
   \label{fig25}
\end{center}
   \end{figure}

To compare with Figure 25, 
Figures 26(a) and (b) show the solutions numerically obtained from Eqs.(\ref{phi-eq}) and (\ref{psi-eq}) in the same manner as Figures 1 and 2. These are also obtained for 
 $\kappa=5.4575$ and $\epsilon=0.2$, where 
the profiles of labels a and b in these Figures are drastically 
diverged to either of positive or negative infinity 
beyond a $\xi$ value for extremely small difference of $\lambda$. 
Figures 26(a) and (b) are respectively obtained for $\phi'(0)=10$ and $50$. 
It may be expected that the zero-crossing solutions and 
zero-converging solution are found in the extremely narrow range 
of $\lambda$, if they exist. 
At this point, it may be remembered that 
zero-converging solution was predicted between labels b and d in Figure 1(a), 
i.e., $0.2<\lambda<0.5$. 
Unfortunately, due to the numerical error, 
we cannot deeply explore the extremely narrow range. 
The zero-crossing solutions deduced in Figures 26(a) and (b) appear to be 
 close to that in Figure 25, suggesting that $\phi'(0)=+\infty$ limit 
of Eqs.(\ref{phi-eq}) and (\ref{psi-eq}) corresponds to Figure 25. 
Those three figures are equally obtained for $\kappa \epsilon=1.0915$.

\setcounter{figure}{25}
\renewcommand{\thefigure}{\arabic{figure}(\textbf{a})}

\begin{figure}
\begin{center}
\vspace{20mm}
\includegraphics[bb=0 0 512 512,width=0.5\hsize]{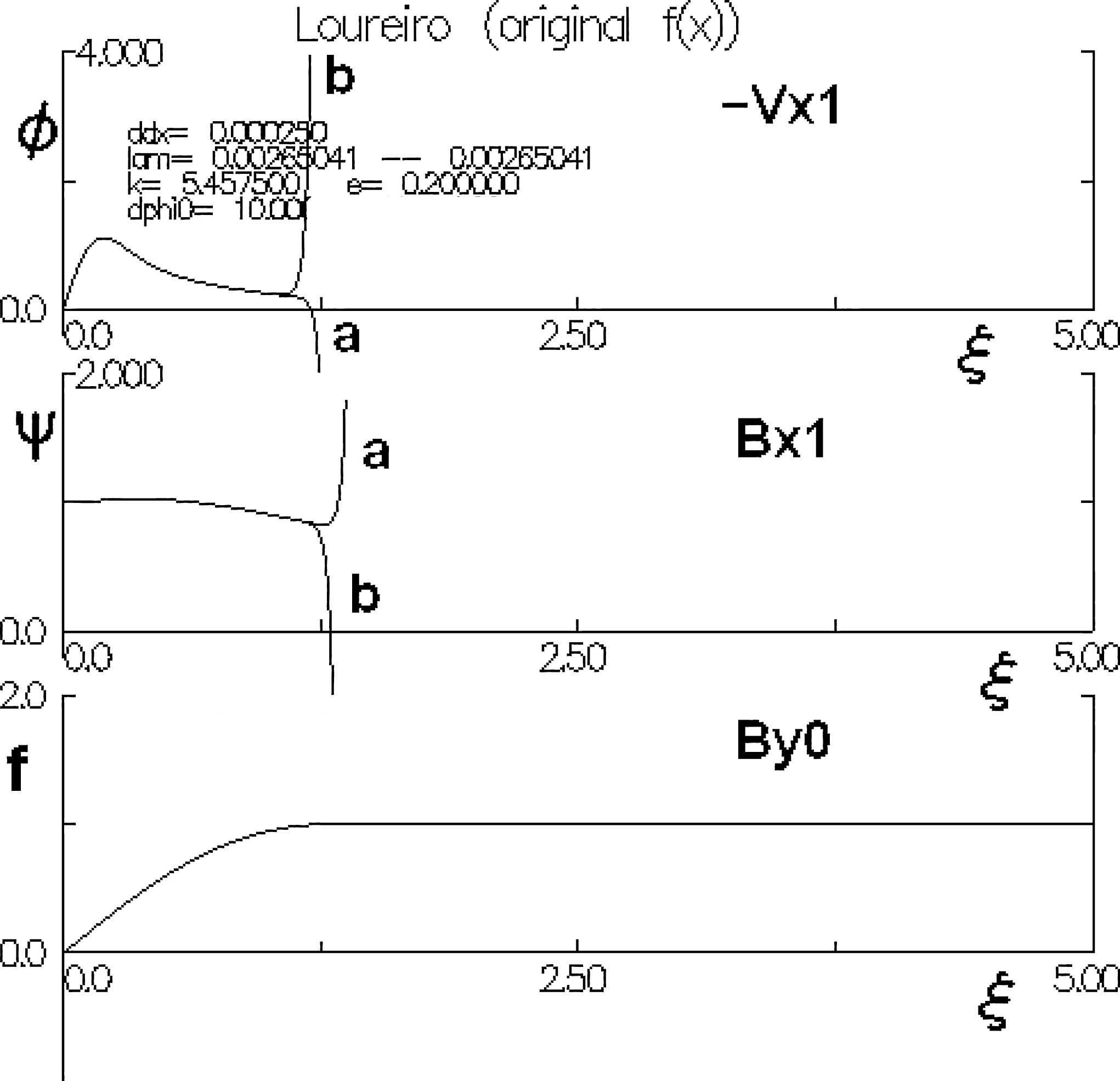}
   \caption{ $\phi$ (upper panel) and $\psi$ (middle panel) obtained for $\kappa=5.4575$, $\epsilon=0.2$, $\phi'(0)=10$, and $\lambda=$  $0.0026504125155326$ (label a), and $0.0026504125155327$ (label b). If present, the zero-converging solution and zero-crossing solutions will exist between these labels. 
 }
   \label{fig26a}
\end{center}
   \end{figure}

\setcounter{figure}{25}
\renewcommand{\thefigure}{\arabic{figure}(\textbf{b})}

\begin{figure}
\begin{center}
\vspace{20mm}
\includegraphics[bb=0 0 512 512,width=0.5\hsize]{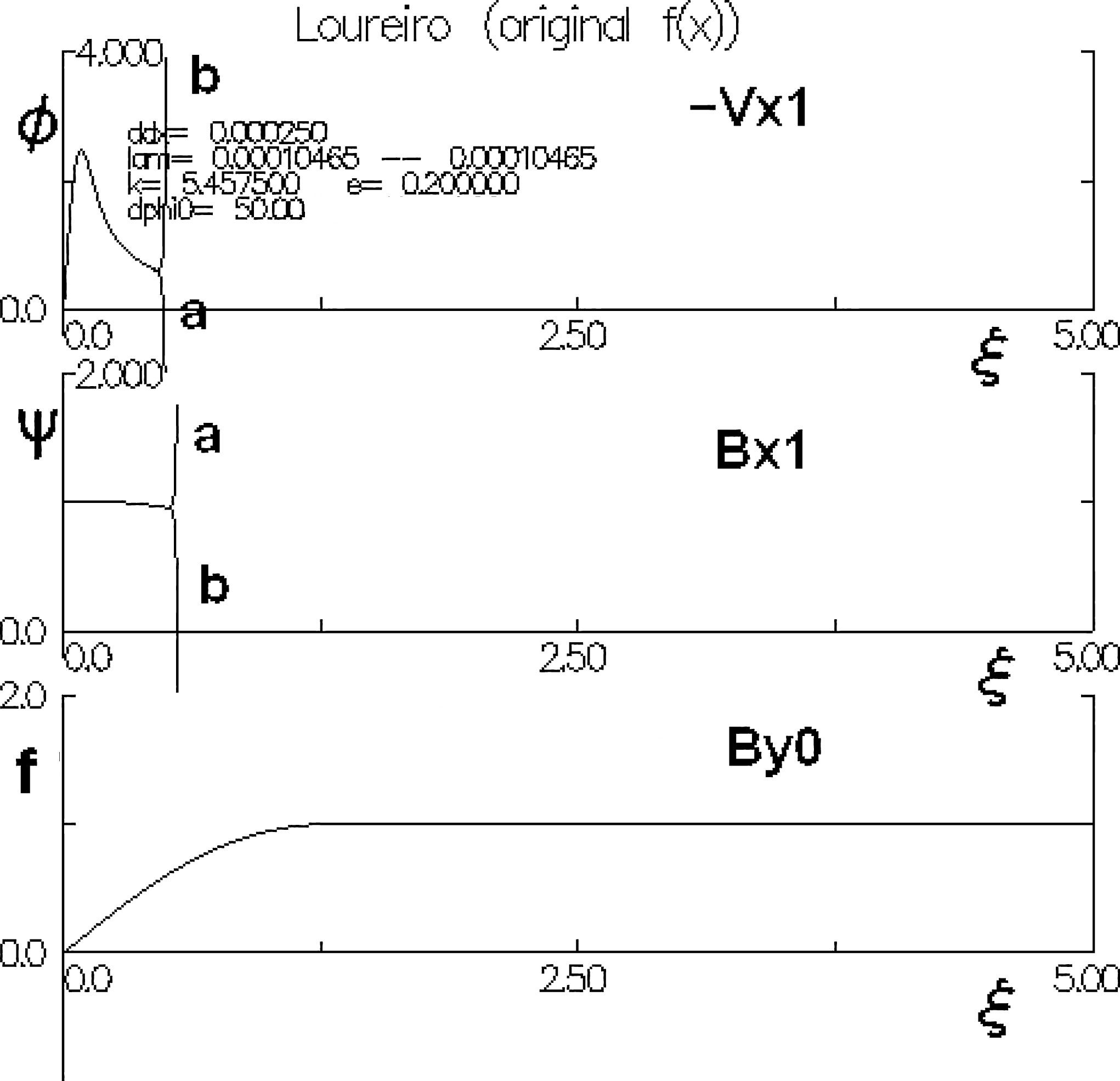}
   \caption{ 
$\phi$ (upper panel) and $\psi$ (middle panel) obtained for $\kappa=5.4575$, $\epsilon=0.2$, $\phi'(0)=50$, and $\lambda=$  $0.000104651516627333$ (label a), and  $0.000104651516627334$ (label b). If present, the zero-converging solution and zero-crossing solutions will exist between these labels. Comparing Figs.26(a) and (b), Fig25 will correspond to the case of the $\phi'(0)=+\infty$ limit. 
 }
   \label{fig26b}
\end{center}
   \end{figure}

For $\kappa \epsilon<1.0915$, as shown in Section 2, 
we can easily find zero-crossing solutions while it is unclear 
whether zero-crossing solutions for $\kappa \epsilon>1.0915$ exist. 
Hence, it is unclear whether the tearing instability in $\kappa \epsilon>1.0915$ is stable or unstable. If it is stable, the critical condition of $\kappa \epsilon(=2\pi \delta_{cs}/\l_{cs})=1.0915$ seems to be similar to that of FKR theory, although the growth rate in FKR theory may be zero. Because, 
as discussed in Section 4.4, 
FKR theory corresponds to the $L_{cs}=+\infty$ limit of LSC theory. 
In other words, FKR theory corresponds to the double limits of 
$\epsilon=2\delta_{cs}/L_{cs}=0$ and $\kappa=\pi L_{cs}/l_{cs}=+\infty$ 
in LSC theory. 

Let us discuss about perspectives for FKR theory from LSC theory, moreover. 
In general, the perturbation theory must start from a rigorous equilibrium 
but the current sheet of finite thickness assumed in FKR theory 
cannot be equilibrium for the null flow field in the resistive MHD, i.e. 
$\epsilon \neq 0$. There are the following strategies to avoid this problem.

First, if $\epsilon=0$ limit, i.e., ideal-MHD limit,  
is taken, the current sheet can be the rigorous equilibrium 
for the null flow field. 
In fact, with the exception of the $\delta_{cs}=+\infty$ limit, 
the $\epsilon=2\delta_{cs}/L_{cs}=0$ limit 
is established for the $L_{cs}=+\infty$ limit 
which has been discussed above. 
In this case, the tearing instability in FKR theory 
will occur for $\kappa \epsilon(=2\pi \delta_{cs}/\l_{cs})=0<1.0915$. 
However, as mentioned above, the growth rate will be close to zero. 
Note that the $\epsilon=0$ limit means 
when the thickness $\delta_{cs}$ of the current sheet 
is extremely thin with respect to $L_{cs}$. 

Second, let us consider when the thickness  $\delta_{cs}$ 
of the current sheet is extremely thick, i.e., 
the case of $\delta_{cs}=+\infty$ limit. 
Such a current sheet of infinite thickness seems to be a tricky consideration 
but ,in fact, it has been discussed \citep{Cross1971} 
though potentially remaining a controversial topic. 
E.g., since Eq.(2.4) in the vicinity of the neutral sheet, 
i.e. $\xi=0$, is almost a linear function with respect to $\xi$, 
if we focus only on the vicinity, 
the equilibrium of FKR theory can be rigorously established therein, 
in resistive MHD with the null flow field. 
As mentioned above, note 
that FKR theory requires $L_{cs}=+\infty$ limit in LSC theory. 
When the $L_{cs}=+\infty$ and $\delta_{cs}=+\infty$ limits are simultaneously 
taken, $\epsilon=2\delta_{cs}/L_{cs}=0$ limit is not necessarily required 
to keep the rigorous equilibrium. 
Moreover, it leads to $\kappa=\pi L_{cs}/l_{cs}=+\infty$ 
with the exception of $l_{cs}=+\infty$. 
It results in $\kappa \epsilon(=2\pi \delta_{cs}/\l_{cs})>1.0915$. 
Hence, the tearing instability in FKR theory will be 
stabilized for the current sheet of infinite thickness.

\section{The first tearing instabilities in MHD simulation}

Figure 27 shows the time variations of the x-profiles of $V_x$ along the x-axis, i.e., $y=0$ in $80<t<280$, in which the first tearing instability occurs, as shown in Figures 12(a) and (b). At this time, the X-point is constantly located at the origin. As shown in Figure 27, since the $V_x$ fluctuation observed during $0<x<30$ at $t=80$ is extremely weak, the $\partial V_x/\partial x$ measured at the origin results in approximately $L_{cs}=2000$, which is much larger than that of the second tearing instability shown in Table 1. Moreover, since the weak $V_x$ fluctuation is strongly localized around the origin, $l_{cs}$ at $t=80$ is approximately $30$, which is not excessively large. Hence, $\kappa$ reaches approximately $200$, which is very large. Unfortunately, Figure 8 does not cover the $\kappa$ range, but the predicted $\lambda_{up}$ will be extremely low. Hence, the first tearing instability develops slowly initially, i.e., $t<160$. However, as the first tearing instability becomes fully developed at $t>160$, the predicted growth rate accelerates because $L_{cs}$ drastically decreases to $200$, resulting in a substantial decrease in $\kappa$. As a result, the first tearing instability slowly proceeds in $t<160$ and gradually accelerates for $160<t<240$. Then, the instability is terminated around $t=240$, where the $L_{cs}$ switches from decreasing to increasing in time. This switching is the same as that observed at $t=356$ in Table 1. Finally, this process leads to the formation of an elongated SP-like sheet, as shown in Figure 12(b).

\setcounter{figure}{26}
\renewcommand{\thefigure}{\arabic{figure}}

\begin{figure}
%\begin{center}
\vspace{40mm}
\includegraphics[bb=0 0 256 256,width=0.25\hsize]{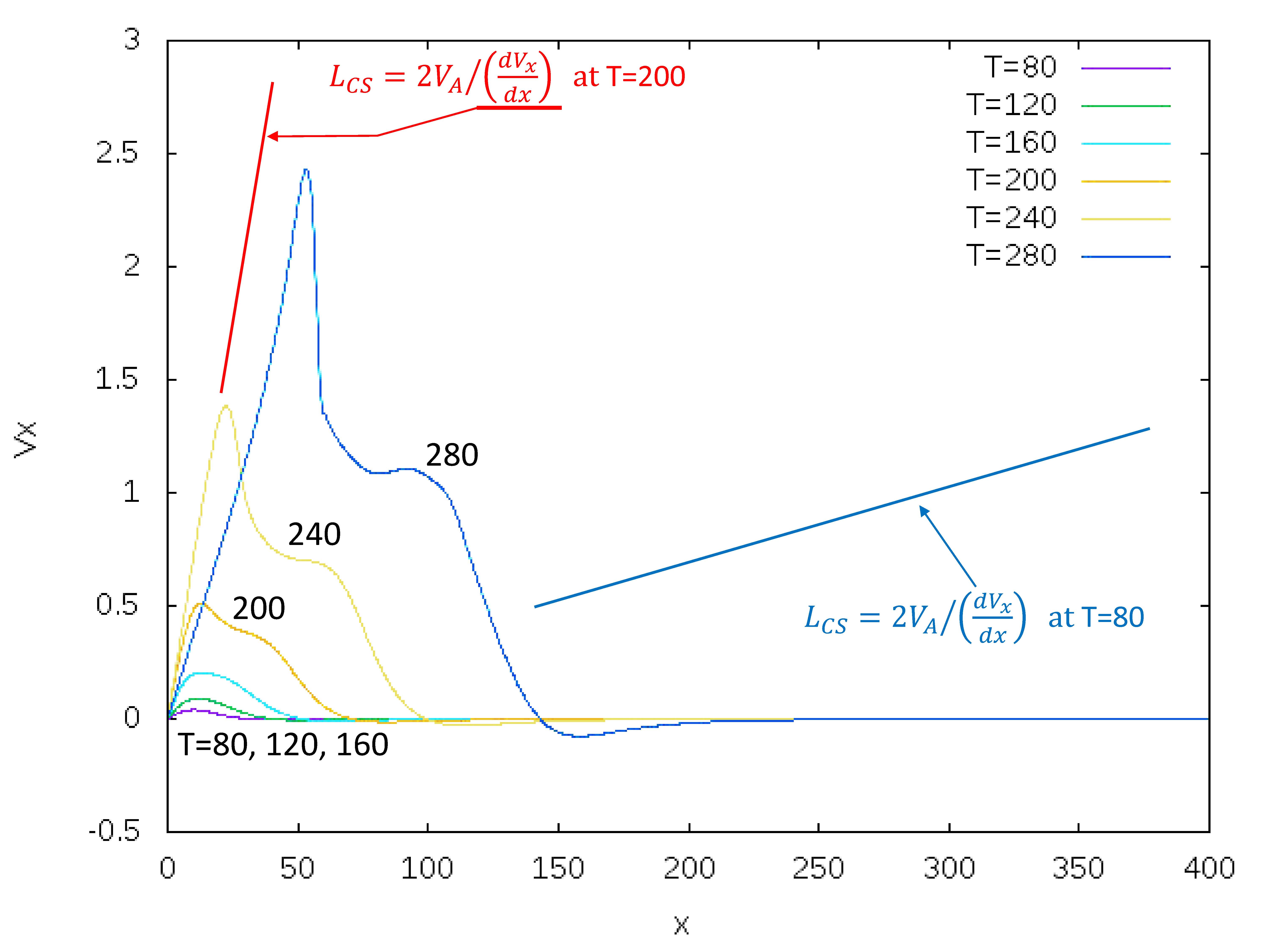}
   \caption{ The first tearing instability in $80<t<280$. The figure format is the same as that in Fig.17. 
$L_{cs}$ drastically changes between $t=80$ (blue slope) 
and $200$ (red slope), leading to the drastic change in the growth rate.  }
   \label{fig27}
%\end{center}
   \end{figure}

\section{ Previous studies without the use of $\Delta'$ index}

Traditionally, $\Delta'$ index is widely used to study FKR and original LSC theories, where the $\Delta'$ index is required to connect the inner and outer regions as a boundary condition. The outer region is solved in ideal-MHD, and 
the inner region is solved in resistive-MHD. 
In those theories, the inner region is assumed to have a discontinuity 
in the vicinity of the neutral sheet, i.e., $\xi=0$. 
In contrast, in this paper, without the use of the $\Delta'$ index, 
LSC theory was numerically solved as an initial value problem. 
It means that no discontinuity is assumed, and hence, 
the inner and outer regions are seamlessly solved in resistive-MHD. 

 Since there are numerous studies 
which have examined tearing instability throughout history, 
there should be previous studies without $\Delta'$ index. 
In fact, it seems that \citet{Cross1971} studied the growth rate of 
double tearing instability using the Fourier analysis technique 
without $\Delta'$ index. 
In the Cross's study, solutions of $B_{1y}$ and $v_{1y}$ were 
numerically found, in which the Fourier series converges 
in higher wave numbers. 
Those converging solutions will have no discontinuity 
at the magnetic resonance surface, i.e., $\xi=0$. 
However, the authors did not discuss about the discontinuity 
assumed in FKR theory. 
The non-use strategy of the $\Delta'$ index 
in the present paper may not be new. 
However, we would like to invoke that it is worth revisiting 
FKR and LSC theories without $\Delta'$ index.

\vspace{20mm}
{\bf Acknowledgements :}

The numerical calculations were performed on parallel computer systems at Kyoto and Nagoya University Data Processing Centers. The data analysis executed in this study was partially assisted by Mr. Katakami, Mr. Nishimura, and Mr. Fukumoto who were undergraduate students in the Research Center for Space and Cosmic Evolution (RCSCE) of Ehime University. 
Also, auxiliary researches to publish this paper were asisted by 
Dr. K. Fujimoto in Scholl of Space and Environment of Beihang University and 
Prof. K. Shibata in Doshisha University 
who moved from Kwasan Observatrory of Kyoto University. 
To check the numerical precision of the MHD simulations executed by 2 step Lax-Wendroff code, HLLD code \citep{shi2017} developed by Dr. S. Zenitani in Kobe University was used. The authors thank them for their assistances.

\bibliographystyle{jpp}

\bibliography{jpp-instructions}

\end{document}